\documentstyle[12pt]{report}
\input epsf

\def\smash{\mbox{$\,\rule{0.3pt}{1.1ex}\!\times\,$}}
\def \wtilde{\widetilde}
\def \tr {\triangleright}
\def \tl {\triangleleft}
\def \tw {\tilde{w}}
\def \ttheta {\wtilde{\theta}}

\def \beq{\begin{equation}}
\def \eq{\end{equation}}
\def \berr{\begin{eqnarray}}
\def \err{\end{eqnarray}}
\def \nn{\nonumber}
\def \a{\alpha}
\def \b{\beta}
\def \g{\gamma}
\def \d{\delta}
\def \r{\rho}
\def \s{\sigma}

\def \Del{\Delta}
\def \w{\omega}
\def \Om{\Omega}
                                 
\def \eps{\epsilon}
\def \vareps{\varepsilon}
\def \om{\omega}
\def \l{\lambda}
\def \dl{\partial}

\def \Alg{{\cal A}}
\def \C{{\cal C}}
\def \G{{\cal G}}
\def \M{{\cal M}}

\def \F{{\cal F}}

\def \U{{\cal U}}
\def \W{{\cal W}}
\def \R{{\cal R}}
\def \H{{\cal H}}
\def \B{{\cal B}}

\def \Par{\mbox{Par}}
\def \ch{\mbox{ch }}
                      
\def \({\left(}
\def \){\right)}
\def \<{\langle}
\def \>{\rangle}
\def \[{\left[}
\def \]{\right]}
\def \obar{\overline}
\def \cobar#1{{\overline{#1}}^c}
\def \robar#1{{\overline{#1}}^r}
\def \bobar#1{{\overline{#1}}^b}
\def \und#1{\underline{#1}}

\def\tens{\mathop{\otimes}}

\def\ttens{\tilde{\tens}}
\def\id{\rm id}
\newcommand \reals{I \! \! R}
\newcommand \compl{C \! \! \! \! {\scriptscriptstyle {}^{{}_|}}\ }
\newcommand \N{I \! \! N}
\newcommand \Z{Z \! \! \! Z}

\def\reps{representations }
\def\rep{representation }
                                              
\def\lform{\hbox{$\sqcup$}\llap{\hbox{$\sqcap$}}}
\def\eqn#1#2{\begin{equation}#2\label{#1}\end{equation}}
\def\proof{\goodbreak\noindent{\bf Proof\quad}}
\def\endproof{{\ $\lform$}\bigskip }

\newtheorem{prop}{Proposition}[section]
\newtheorem{theorem}[prop]{Theorem}
\newtheorem{lemma}[prop]{Lemma}
\newtheorem{definition}[prop]{Definition}

\newtheorem{conjecture}[prop]{Conjecture}

\def\Sq{S_q^{N-1}}
\def \dN{d^{N}\!}

\begin{document}

\begin{titlepage}
\begin{center}
\today          \hfill   LBNL-40356\\
                \hfill   UCB-PTH-97/28\\

\vskip .5in
{\large \bf Quantum Groups, Roots of Unity \\
           and \\ \vskip .05in
    Particles on quantized Anti--de Sitter Space}
\footnote{This work was supported in part by the Director, Office of Energy
Research, Office of High Energy and Nuclear Physics, Division of High
Energy Physics of the U.S. Department of Energy under Contract
DE-AC03-76SF00098 and in part by the National Science Foundation under
grant PHY-9514797.}

\vskip .5in

Harold Steinacker\footnote{email address: hsteinac@physics.berkeley.edu}

\vskip .5in

{\em Department of Physics\\
University of California, Berkeley, California 94720\\
and\\
Theoretical Physics Group\\
Ernest Orlando Lawrence Berkeley National Laboratory\\
University of California, Berkeley, California 94720}

\vskip .5in

A dissertation submitted in partial satisfaction of the\\
requirements for the degree of\\
Doctor of Philosophy in Physics.

\vskip .5in

Committee in charge:\\[.2in]
\parbox{6cm}{Professor Bruno Zumino, Chair\\
Professor Korkut Bardakci\\
Professor Nicolai Y. Reshetikhin}

\end{center}

\pagebreak

\begin{abstract}

Quantum groups in general and the quantum Anti--de Sitter
group $U_q(so(2,3))$
in particular are studied from the point of view of quantum field theory.
We show that if $q$ is a suitable root of unity, there exist
finite--dimensional, unitary representations corresponding to
essentially all the
classical one--particle representations with (half)integer spin,
with the same structure at low energies as in the classical case.
In the massless case
for spin $\geq 1$, the "naive" \reps are unitarizable only after
factoring out a subspace of "pure gauges", as classically.
Unitary many--particle representations are defined, with the correct
classical limit.
Furthermore, we identify a remarkable element $Q$ in the center of
$U_q(g)$, which plays the role of a BRST operator
in the case of $U_q(so(2,3))$ at roots of unity, for any spin $\geq 1$.
The associated ghosts are an intrinsic part of the indecomposable
representations. We show how to define an involution on
algebras of creation and anihilation operators at roots of unity,
in an example corresponding to non--identical particles.
It is shown how nonabelian gauge fields appear naturally
in this framework, without having to define connections on fiber bundles.
Integration on Quantum Euclidean
space and sphere and on Anti--de Sitter space is studied as well.
We give a conjecture how $Q$ can be used in general
to analyze the structure of indecomposable representations,
and to define a new, completely reducible associative (tensor) product of
representations at roots of unity, which generalizes the standard
"truncated" tensor product as well as our many--particle representations.
\end{abstract}
\end{titlepage}

%THIS FILE (PAGE ii) CONTAINS THE LBL DISCLAIMER

\renewcommand{\thepage}{\roman{page}}
\setcounter{page}{3}
\mbox{ }
\vfill
%\vskip 1in

\begin{center}
{\bf Disclaimer}
\end{center}

\vskip .2in

\begin{scriptsize}
\begin{quotation}
This document was prepared as an account of work sponsored by the United
States Government. While this document is believed to contain correct
information, neither the United States Government nor any agency
thereof, nor The Regents of the University of California, nor any of their
employees, makes any warranty, express or implied, or assumes any legal
liability or responsibility for the accuracy, completeness, or usefulness
of any information, apparatus, product, or process disclosed, or represents
that its use would not infringe privately owned rights.  Reference herein
to any specific commercial products process, or service by its trade name,
trademark, manufacturer, or otherwise, does not necessarily constitute or
imply its endorsement, recommendation, or favoring by the United States
Government or any agency thereof, or The Regents of the University of
California.  The views and opinions of authors expressed herein do not
necessarily state or reflect those of the United States Government or any
agency thereof, or The Regents of the University of California.
\end{quotation}
\end{scriptsize}

\vskip 2in

\begin{center}
\begin{small}
{\it Lawrence Berkeley Laboratory is an equal opportunity employer.}
\end{small}
\end{center}

\pagebreak

\renewcommand{\thepage}{\arabic{page}}
\setcounter{page}{1}

\section*{Acknowledgements}
%\addcontentsline{toc}{part}{Acknowledgements}

This thesis would not exist without the help of many
people. First and foremost, it is a pleasure to thank my advisor
Professor Bruno Zumino, who introduced me to this subject,
for his kindness and support, for sharing his amazing intuition and
insights, and especially for creating an atmosphere which is
productive and pleasant at the same time.
I consider myself very lucky having been his student.

I would also like to thank my colleagues Chong--Sun Chu, Pei-Ming Ho and
Bogdan Morariu for their friendship, and for uncountable
stimulating discussions.
The same goes for my fellow graduate students
Nima Arkani--Hamed, Luis Bernardo, Kamran Saririan and Zheng Yin,
who contributed much to the unique environment in Berkeley.
I also want to thank
Paolo Aschieri, Chris Chryssomalakos, Joanne Cohn, Julia Kempe,
Michael Schlieker, Peter Schupp, Paul Watts and Miriam Yee, for interesting
conversations and a good time.

Needless to say, I learned a lot from many Professors in Berkeley,
notably Korkut Bardakci and Nicolai Reshetikhin who were also on my
thesis committee, as well as Mary K. Gaillard, among others.
Professor Reshetikhin was able to explain to me
some of the more mysterious aspect of the subject.

Special thanks go to Laura Scott and Anne Takizawa, who made the bureaucratic
matters on campus almost a pleasure, as well as their counterparts
Barbara Gordon, Mary Kihanya and Louanne Neumann at LBL.

Generally, I am grateful for having had the opportunity to study in Berkeley,
which would not have been possible without the help of
Professors Josef Rothleitner and Dietmar Kuhn in Innsbruck,
and Gerald B. Arnold in Notre Dame.

Last but by no means least, I thank my parents Christoph and Irmgard
Steinacker for their support, faith and confidence.

This work was supported in part by the Director, Office of Energy
Research, Office of High Energy and Nuclear Physics, Division of High
Energy Physics of the U.S. Department of Energy under Contract
DE-AC03-76SF00098 and in part by the National Science Foundation under
grant PHY-9514797.

\tableofcontents
\pagebreak

\newpage
\renewcommand{\thepage}{\arabic{page}}
\setcounter{page}{1}

\chapter*{Introduction}
\addcontentsline{toc}{part}{Introduction}

The topic of this thesis is the study of quantum groups and quantum spaces
from the point of view of Quantum Field Theory.

The motivation behind such
an endeavour is easy to see. Quantum field theory (QFT)
is a highly successful
theory of elementary particles, with an embarrassing "fault":
except for some special cases, it cannot be defined without some sort of
a regularization of the underlying space,
which at present is little more than a
recipe to calculate divergent integrals. Physically, it is in fact
expected that space--time will not behave like a classical manifold below the
Planck scale, where quantum gravity presumably modifies its structure.
The hope is that this somehow provides a regularization for QFT.
From a mathematical point of view, it also seems that there should
exist some rigorous theory behind such "naive" quantum field theories,
given the rich mathematical structures apparently emerging from them.
And of course, it would be highly desirable to put the physically
relevant quantum field theories on a firm theoretical basis.

In view of this, it seems very natural that a consistent theory
of elementary particles should not be based on concepts of classical
geometry, but rather on some kind of
"fuzzy", or "quantized" space--time. With very little  experimental
guidance, finding the
correct description may seem rather hopeless. The approach we will
pursue here relies heavily on mathematical guidance,
given the "unreasonable usefulness
of mathematics in physics" (Wigner).
                                                               
With the development of Non--Commutative Geometry
in recent years \cite{connes}, a possible candidate for a new framework has
emerged. It is a generalization of the (rather old) idea that a manifold $\M$
can be described by the algebra of functions $Fun(\M)$ on it.
In Non--Commutative Geometry,  one considers instead some non--commutative
algebras replacing $Fun(\M)$, with sufficiently rich additional structures.
A "quantum deformation" or simply "deformation" of a classical manifold
is essentially a (noncommutative) algebra
with a deformation parameter $h$, such that the classical algebra of
functions on the manifold is obtained in the limit $h\rightarrow 0$.

This idea is in fact very familiar to physicists: Quantum Mechanics
can be viewed as a noncommutative geometry on phase space,
and the Planck constant
plays the role of the deformation parameter.
This example also shows that while the limit $h\rightarrow 0$ may be smooth
in some sense, the physical interpretation may be very different.
Furthermore if $h$ is dimensional, it is expected that
the "quantum" case should behave
classically at large enough scales.

This is a new and vast field of research, and to make any progress,
one clearly has to choose a particular approach. A simple example of a
theory of elementary particles on a noncommutative space was
proposed by Connes \cite{Co} and Connes \& Lott \cite{CoL}. It is based
on the space $M\times \Z_2$, where $M$ is ordinary Minkowski space and
$\Z_2$ is considered as a noncommutative space with a connection,
which can be interpreted as a Higgs field. This leads to a
new approach to the standard
model. Fr\"ohlich and collaborators  \cite{CFF} introduced gravity in this
context.
                    
Incidentally, it has been pointed out that string theory, a candidate for a
theory underlying quantum field theory including gravity, seems to
predict some noncommutativity of certain coordinate algebras
\cite{witten}. Other recent developments \cite{susskind}
also suggest some
relevance of Non-Commutative geometry to M--theory or string theory, which is
traditionally formulated in the language of classical geometry.

Of course one would really like to consider truly noncommutative spaces.
There have been many approaches to "quantize" physically relevant spaces
like Minkowski space, fiber bundles, and many others.
While many interesting examples have been found, a clear guideline is
missing.
                      
At this point, we want to emphasise (without the need to do so)
the importance of Lie groups in
elementary particle physics, notably the Poincare group
which dictates the behaviour of free particles,
and internal symmetry groups which may strongly
constrain their interactions.

Quantum groups are remarkable examples of
Noncommmutative Geometry, since they can be viewed as
deformations of classical Lie
groups resp. their manifolds. Their mathematical structure is well studied
and even richer than that of classical Lie groups. They depend on a deformation
parameter $q=e^h$, where $h=0$ corresponds to the classical case. Furthermore,
they act naturally on associated quantum spaces.
                                
Thus it seems that
quantum groups, which combine the features of both Lie groups and
Non--Commutative  Geometry in an analytic way,
should be a powerful guide towards a  realization of the above ideas.
In this thesis, we want to follow this approach and see where it leads to.

It is fair to say that quantum groups are analytic "deformations"
of classical groups, for "generic" $q$. However when $q$ is a root of unity,
their structure is in many ways very different,
and one is facing a truly new and very rich mathematical object.
One of the main points we want to make
is that the root of unity case seems to be the most interesting one
from a QFT point of view, beyond its known relevance to Conformal
Field Theory \cite{sierra,mack_schom}.
                                                                
We will mainly study the Anti--de Sitter group
$SO_q(2,3)$ and its representation theory,
which will play the role of the Poincare group. This choice
is vindicated by its simplicity and the wealth of interesting
features found, which constitute the main part of this thesis.
In the classical case, the Poincare group can be obtained
from $SO(2,3)$ by a contraction; however
we will not do a corresponding contraction in the quantum case
\cite{lukierski}, since our main results would all break down.

This thesis is organized as follows. Chapter 1 is a brief, general
introduction to quantum groups and their \rep theory, with
emphasis on those aspects which will be important later. Whenever possible,
a short explanation is included on how these facts can be obtained.
We will mainly work with the "quantized universal enveloping algebra"
$U_q(g)$. Most of this chapter is well known, but it also contains some
new results and definitions.
                                                   
In chapter 2, we consider quantum spaces associated to quantum groups,
and study integration on quantum Euclidean space, sphere, and
on quantum Anti--de Sitter (AdS) space. We point out that
quantum AdS space has an intrinsic length scale, above which it looks
like a classical manifold.

Chapter 3 starts with a brief review of the unitary \reps of the
classical AdS group corresponding to elementary particles, as well as
a discussion of massless particles, (abelian) gauge theories and BRST
from a group theoretic point of view. We then study these issues
for the quantum AdS group. In particular, we show that for suitable
roots of of unity $q$, there are
{\em finite--dimensional}, unitary \reps corresponding to all the
classical ones, with the same structure at low energies\footnote{For
the singleton representations, this was already shown in \cite{dobrev}.}.
In the massless case for spin $\geq 1$, the "naive" \reps
contain a subspace of "pure gauges" which must be
factored out to get unitary, irreducible representations, as classically.
A definition of unitary many--particle \reps is given.
Furthermore, we identify a remarkable element $Q$ of the center of
$U_q(g)$ which plays the role of a (abelian) BRST operator
in the case of the AdS group at roots of unity, for any spin $\geq 1$.
The corresponding ghosts are an intrinsic part of the indecomposable
representations.

In chapter 4, we give a conjecture that $Q$ can be used for any
group to understand the structure of
the tensor product at roots of unity, and to define a new,
associative (tensor) product of irreducible
representations at roots of unity, which generalizes the well--known
"truncated" tensor product used in conformal field theory, as well as the
many--particle \reps mentioned above.
We then show how one can define an involution on algebras
of creation and anihilation
operators, for $q$ a root of unity; lacking a symmetrization
postulate at present,
we have to work with a version corresponding to non--identical particles.
It is shown how all this might be
used towards constructing a quantum field theory.
The main missing piece to achieve this goal is a way to define identical
particles, i.e. a symmetrization postulate.
Finally, we point out that nonabelian gauge fields appear naturally
in this framework, without having to define something like
connections on fiber bundles.

%          24-MAR-1997 (harold) 
%-------------------------------------------------------------------
\chapter{Quantum Groups}

\section{Hopf Algebras and Quantum Groups}  \label{sec:quantum_groups}

In this thesis, we will be concerned with the \rep theory of 
quantum groups in the Drinfeld--Jimbo formulation, which is a certain
Hopf algebra with additional structure to be explained below.
The most economical approach to this goal would be
to start with these given mathematical objects, 
and study its properties from the point of view we have in mind.
However a reader who  is not very familiar with quantum groups
would be left in the dark wondering where all this comes from, and
quite possibly develop some misconceptions. Therefore we first give a 
brief review of the underlying mathematical structure. For more
details, the reader is referred to \cite{drinfeld,FRT} or a number 
of existing reviews, such as \cite{CH_P,jantzen}.

There are at least two ways to introduce quantum groups. 
One is to consider the universal enveloping algebra of a simple Lie group,
and discover a new structure on it, 
namely that of a quasitriangular Hopf algebra. 
The other, more geometric approach is to "quantize" the space of 
functions on a (compact) Lie group,
which turns out to have a remarkable Poisson--structure. 
These two approaches are dual to each other, and they
originated in the study of certain integrable models. 

\subsection{Hopf Algebras}

The mathematical language to describe both points of view is that of
a Hopf algebra. The most familiar example of a Hopf algebra $\Alg$ is the
space of functions $Fun(G)$ on a (compact) Lie group $G$. This is a
{\em commutative} algebra by pointwise multiplication, 
but this has nothing to do with the 
group structure. The group multiplication is encoded in $\Alg$ as a {\em 
coproduct}, which is a map $\Delta :\Alg \rightarrow \Alg \tens \Alg$,
where $(\Delta(f))(x,y) = f(x\cdot y)$ for $f \in Fun(G)$. The inverse is 
encoded as antipode $S:\Alg \rightarrow \Alg$ where $(S f)(x) =f(x^{-1})$ in 
the case of $Fun(G)$, and the unit element $e \in G$ becomes 
the counit $\eps :\Alg \rightarrow \compl$, with $\eps(f) = f(e)$ for 
$Fun(G)$.  In this way, all the structure of $G$ has been 
encoded in $Fun(G)$. In general,
a Hopf algebra is an algebra $\Alg$ with coproduct, antipode and counit
and the following compatibility conditions:
\berr
(\Delta \tens \id)\Delta (a) & = & (\id \tens \Delta )\Delta (a),  
   \quad \mbox{(coassociativity),}\\
\cdot (\eps \tens \id)\Delta (a) & = & \cdot (\id \tens \eps)\Del (a) = a,
  \quad  \mbox{(counit),}\\
\cdot (S \tens 1)\Delta (a) & = & \cdot (\id \tens S)\Delta (a)
= 1 \epsilon (a), 
  \quad \mbox{(coinverse),} \\
\Delta (a b) & = & \Delta (a) \Delta (b),\\
\epsilon (a b) & = & \epsilon (a) \epsilon (b),\quad\mbox{and}\\
\Delta (1) & = & 1 \otimes 1,\quad \epsilon (1) = 1,
\label{hopf_algebra}
\err
for all $a,b \in \Alg$. 
This implies
\berr
S(a b) & = & S(b) S(a), \quad \mbox{(antihomomorphism)},\\
S(1) & = & 1, \\
\Delta (S(a)) & = & \tau (S \otimes S)\Delta (a), \quad
\mbox{with}\quad\tau (a \otimes b) \equiv b \otimes a,\\
\epsilon (S(a)) & = & \epsilon (a).
\err
We will use Sweedler's \cite{SW} notation for the
coproduct:
\beq
\Delta (a) \equiv a_{(1)} \otimes a_{(2)}\quad \mbox{(summation
is understood).}
\label{sweedler}
\eq
$\Alg$ is not required to be commutative, and
in general it is non-cocommutative, i.e. 
$\Del' \equiv \tau \circ \Del \neq \Del$.

\subsection{$U_q(g)$ and Quasitriangular Hopf Algebras}  \label{subsec:U_q}

The fastest way to introduce quantum groups is to simply write down 
a certain deformation of the universal enveloping algebra of a 
simple Lie algebra $g$ in a Chevalley basis with a complex parameter $q$,
and study its properties. 
We will mainly work in this framework, which was introduced by Drinfeld 
\cite{drinfeld} and Jimbo \cite{jimbo}. 

Let $q \in \compl$ and $A_{ij}=2\frac{(\a_i, \a_j)}{(\a_j,\a_j)}$ be
the Cartan matrix of a classical simple Lie algebra ${g}$ of rank $r$,
where $( , )$ is the Killing form and $\{\a_i,\quad i=1,...,r \} $
are the simple roots.
Then the {\em quantized universal enveloping algebra}
$\U \equiv U_q({g})$ is the Hopf algebra with generators
$\{ X^{\pm}_i, H_i; \quad i=1,...,r\}$ and relations \cite{FRT,jimbo,drinfeld}
\berr
\[H_i, H_j\]      &=& 0                      \label{CR_HX} \\
\[H_i, X^{\pm}_j\] &=& \pm A_{ji} X^{\pm}_j,     \\
\[X^+_i, X^-_j\]   &=& \d_{i,j} \frac{q^{d_i H_i}
                          -q^{-d_i H_i}}{q^{d_i}-q^{-d_i}}
                    = \d_{i,j} [H_i]_{q_i}      \\
\sum_{k=0}^{1-A_{ji}} &\ &
           \[\begin{array}{c} 1-A_{ji} \\ k\end{array}\]_{q_i}(X^{\pm}_i)^k
         X^{\pm}_j (X^{\pm}_i)^{1-A_{ji}-k} = 0, \quad i\neq j
\label{UEA}
\err
where $d_i = (\a_i, \a_i)/2, \quad q_i = q^{d_i}, \quad [n]_{q_i}
           = \frac{q_i^n-q_i^{-n}}{q_i-q_i^{-1}}$ and
\beq
\[ \begin{array}{c} n\\m \end{array} \]_{q_i} =
             \frac{[n]_{q_i}!}{[m]_{q_i}! [n-m]_{q_i} !} .  \label{binom}
\eq
The last of (\ref{UEA}) are the deformed Serre relations.
As algebra, it can be shown \cite{drinfeld_2}
that if $q$ is considered as a formal variable, 
$\U$ is essentially\footnote{without going into mathematical detais here}
the same as the classical, undeformed enveloping algebra 
with a formal variable $q$ and a different choice of generators.
However it is {\em not} equivalent as Hopf algebra:
the comultiplication on $\U$ is defined by
\berr
\Del(H_i)          &=& H_i \tens 1 + 1 \tens H_i \nonumber \\
\Del(X^{\pm}_i) &=&  X^{\pm}_i \tens q^{d_i H_i/2} + q^{-d_i H_i/2}
\tens  X^{\pm}_i,
\err
and antipode and counit are
\berr
S(H_i)    &=& -H_i, \nonumber \\
S(X^+_i)  &=& -q^{d_i} X^+_i, \quad S(X^-_i)  = -q^{-d_i} X^-_i, \nonumber \\
\eps(H_i) &=& \eps(X^{\pm}_i)=0.
\err
The classical case is obtained by taking $q=1$. The consistency of this
definition can be checked explicitely.

The Cartan--Weyl involution is defined as
\beq
\theta({X_i^{\pm}}) = X_i^{\mp}, \quad \theta({H_i}) = H_i,
\eq
extended as a linear anti--algebra map (involution). In particular,
$\theta(q) = q$ for any $q \in \compl$. It is obviously consistent with the
algebra, and one can check that
\berr
(\theta\tens\theta) \Del(x) = \Del(\theta(x)), \label{theta_del}\\
S(\theta(x)) = \theta(S^{-1}(x)).
\err

The conventions we use are those of \cite{FRT} except for
a replacement $q \rightarrow q^{-1}$
for reasons explained in the next section, and agree 
up to normalization (see below) with those of 
\cite{drinfeld,kirill_resh}.
They differ from e.g. \cite{CH_P} by some redefinitions.
In the mathematical literature, usually a rational version of the
above algebra, i.e. using $q^{d_i H_i}$ instead of $H_i$ 
is considered. Since
we are mainly interested in specific representations,
we prefer to work with $H_i$. Furthermore, if $q$ is a root of
unity, one has to specify if one includes the "divided powers" 
$(X_i^{\pm})^{(k)} = \frac{(X_i^{\pm})^k}{[k]_{q_i}!}$ 
("restricted specialization")
or not ("unrestricted specialization"); the \rep
theory is quite different in these cases. 
We will mostly work in the unrestricted specialization, 
however since we are really only interested
in certain representations, it will become clear from the context 
what is appropriate.

Often the following operators
are often more useful:
\beq
h_i = d_i H_i, \quad e_{\pm i} =  \sqrt{[d_i]}X^{\pm}_i,
\label{little_h}
\eq
Then the first two relations in (\ref{UEA}) become 
\berr
\[h_i, e_{\pm j}\] &=& \pm (\a_i, \a_j) e_{\pm j}, \label{CR_he} \\
\[e_i, e_{-j}\]       &=&  \d_{i,j} [h_i]_q .
\err
In order to have the standard Physics normalization for angular momenta, 
the normalization of the Killing form will be chosen so that 
the short roots 
%(resp. all in the simply laced cases) 
have length 
$d_i = \frac 12$, i.e. $(\a_i,\a_i)=1$, and the long roots have length 1.
In any case, a rescaling of the Killing form can be absorbed by a 
redefinition of $q$.

One could also define another Hopf algebra  with reversed coproduct
$\Del'(x) = \tau \circ \Del(x)$, and $S^{-1}$ instead of $S$. 
However this is 
essentially the same. The reason is that $\U$ has the
very important property of being {\em quasitriangular}, i.e. there
exists a universal ${\R}$ $\in \U \tens \U$ with the following properties:
\berr
(\Del\tens \id) \R  &=& \R_{13} \R_{23},  \label{qtr_1}\\
(\id \tens \Del) \R &=& \R_{13} \R_{12}  \label{qtr_2}\\
\Del'(x) &=& \R \Del(x) \R^{-1} \label{qtr_3}
\err
for any $x \in \U$, where lower indices denote the position of the 
components of $\R$
in the tensor product algebra $\U \tens \U \tens \U$ :
if $\R \equiv a_{i} \tens b_{i}$  (summation is
understood), then e.g. $\R_{13}\equiv a_{i} \tens 1
\tens b_{i}$.
By considering $(\Del'\tens \id) \R = \R_{23} \R_{13}$, 
one obtains the Quantum Yang--Baxter equation
\eqn{YBE}{
\R_{12} \R_{13} \R_{23} = \R_{23} \R_{13} R_{12} . 
}
Furthermore, the following properties are a consequence of 
(\ref{qtr_1}) to  (\ref{qtr_3}):
\berr
(S \tens \id)\R & = & \R^{-1},\label{R_properties}\\
(\id \tens S)\R^{-1} & = & \R, \\
(\epsilon \tens \id)\R & = & (\id \tens \epsilon )\R = 1.
\err
The construction of $\R$ and the proof of the relations (\ref{qtr_1}) to  
(\ref{qtr_3})
is based on the so--called quantum double construction due to 
Drinfeld \cite{drinfeld}. It turns out that 
the Borel subalgebras $\B^-$ and $\B^+$, generated by $\{H_i,X_j^-\}$ and
$\{H_i,X_j^+\}$ respectively, are Hopf--subalgebras which are dually paired
(see below).
If $\{a_i\}$ is a basis of $\B^-$ and $\{b_i\}$
the dual basis of $\B^+$, then $\R = a_i \tens b_i$, after factoring out
a copy of the Cartan subalgebra which has been counted twice. The relations
(\ref{qtr_1}) to  (\ref{qtr_3}) are then easy to see.

This universal $\R$ is the essential feature of a quantum group, 
and we will make extensive use of it. It incorporates the additional structure
of Lie groups which is not used in the classical theory, namely the existence
of a certain Poisson structure compatible with the group structure.
Furthermore, all this is combined into objects which are holomorphic
in $q$. Therefore one should expect that there is a lot to say about 
this rich structure.

\subsection{$Fun(G_q)$ and dually paired Hopf Algebras}  
       \label{subsec:dual_hopf}

Before studying $\U$ any further,
let us now sketch the second approach to quantum groups; 
for a general review see \cite{CH_P}. It is 
based on the observation that any (compact) Lie group $G$ with Lie algebra 
$g$ is actually a (coboundary) Poisson--Lie group, i.e. 
there is a particular Poisson structure on the group manifold
which can be written in terms of a "classical r-matrix" $r \in g \tens g$,
and enjoys certain compatibility conditions.
%To be explicit,
%$r=\sum_i d_i H_i\tens H_i + \sum_{\a} 2d_{\a} X_{\a}^+ \tens X_{\a}^-$
%where $\a$ are the positive roots.
$r$ can again be obtained from a "double construction" 
\cite{drinfeld}; 
it is {\em not} given by the structure constants
of $G$, it is truly an {\em additional} structure on $G$. 
Now as in Quantum Mechanics on a phase space, this Poisson structure 
can be quantized, giving rise to a non--commutative algebra $Fun(G_q)$
which replaces the commutative algebra $Fun(G)$. If one writes $q=e^h$,
$h$ plays the role of the Planck constant, and the classical case
$Fun(G)$ is obtained in the limit $h \rightarrow 0$, i.e. 
$q \rightarrow 1$. This is the origin of the name "Quantum group",
and even though this quantization procedure may be formal, 
the final result is known to exist. Upon this quantization, the 
"classical r-matrix" $r$ turns into the universal $\R \in \U\tens \U$.

These two approaches are in fact dual to each other. This means that 
$\U_{q^{-1}}\equiv U_{q^{-1}}(g)$ and $\Alg=Fun(G_q)$ are dually paired 
Hopf algebras (notice the replacement $q \rightarrow q^{-1}$). In general, 
two Hopf algebras $\U$ and $\Alg$ are said to be dually paired
if there exists a non-degenerate inner product $<\;,\;>:$
$\U \tens \Alg \rightarrow \compl$, such that:
\berr
<x y,a> & = & <x \tens y, \Delta (a)> \equiv <x,a_{(1)}><y,a_{(2)}>,
 \nn\\
<x,a b> & = & <\Delta (x),a \tens b>, \nn \\
<S(x),a> & = & <x,S(a)>, \nn\\
<x,1> & = & \eps (x),\quad \mbox{and}\quad <1,a>=\eps (a),
\label{hopf_dual}
\err
for all  $x,y \in \U$ and $a,b \in \Alg$.

The algebra $Fun(G_q)$ can be written down explicitely \cite{FRT} 
if it is written as a pseudo matrix group \cite{woron}, generated by
the elements of a $N \times N$ matrix 
$A \equiv ({A^i}_j)_{i,j = 1...N} \in M_{N}(Fun(G_q))$\footnote{This 
corresponds to $GL_{q}(N)$ unless there are explicit
or implicit restrictions on the matrix elements of $A$.}. 
The coproduct on $Fun(G_q)$ is defined as classically,
\beq
\Delta A = A \dot{\tens} A, \quad\mbox{i.e.}\quad\Delta ({A^i}_j)=
{A^i}_k\tens{A^k}_j, \label{coprod_A}
\eq
and $S({A^i}_j) = {(A^{-1})^i}_j$, $\eps({A^i}_j) = {\delta^i}_j$.
Now if $<\; ,\;>$ is a dual pairing of $\U_{q^{-1}}$ with $Fun(G_q)$,
then this implies that ${\pi^i}_j \equiv < .\; , {A^i}_j>$ is a
representation of $\U_{q^{-1}}$, i.e.
\beq
\begin{array}{lr}
{\pi^i}_j : \U_{q^{-1}} \rightarrow \compl, & \\
{\pi^i}_j(xy) = \sum_{k}^{} {\pi^i}_k(x) {\pi^k}_j(y), &\quad
\forall x,y \in \U_{q^{-1}};
\label{rep_1}
\end{array}
\eq
we will say much more about \reps in a later section.
In this representation,
the universal $\R  \in \U_{q^{-1}} \tens \U_{q^{-1}}$ gives the numerical 
$R$-matrix:
\beq
<\R,{A^i}_k \tens {A^j}_l> = {R^{ij}}_{kl}.
\eq
Now the definition of a dual pairing (\ref{hopf_dual}) and 
(\ref{qtr_3}) imply \cite{drinfeld,FRT}
\berr
<x , {A^j}_s {A^i}_r>& = & <\Del x ,{A^j}_s \tens {A^i}_r> \nn\\
&=&< \tau \circ \Del x,{A^i}_r \tens {A^j}_s> \nn \\
&=&<\R (\Del x) \R^{-1}, {A^i}_r \tens {A^j}_s>\nn\\
&=&< x , {R^{ij}}_{kl} {A^k}_m {A^l}_n {(R^{-1})^{mn}}_{rs}>,
\err
for any $x \in \U_{q^{-1}}$, i.e. the matrix elements of $A$
satisfy the commutation relations
\beq
{R^{ij}}_{kl} {A^k}_m {A^l}_n  =   {A^j}_s {A^i}_r {R^{rs}}_{mn},
\eq
which can be written more compactly in tensor product notation as follows:
\berr
R_{12} A_1 A_2 & = &   A_2 A_1 R_{12}; \label{RTT} \\
R_{12} = (\pi_1 \tens \pi_2) (\R)&  \equiv
&  <\R , A_{1} \tens A_{2}>.
\err
Starting from this formalism, one can introduce differential forms
etc. and study the noncomutative differential geometry of quantum groups,
see \cite{W3,zumino_dg,schupp_th,watts}.
So far, we are considering all algebras over $\compl$ without 
any reality structure, which we will discuss below.

Now one can recast the commutation relations of $U_{q^{-1}}(g)$ 
into a more compact form \cite{FRT} .
For a \rep $\pi$, define matrices
\berr
L^{+}_{\pi} & \equiv & (\id \tens \pi)(\R ) , \nn\\
SL^{-}_{\pi} & \equiv & (\pi \tens \id) (\R ), \nn\\
L^{-}_{\pi} & \equiv & (\pi \tens \id) (\R^{-1}). 
\end{eqnarray}
Then the commutation relations for these matrices follow from
the quantum Yang-Baxter equation, e.g.
\berr
0 & = & (\id\tens\pi\tens\pi) 
          (\R_{23}\R_{13}\R_{12}  -  \R_{12}\R_{13}\R_{23}) \\
  & = & R_{12} L^+_2 L^+_1 - L^+_1 L^+_2 R_{12} 
\err
and similarly
\berr
R_{12} L^-_2 L^-_1&=& L^-_1 L^-_2 R_{12},\\
R_{12} L^+_2 L^-_1&=& L^-_1 L^+_2 R_{12}.
\err
The coproduct is now
\beq
\Delta L^{\pm} = L^{\pm}\dot{\tens} L^{\pm}, 
\eq
and $\eps(L^{\pm})=I$, $S(L^{\pm})= (L^{\pm})^{-1}$.
The $X_i^{\pm}$ can be extracted from the upper resp. lower triangular 
matrices $L^{\pm}$ \cite{FRT}. 

One can also turn the logic around and show that there exists a 
dual pairing between $U_{q^{-1}}(g)$ and the Hopf algebra 
defined by (\ref{RTT}) and 
(\ref{coprod_A}) (with some suitable additional constraints depending on 
the  group, cp. \cite{majid}),
which can be seen to be a quantization of $Fun(G)$.

\subsection{More Properties of $U_q(g)$}  \label{subsec:properties}

Let us describe $\U$ in more detail.
In the classical case, the Weyl group $\W$ acting on weight space
by the reflections 
$\sigma_i$ along the simple roots can be "lifted" to an
action of the braid group\footnote{this has nothing to do with 
the \reps of the braid group obtained from $\R$.}
with generators $T_i$ on representations of $g$, 
in particular on $g$ itself with the adjoint representation. The 
relations of the braid group are $(T_i T_j)^{m_{ij}}=1$ if 
$(\sigma_i \sigma_j)^{m_{ij}}=1$, 
but the square of $T_i$  is not required to be 1 any more. 
The same can be done for $\U$ \cite{lusztig}: 
there exist algebra automorphisms of $\U$ defined as
\berr
T_i(H_j)   &=& H_j - A_{ij} H_i, \quad T_i X_i^+ = -X_i^- q_i^{H_i},
                      \nonumber \\
T_i(X_j^+) &=& \sum_{r=0}^{-A_{ji}}(-1)^{r-A_{ji}}
           q_i^{-r}(X_i^+)^{(-A_{ji}-r)}X_j^+ (X_i^+)^{(r)}  
\label{braid_action}
\err
where $(X_i^{\pm})^{(k)} = \frac{(X_i^{\pm})^k}{[k]_{q_i}!}$, and
similarly for lowering operators\footnote{the $T_i$ can actually
be implemented as $\om_i (...) \om_i^{-1}$ in an extension of $\U$,
see \cite{kirill_resh,lev_soib}; we will come back to that.}.
If $\omega = \sigma_{i_1} ... \sigma_{i_N}$ is a reduced expression for
the longest element of the Weyl group, then
$\{\b_1=\a_{i_1}, \b_2=\sigma_{i_1}(\a_{i_2}), \dots ,\b_N=\sigma_{i_1} ...
\sigma_{i_{N-1}} (\a_{i_N})\}$ is an ordered set of positive roots.
Now one can define root vectors of $\U$ as
$X_{\b_r}^{\pm} = T_{i_1} ... T_{i_{r-1}}(X_{i_r}^{\pm})$.
This can be used to obtain a Poincare--Birkhoff--Witt basis of
$\U = \U^- \U^0 \U^+$ where $\U^{\pm}$ is generated by the
$X^{\pm}_i$ and $\U^0$ by the $H_i$:
for $\underline{k}=(k_1, ..., k_N)$, let
$X^+_{\underline{k}} = (X_{\b_N}^+)^{k_N} ... (X_{\b_1}^+)^{k_1}$
and similarly for $X^-_{\underline{k}}$. 
Then the $X^{+}_{\underline{k}}$ form
a P.B.W. basis of $\U^{+}$, and similarly for $\U^-$ \cite{lusztig_90}.
%(assuming $q^4\neq 1$).

Using this, one can find explicit formulas for $\R = \R(q)$ 
\cite{tolstoi,kirill_resh}. They are somewhat complicated however, 
and all we need for now is the following form:
\beq
\R=q^{\sum  (\a^{-1})_{ij} h_i\tens h_j}
  \(\sum c_{k,k'}(q) \tilde{X}^+_{\underline{k}}\tens 
                                      \tilde{X}^-_{\underline{k'}}\)
\label{R_formula}
\eq
where\footnote{Since we will work with "nice"
representations, $\R$ converges.} $(\a)_{ij} = (\a_i,\a_j)$, 
$\tilde{X}^{\pm}_{\underline{k}}$ is defined like $X^{\pm}_{\underline{k}}$
with $X_i^{\pm}$ replaced by 
$\tilde{X}_i^{\pm}\equiv q_i^{\pm\frac 12 H_i} X_i^{\pm}$,
and $c_{k,k'}(q) \in \compl$ are rational functions of $q$.
Furthermore, the coefficients in (\ref{R_formula}) are uniquely determined by 
the properties (\ref{qtr_1}) to  (\ref{qtr_3}) \cite{drinfeld_2,tolstoi}. 
Using this, is easy to see that 
\berr
\R(q^{-1}) &=& \R^{-1}(q),  \label{R_q-1}\\
(\theta\tens\theta) \R_{12} &=& \R_{21}.  \label{R_theta}
\err
For the case of $U_q(sl(2))$, the explicit form is
\beq
\R(q) = q^{\frac12 H\tens H} \(\sum_{l\geq 0} q^{\frac 12 l(l-1)} 
       \frac{(1-q^{-2})^l}{[l]_{q}!} (q^{\frac 12 H} X^+)^l\tens 
      (q^{-\frac 12 H} X^-)^l\)
\label{R_sl(2)}
\eq
where we have set $d=(\a,\a)/2=1$, i.e. $q_i=q$.

It is easy to check that the square of the antipode is an inner automorphism
\beq
S^2(x) = q^{2\tilde{\rho}} x q^{-2\tilde{\rho}},
\label{S2_inner}
\eq
where
$\tilde{\rho} = \sum_{\a>0} h_{\a}$ and
$h_{\a} = \sum n_i h_i$ if $\a = \sum n_i \a_i$ \cite{drinfeld}.
As shown in \cite{drinfeld}, there is another
element in $\U$ with that property, namely $u \equiv S\R_2 \R_1$. 
Therefore 
\beq
v\equiv (S\R_2) \R_1 q^{-2\tilde{\rho}}
\label{v}
\eq
commutes with any element in $\U$ and will be called Drinfeld--Casimir.
It satisfies
\berr
\Del(v) &=& (\R_{21} \R_{12})^{-1} v\tens v, \label{v_coprod}  \\
v^{-1} &=& q^{2\tilde{\rho}} \R_2 S^2(\R_1), \label{v_inv}\\
S(v)&=&v
\err
where $\R_{12} = \R$ and $\R_{21} = \tau \circ \R$.
Furthermore, it is easy to check from (\ref{R_theta}) that 
\beq
\theta(v)=v.
\eq
The value of $v$ on a highest weight \rep was first 
determined in \cite{resh_1},
and can be obtained easily from (\ref{R_formula}):
if $w_{\l}$ is a h.w. vector  with 
$X^+ \cdot w_{\l} = 0$ and  $h_i \cdot w_{\l} = (\l,\a_i) w_{\l}$ 
(see section \ref{sec:rep_theory}), 
then
\beq
v \cdot w_{\l} = q^{-c_{\l}} w_{\l},
\label{v_h_w}
\eq
where $c_{\l}=(\l,\l+2\rho)$ is the value of the {\em classical} 
quadratic Casimir on $w_{\l}$. 

Later we will need analogs of $\R$ for $\U^{\tens l}$ with $l>2$. 
If $\Del_{(l)}(x) \in \U^{\tens^l}$ is the (unique) $l$ -- fold coproduct of 
$x \in \U$, let 
\berr
\R^{(a)}_{12...l} &\equiv& (\R^{(a)}_{12...(l-1)}\tens 1)(\Del_{(l-1)}
  \tens \id)\R_{12},   \label{R_l_a}\\
\R^{(b)}_{12...l} &\equiv& (1 \tens\R^{(b)}_{12...(l-1)})
       (\id \tens \Del_{(l-1)}) \R_{12} 
\label{R_l_b}
\err
for $l>2$. They have similar properties as $\R$:
\begin{lemma}
\beq
\R^{(a)}_{12...l} = \R^{(b)}_{12...l} \equiv \R_{12...l},
\label{R_l_equal}
\eq
and 
\berr
\Del'_{(l)}(x) &=& \R_{12...l} \Del_{(l)}(x) \R^{-1}_{12...l},  
                                                   \label{qtr_l_3}\\
\R_{12...l}(q^{-1}) &=& \R_{12...l}^{-1}(q),     \label{R_l_q} \\
(\theta\tens \dots \tens\theta) \R_{12...l} &=& \R_{l...21} \label{R_l_theta}.   
\err
\label{R_l_lemma}
\end{lemma}
\begin{proof}
(\ref{R_l_equal}) follows by straightforward induction:
For $l=3$, it reduces to the Yang--Baxter equation.
By the induction assumption, we have
\berr
\R^{(a)}_{12...l}&=& (\R_{12...(l-1)}\tens 1)(\id\tens\Del_{(l-2)}\tens \id)
     (\Del\tens 1)\R_{12} \nn\\
  &=& (\R^{(b)}_{12...(l-1)}\tens 1)(\id\tens\Del_{(l-2)}\tens \id)
     \R_{13} \R_{23} \nn\\
  &=& (1\tens \R_{12...(l-2)}\tens 1) (\id\tens\Del_{(l-2)}\tens\id)
   \R_{12}\R_{13} \R_{23}.
\err
Similarly, 
\berr
\R^{(b)}_{12...l}&=& (1\tens \R^{(a)}_{12...(l-1)})
                  (\id\tens\Del_{(l-2)}\tens \id)(\id\tens\Del)\R_{12} \nn\\
    &=& (1\tens \R_{12...(l-2)}\tens 1) (\id\tens\Del_{(l-2)}\tens\id)
   \R_{23}\R_{13} \R_{12},
\err
and (\ref{R_l_equal}) follows using the Yang--Baxter equation.
Applying $(\Del_{(l-1)}\tens \id)$ to (\ref{qtr_3}), one obtains 
$(\Del_{(l-1)}x_{(2)})\tens x_{(1)} = ((\Del_{(l-1)}\tens\id)\R_{12}) 
\Del_{(l)}(x) ((\Del_{(l-1)}\tens\id)\R_{12}^{-1})$,
and by induction and using (\ref{R_l_equal}) it follows 
\berr
(\Del'_{(l-1)}x_{(2)})\tens x_{(1)} &=& 
     (\R_{12...(l-1)}\tens 1)((\Del_{(l-1)}\tens\id)\R_{12}) \Del_{(l)}(x) \\ 
&\cdot& ((\Del_{(l-1)}\tens\id)\R_{12}^{-1})(\R^{-1}_{12...(l-1)}\tens 1)\nn\\
     &=& \R^{(a)}_{12...l} \Del_{(l)}(x) (\R^{(a)}_{12...l})^{-1},
\err
which shows (\ref{qtr_l_3}). 
 
For illustration let us also show (\ref{R_l_theta}). From (\ref{R_l_a}), 
one gets by induction 
\berr
(\theta\tens\dots\tens\theta)\R^{(a)}_{12...l} &=& 
  ((\Del_{(l-1)} \tens \id)\R_{21})(\R_{(l-1)...21}\tens 1)  \nn\\
  &=&  (\R_{(l-1)...21}\tens 1) ((\Del'_{(l-1)} \tens \id)\R_{21})  \nn\\
  &=&  \R^{(b)}_{l...21} = \R_{l...21},
\err
using the flipped (\ref{qtr_l_3}) and (\ref{R_l_equal}). 
Similarly one can see (\ref{R_l_q}).
\end{proof}

All this will become more obvious in section \ref{sec:weyl}.

\subsection{Reality Structures}  \label{subsec:reality}

So far we considered all algebras over $\compl$, in particular $\U$ is the 
universal enveloping algebra of a complex Lie algebra. 
If one wants to consider 
e.g. $SO_q(5)\equiv U_q(so(5;\reals))$ or 
$SO_q(2,3)\equiv U_q(so(2,3;\reals))$,
one has to introduce a star--structure, an {\em antilinear} involution 
$\obar{x}$, as classically.  This star--structure must be 
such that if $x\in \U$ is a group element for $q=1$, e.g. 
$x=\exp(y)$ for $y$ an element in the Lie algebra,
then $\obar{x} = x^{-1}$, i.e. $\obar{x}$ becomes the adjoint of $x$ 
in a unitary 
\rep of $\U$ (this will be considered in detail below).

There are several possible star--structures on $\U$. 
One has to distinguish the cases $q\in\reals$ and $|q|=1$. If $q\in \reals$, 
then a natural definition is
\berr
\robar{x} &\equiv& \theta(x^{\ast}) \\
\robar{x\tens y} &\equiv& \robar{x} \tens\robar{y}  \label{robar_tens},
\err
with 
\beq
\robar{\Del(x)} = \Del(\robar{x})
\eq
and $\robar{Sx} = S^{-1} \robar{x}$,
%and $\robar{\theta(x)} = \theta(\robar{x})$
which is a standard
Hopf algebra star--structure ($x^{\ast}$ is the 
complex conjugate of $x\in \U$,
where the $X_i^{\pm}$ are considered real). Using the uniqueness
of $\R$ (see the discussion below (\ref{R_formula})), one can see that
\beq
\robar{\R_{12}} = \R_{21} = \tau \circ \R .
\eq

If $|q|=1$, a natural definition is
\berr
\cobar{x} &\equiv& \theta(x^{\ast}) \label{cbar} \\
\cobar{x\tens y} &\equiv& \cobar{y} \tens\cobar{x} \label{cbar_tens}
\err
with 
\beq
\cobar{\Del(x)} = \Del(\cobar{x}) \label{cbar_del}
\eq
and $\cobar{Sx} = S \cobar{x}$, which is a {\em nonstandard}
Hopf algebra star--structure; notice that $\cobar{q}=q^{-1}$.
In this case, $\cobar{\R} = \R^{-1}$, and more generally 
$\cobar{\R^{(a)}_{12...l}} = (\R^{(b)}_{12...l})^{-1}$ with the obvious 
extention of (\ref{cbar_tens}) to several factors, thus
\beq
\cobar{\R_{12...l}} = \R_{12...l}^{-1}  \label{cobar_R_l}
\eq
using Lemma \ref{R_l_lemma}.

Furthermore, $\cobar{(S \R_2) \R_1} = \R^{-1}_2 S\R^{-1}_1 = \R_2 S^2 \R_1 =
(S \R_2 \R_1)^{-1}$ (see \cite{CH_P}), so
\beq
\cobar{v} = v^{-1}.   \label{cobar_v}
\eq

Both $\robar{x}$ and $\cobar{x}$ correspond to the {\em compact} case, 
such as $SO_q(5)$; however we will mainly be interested 
in the case of $|q|=1$.
Having applications in QFT in mind, one might then be worried about 
(\ref{cbar_tens}). We will
see in sections \ref{sec:creat_anihil_op}, \ref{sec:inner_prod_g0} 
and to some extent in section \ref{subsec:invar_forms}
how this can be used consistently with a 
many--particle interpretation. In the classical case, the coproduct on 
$\U$ is cocommutative, and it does not make any difference whether its 
components are flipped by the reality structure or not.

Reality structures corresponding to noncompact groups can be obtained 
from $\cobar{x}$ by conjugation with elements of the Cartan subalgebra.
We will only consider star--structures of the form 
$\obar{X^+_i} = \pm X^-_i$ and $\obar{H_i} = H_i$.
For example, $SO_q(2,1)$ is the algebra $U_q(sl(2,\compl))$ with 
$\obar{H}=H$ and $\obar{X^{\pm}} = - X^{\mp}$, which can be
realized as $\obar{x} = (-1)^{-H/2} \cobar{x} (-1)^{H/2}$.
Then for $|q|=1$, again $\obar{\R} = ((-1)^{-H/2}\tens (-1)^{-H/2})\R^{-1}
 \cdot$ $((-1)^{H/2}\tens (-1)^{H/2}) = \R^{-1}$.
The cases $SO_q(2,1)$ and $SO_q(2,3)$ will be considered in much more detail 
below. We will only consider star--structures with
\beq
\obar{\R}=\R^{-1}    \label{obar_R}
\eq
for $|q|=1$.

\section{Representation Theory}  \label{sec:rep_theory}

We will only consider the \rep theory of $\U$.
The main advantage of this point of view is that the representation theory
of $\U$ can be formulated in the familiar language of ordinary semi--simple 
Lie algebras. 

Let us first collect the basic definitions.
We will (essentially) only consider finite--dimensional representations. 
A \rep of  $\U$ on a vectorspace $V$ is a map $\U \rightarrow GL(V)$
such that $(xy)\cdot v = x\cdot(y\cdot v)$ and $1\cdot v=v$ 
(sometimes we will omit the $\; \cdot \;$). Then one can 
as usual diagonalize the Cartan subalgebra, and every \rep is a sum of 
weight spaces. A vector $v_{\l}$ has weight $\l$ if 
$h_i v_{\l} = (\l,\a_i) v_{\l}$. Then  $X_i^{\pm}$ are 
rising and lowering operators as in the classical case, since
(\ref{CR_HX}) and (\ref{CR_he}) are undeformed.
We will mainly (but not always) consider the case of {\em integral} weights,
i.e. $(\l, \b_i) \in d_i \Z$. 
Classically, all irreducible \reps are 
highest weight \reps $V$ with {\em dominant} integral highest weight 
$\l$, i.e. $V=\U w_{\l}$ and $(\l,\a_i)>0$ for all simple roots $\a_i$. 
Finally, let $Q = \sum \Z \a_i$ be the root lattice and 
$Q^+ = \sum \Z_+ \a_i$
where $\Z_+ =\{0,1,2,...\}$.
We will write 
\beq
\l \succ \mu \quad \mbox{if} \quad \l - \mu \in Q^+.
\eq

Given two representations $V_1, V_2$ of a Hopf algebra, 
the tensor product \rep
is naturally defined as $x\cdot (v_1\tens v_2) \equiv \Del(x)(v_1\tens v_2)
 = (x_{(1)}\cdot v_1)\tens(x_{(2)}\cdot v_2)$.

So far, we have not specified $q$ at all. For the \rep theory however,
there are two very different cases. One is if $q$ is "generic", i.e.
{\em not} a root of unity, and the other is if $q$ is a root of unity.

If $q$ is generic, then the representation theory is essentially the same
as in the classical case, in the sense that all the important theorems
have a perfect analog. This is quite intuitive, since everything 
will be holomorphic in $q$, which as always is a very strong constraint.
We will quote the main results in a moment. If $q$ is a root of 
unity however, the \rep theory changes completely, and essentially 
none of the classical theorems continue to hold. Roughly speaking this
happens because poles resp. zeros occur in various quantities. While
it is more complicated and therefore often discarded, this is the truly
interesting case. The main objective of this thesis is to point out that many
of its features seem to be very relevant to Quantum Field Theory,
and not only to Conformal Field Theory. In any case,
the root of unity case is not well enough understood, and deserves 
further study.

Consider first the case of generic $q$. 
Then the basic results are as follows:
\begin{itemize}
\item Any finite--dimensional \rep of $\U$ is completely reducible,
   i.e. it decomposes into a direct sum of irreducible \reps (irreps).
\item The irreps are highest weight \reps
    with dominant integral highest weight $\l$, and the \rep space
     is the same as classically.
\item The fusion rules are the same as classically. 
\end{itemize}
So the Weyl group acts on the weights of an irrep, and in fact
a braid group action can be defined on any representation 
\cite{lusztig_book,jantzen,kirill_resh}.
Complete reducibility was first proved by Rosso \cite{rosso}.

These results are not hard to understand. The first two would be
obvious if one could apply the fact that as {\em algebra}, 
$\U=U_q(g)$
is nothing but the classical enveloping algebra \cite{drinfeld_quasi}
(with a {\em formal} variable $q$ however, and the correspondence may
not be realized for a given $q \in\compl$).
Since similar issues will arise later, we want to explain
here why a \rep $V$ of  $\U$ which is irreducible for $q=1$ can become 
reducible only for $q$ a root of unity.
Such a \rep must have a dominant integral highest weight $\l$. 
If it contains a submodule at $q_0$ with a highest weight vector 
with weight $\l_0$
(which is dominant again by virtue of the Weyl group resp. braid group 
action), then the Drinfeld
Casimir $v$ must be the same on this submodule, so
$q_0^{c_{\l}} = q_0^{c_{\l_0}}$ using (\ref{v_h_w}).
However $\l \succ \l_0$, and we have
\begin{lemma} If $\l, \l_0$ are dominant weights with $\l \succ \l_0$, then
\beq
c_{\l} > c_{\l_0}.
\eq
\label{casimir_ord}
\end{lemma}
\begin{proof}
Notice first that
$c_{\l} \equiv (\l,\l+2\rho) = c_{\tilde\sigma_i(\l)}$, where 
$\tilde\sigma_i(\l)=\l-\frac{(\l+\rho,\b_i)}{d_i}\b_i$ is the modified action
of the Weyl reflection along any root $\b_i$ with reflection center $-\rho$.
Since  $\l \succ \l_0$ and both are dominant, 
$\l_0$ is contained in the convex hull of $\l$ and the $\tilde\sigma_i(\l)$.
But the Killing form is Euclidean and therefore convex, and the 
statement follows.
\end{proof}

Therefore if this $V$ is not irreducible at $q_0$, $q_0$ must be a phase, 
and in fact a root of unity (we assume $q_0 \neq 0$).

Complete reducibility can be understood
using the concept of invariant sesquilinear forms.  

\subsection{Invariant Forms, Verma Modules and Unitary 
Representations}   \label{subsec:invar_forms}

A bilinear form $(\;,\;)$ on a \rep $V$ is linear in both arguments, while
a sesquilinear form is linear in the second argument
and antilinear in the first. 

A bilinear form is called {\em invariant} if
\beq
(u,x\cdot v) = (\theta(x)\cdot u,v)    \label{bili_invariant}
\eq
for $u,v \in V$; this is sometimes called covariant \cite{deconc_kac}.
This can be considered for any $q\in\compl$.

For $q\in\reals$ or $|q|=1$, consider a star--structure as in 
section \ref{subsec:reality}
and denote it by $\obar{x}$, so
$\obar{X^{\pm}_i} = \pm X^{\mp}_i$ and $\obar{H_i} = H_i$.
Then a sesquilinear form $(\; ,\;)$ is called {\em invariant} if
\beq
(u,x\cdot v) = (\obar{x}\cdot u,v)    \label{invariant}
\eq
for $u,v \in V$; this is also sometimes called covariant. It 
is {\em hermitian} if 
\beq
(u,v)^{\ast} = (v,u).
\eq
A hermitian sesquilinear form is called an {\em inner product}. Note that
we always consider $q$ to be a complex number, so $\obar{q} = q^{\ast}$; in
the literature, $q$ is often treated as an indeterminate, and our definitions
agree with those of e.g. \cite{deconc_kac} only for $|q|=1$, 
which is the case
we are mainly interested in.
Finally, a \rep $V$ is {\em unitary} or {\em unitarizable} if there exists 
a positive definite 
invariant inner product on $V$.

Given a highest weight (h.w.) representation $V(\l)$ with h.w. 
vector $w_{\l}$,
there is a unique invariant inner product $(\; ,\;)$ on $V(\l)$ 
for $q\in\reals$ or $|q|=1$, resp. an invariant symmetric bilinear form 
for any $q\in\compl$. 
Uniqueness is clear, since one can express any 
$(\U \cdot w_{\l},\U\cdot w_{\l})$ in terms of 
$(w_{\l}, \U^0 \cdot w_{\l})$ or
$(\U^0 \cdot w_{\l}, w_{\l})$ using invariance and the commutation relations.
These two results agree and $(\; ,\;)$ is hermitian, 
because the star--structure is consistent with 
the commutation relations; notice that $[h_i]_q \in \reals$. 

Thinking of applications in Quantum Mechanics, the importance of unitarity
is obvious. But invariant sesquilinear forms 
(or bilinearforms) are also very useful as technical tools, 
due to the following well--known observation: if a highest weight 
\rep $V(\l)$ is not irreducible, it contains a submodule. Now
all these submodules are null spaces w.r.t. the sesquilinear form,
i.e. they are orthogonal to any state in $V(\l)$. Therefore one can 
consistently 
factor them out, and obtain a sesquilinear form on the quotient space.
To see that they are null, let 
$v_{\mu} \in V(\l)$ be in some submodule, i.e. 
$w_{\l} \notin \U \cdot v_{\mu}$. 
Now  for  any $v \in \U\cdot w_{\l}$, it follows 
$(v_{\mu}, v) \in (\U v_{\mu}, w_{\l}) = 0$, using invariance and 
the fact that there is only one vector with weight $\l$ in the h.w. \rep 
$V(\l)$, namely $w_{\l}$. Conversely,

\begin{lemma} \label{irred_nondeg}
Let $w_{\l}$ be the highest weight vector of an  irreducible highest weight
\rep $L(\l)$ with invariant inner product. If $(w_{\l}, w_{\l}) \neq 0$, then 
$(\; ,\;)$ is nondegenerate, i.e.
\beq         \label{irrep_lemma}
det(L(\l)_{\eta}) \neq 0
\eq
for every weight space with weight $\l - \eta$ in $L(\l)$.
\end{lemma}
\begin{proof}
Assume to the contrary that there is a vector $v_{\mu}$ which is 
orthogonal to all vectors of the same weight, 
and therefore to all vectors of any weight. 
Because $L(\l)$ is irreducible, there exists an $u \in \U$ such that 
$w_{\l} =  u\cdot v_{\mu}$. But then 
$(w_{\l},w_{\l}) = (w_{\l}, u\cdot v_{\mu})
= (\obar{u}\cdot w_{\l},v_{\mu}) =0$, which is a contradiction.
\end{proof}

\paragraph{Verma modules}
For any weight $\l$, there exists a "universal" highest weight module,
the {\em Verma module} $M(\l)$.
It is the (unique) $\U$ - module having a h.w.
vector $w_{\l}$ such that
the vectors $X^-_{\underline{k}} w_{\l}$ form a basis of 
$M(\l)$ \cite{deconc_kac}, where the $X^-_{\underline{k}}$ are a P.B.W. basis
of $\U^-$. This is the only infinite--dimensional 
\rep we will consider, and only as a technical tool. The importance
of Verma modules lies in the fact that all highest weight \reps can be
obtained from $M(\l)$ by factoring out an appropriate submodule.
In particular, the (unique) irrep $L(\l)$ with h.w. $\l$ is obtained
from $M(\l)$ by factoring out its maximal proper submodule.
Since it is a highest weight module,
one can define a unique invariant inner product $(\ ,\ )$ on a Verma module
for $|q|=1$ and $q\in\reals$,
and its maximal proper submodule is precisely the corresponding null subspace
(see \cite{deconc_kac} on how to define
analogous forms for generic $q$).

\paragraph{Forms on tensor products}
Now let $V_i$ be h.w. \reps of $\U$ for any $q \in \compl$ with dominant 
integral highest weight $\mu_i$, such that the $V_i$ are 
irreducible as long as $q$ is not a root 
of unity. Therefore on each $V_i$ there is an invariant inner product
$(\; ,\;)_i$ for $q\in\reals$ and for $|q|=1$, which is non--degenerate 
if $q$ is not a root of unity. 
It is important to realize that the \rep {\em space} $V_i$ is the
same for all $q$ (in particular for $q=1$), 
only the action of $\U$ on it depends on $q$, and is in fact
analytic (one way to see this is to use a P.B.W. basis, 
another is to construct the $V_i$ by taking suitable 
tensor products, as we will see in a moment).
Let
\beq
V \equiv V_1 \tens ... \tens V_r,
\eq
and for $a=a_1 \tens ... \tens a_r \in V_1 \tens ... \tens V_r$ 
and $b \in V$, define 
\beq
(a, b)_{\tens} \equiv (a_1,b_1)_1 ... (a_r,b_r)_r.
\eq
We claim that for $q\in\reals$, $(\; , \;)_{\tens}$ is a positive--definite 
inner product:

$(\; , \;)_{\tens}$ is invariant because of (\ref{robar_tens}) 
for $q\in\reals$, and it is certainly hermitian and positive definite 
if the $(\; ,\;)_i$ are. 
Let $M^{(i)}_{k_i,l_i}$ be the hermitian matrix of $(\; ,\;)_i$ 
in some basis of $V_i$.
Since the $(\; ,\;)_i$ are determined by the algebra alone,
the $M^{(i)}_{k_i,l_i}$ are
certainly continuous (and can be extended to analytic objects), so
their eigenvalues are real and continuous. Since $V_i$ remains
irreducible for $q\in\reals$ as shown above and the eigenvalues 
are positive for $q=1$, they cannot vanish for $q\in\reals$ because of Lemma 
\ref{irred_nondeg}. So $(\; ,\;)_{\tens}$ is indeed a positive--definite 
invariant inner product.
Similarly, $(\;,\;)_{\tens}$ is an invariant bilinear form for any 
$q \in\compl$ if it is built from bilinear forms on the $V_i$.

Now for $q\in \reals$, 
one can use the Gram--Schmidt orthogonalization method as usual, and
{\em $V$ is the direct sum of orthogonal highest weight irreps 
$V_{\l_l}$} with the same highest weights $\l_l$ and multiplicities 
$m_{\l_l}$ as classically. This implies that for $q\in\reals$, the Drinfeld
Casimir $v$ satisfies the characteristic equation
\beq
\prod_{\l_l} (v-q^{-c_{\l_l}}) =0,
\label{char_eq}
\eq
where the product is over all different highest weights counted {\em once},
and not with multiplicity $m_{\l_l}$.
Since $v$ is analytic, (\ref{char_eq}) holds for all $q\in\compl$,
and one can write down the projectors on the eigenspaces of $v$ as
\beq
P_{\l_l} = \frac{\prod_{\l_l' \neq \l_l}(v-q^{-c_{\l_{l'}}})}
               {\prod_{\l_l' \neq \l_l}(q^{-c_{\l_{l}}}-q^{-c_{\l_{l'}}})},
\label{projectors}
\eq
with $\sum P_{\l_l}=1$ and $P_{\l_l} P_{\l_k} = \d_{\l_l,\l_k} P_{\l_l}$. 
They only singularities are isolated poles in $q$, and it follows
that for generic $q$, the image of $P_{\l_l}$ consists of $m_{\l_l}$ 
copies
of the highest weight irrep $V_{\l_l}$ with h.w. $\l_l$. In fact using 
Lemma \ref{casimir_ord}, this may break down only at
roots of unity. Thus we have shown complete reducibility of $V$ 
for $q$ not a root of unity.

This argument illustrates the use of
inner products. From now on, we will only consider the case $|q|=1$. Then 
one can define another invariant sesquilinearform on 
$V = V_1 \tens ... \tens V_r$, namely
\beq
(a, b)_{\R} \equiv (a, \R_{12...l} b)_{\tens}
\label{inner_R}
\eq
with $\R_{12...l}$ as in  Lemma \ref{R_l_lemma}. Indeed 
using (\ref{cbar_del}) and (\ref{cbar_tens}),
\berr
(\obar{x}\cdot a,b)_{\R} &=& (\Del_{(r)}(\obar{x}) (a_1\tens\dots\tens a_r), 
               \R_{1...r} (b_1\tens\dots\tens b_r))_{\tens}    \nn\\
&=& (a_1\tens\dots\tens a_r, 
          \Del'_{(r)}(x) \R_{1...r} (b_1\tens\dots\tens b_r))_{\tens}  \nn\\
  &=& (a_1\tens\dots\tens a_r, 
          \R_{1...r} \Del_{(r)}(x) (b_1\tens\dots\tens b_r))_{\tens}   \nn\\
  &=& (a, x\cdot b)_{\R},
\err
since the $(\; ,\;)_i$ are invariant w.r.t. $\obar{x}$, and $\Del'_{(r)}$
is the flipped $r$ -- fold coproduct. While it is not positive 
definite in general,
$(\; ,\;)_{\R}$ is {\em nondegenerate} if  $q$ is not a root of unity,
which will be very important later.
To see this, let again $M^{(i)}_{k_i,l_i}$ be the invertible matrix 
of the inner products $(\; ,\;)_i$ on the irreps $V_i$.
Then the matrix of $(\; ,\;)_{\R}$ is
$\sum_{k'} M^{(1)}_{k_1,k'_1}\dots M^{(r)}_{k_r,k'_r} 
\R_{l_1,...,l_r}^{k'_1,...,k'_r}$, 
which is invertible, because $\R_{12...r}$ is invertible.
In fact, $(\; ,\;)_{\R}$ {\em remains nondegenerate at roots of unity 
as long as 
all the $V_i$ remain irreducible}, since then $\R_{12...r}$ exists 
and is invertible on these representations, as we will see in 
section \ref{subsec:root_of_1}.

In the classical limit $q\rightarrow 1$, $(\; ,\;)_{\R}$ reduces to 
$(\; ,\;)_{\tens}$ since $\R \rightarrow 1\tens 1$, however it is not
hermitian unless $q=1$ (remember $|q|=1$). In chapter \ref{chap:brst},
we will show how one can define
a (hermitian) inner product on a "part" of $V$ using a BRST operator
for $q$ a root of unity, and in fact a many--particle {\em Hilbert space},
with the "correct" classical limit.
 
Therefore we have to study the root of unity case. But first, we briefly 
discuss the $\hat R$ --matrix:

\subsection{$\hat R$-- Matrix and Centralizer Algebra}
      \label{subsec:centralizer}

For a (finite--dimensional) \rep $V$ of $\U$, consider the $n$--fold
tensor product of $V$ with itself,
\beq
V^{\tens^n} \equiv V \tens ... \tens V.
\eq
This carries a natural \rep of $\U$ using the $n$--fold 
coproduct $\Del_{(n)}$.
Classically, the symmetric group (or its group algebra) generated by
$\tau_{i,i+1}$ which interchanges the factors in position $i$ and $i+1$
commutes with the action of $U(g)$ on $V^{\tens^n}$.
The maximal such algebra commuting with the \rep is called the centralizer
algebra. In the quantum case, there is an analog of this, namely
\beq
\hat R_{i,i+1}\equiv\id\tens ... \tens(\tau\circ(\pi\tens\pi)\R)
     \tens ... \tens\id
\label{R_hat}
\eq
where the nontrivial part is in positions $i$ and $i+1$, and
$\pi$ is the representation on $V$. Notice that such a definition
only makes sense for identical representations. It follows from (\ref{qtr_3})
and coassociativity
that $\hat R_{i,i+1}$ commutes with the action of $\U$ on $V^{\tens^n}$.
Therefore \reps of $\U$ on this space fall into \reps of the centralizer
algebra. This is familiar from quantum field theory, where bosons and
fermions are totally symmmetric resp. antisymmetric \reps of the
permutation group. The Yang--Baxter equation now becomes
\beq
\hat R_{i,i+1} \hat R_{i+1,i+2} \hat R_{i,i+1} =
    \hat R_{i+1,i+2} \hat R_{i,i+1} \hat R_{i+1,i+2}
\eq

Acting on $V\tens V$, (\ref{v_coprod}) becomes
$\Del(v) = (\hat R)^{-2} (v\tens v)$. Now $v$ is diagonalizable for 
generic $q$
with eigenvalues $q^{-c_{\l}}$, because of complete reducibility.
Therefore $(\hat R)^2$ is diagonalizable,
with nonzero eigenvalues. This implies (e.g. using the Jordan normal form,
cp. section \ref{sec:tensor_brst})
that $\hat R$ {\em is diagonalizable
for generic} $q$, with eigenvalues $\pm q^{\frac 12 (c_{\l}-2c_{\mu})}$
where $\mu$ is the highest weight of $V$ if $V$ is irreducible.
Such a result was first obtained in \cite{resh_1} using a different method.

The centralizer algebra provides a connection between quantum groups
and conformal field theory
\cite{fuchs,sierra}. For small representations, it
can be described explicitely, and again the root of unity case is 
very different from the generic case.

\subsection{Aspects of Representation Theory at Roots of Unity} 
    \label{subsec:root_of_1}

Generally speaking, the root of unity case is somewhat more complicated than
the case of generic $q$, but also more interesting and probably more
relevant to physics. Unfortunately, the general
mathematical literature on this subject is not very accessible to
physicist\footnote{\cite{deconc_kac} is among the more readable sources.}.
The rank one case (i.e. $U_q(sl(2))$ and its real structures) however is quite
instructive and is discussed in \cite{keller,pasquier}. 
In this section we will only mention a few important features, and
many of the later sections will be 
devoted to studying certain aspects  in more detail.
In general, it is probably fair to say that the root of unity case is not 
well enough understood.

First, the subalgebras of $U_q(g)$  generated by $X_i^{\pm}$ and $H_i$ are 
nothing but $U_q(sl(2))$ algebras (the coalgebar structure is not the same,
however), with $q_i$ instead of $q$. 
Let 
\beq
q=e^{2\pi i n/m}   \label{q_root}
\eq
with $\gcd(m,n)=1$, and let $M=m$ if $m$ is odd, and $M=m/2$ if $m$ is even. 
Similarly for $q_i= e^{2\pi i d_i n/m}$, let $M_{(i)}=m$ if $d_i=\frac 12$, 
and $M_{(i)}=M$ if $d_i=1$ (recall our normalization conventions 
in section \ref{subsec:U_q}). Then $M_{(i)}$ is the smallest integer such that
\beq
[M_{(i)}]_{q_i} =0.
\eq

\paragraph{Highest weight vectors and irreducible \reps}
The following crucial formula can be checked easily \cite{keller}:
\beq
[X_i^+, (X_i^-)^k] = (X_i^-)^{k-1} [k]_{q_i} [H_i-k+1]_{q_i}.
\label{keller_eq}
\eq
In particular, this shows that $(X_i^-)^{M_{(i)}}$ {\em is central in} $\U$,
and so is $(X_i^+)^{M_{(i)}}$. Now 
if $w_{\l}$ is a highest weight vector, then $(X_i^-)^{M_{(i)}} \cdot w_{\l}$ 
is either zero or again a highest weight vector. In the latter case, the
\rep contains an invariant submodule. In particular, 
\beq
(X_i^-)^{M_{(i)}}\cdot w_{\l}=0 
\eq
on all irreps with highest weight vector $w_{\l}$. 
Due to the braid group action
(\ref{braid_action}) resp. algebra automorphism, similar statements apply 
to all root vectors $X^{\pm}_{\b_r}$, and considering the P.B.W. basis
of $\U$, it follows that all irreducible 
highest weight representations are finite--dimensional at roots of unity.
This is very different from the generic case. 

Another important feature is the existence of non--trivial one--dimensional
\reps at roots of unity, namely 
$w_{\l_0}$ with weight $\l_0 = \sum \frac m{2n} k_i \a_i$
for integers $k_i$. It is easy to check from (\ref{UEA}) that this is indeed
a \rep of $\U$. By tensoring
any \rep with $w_{\l_0}$, one obtains another \rep with identical structure, 
but all weights shifted by $\l_0$. 

There exist also "cyclic" \reps with $(X_i^-)^{M_{(i)}} =\mbox{const}$ if
$q^{H_i}$ is used instead of $H_i$, see \cite{CH_P}.

Assume  $V(\l)$ is a highest weight module of $\U$ which is analytic in $q$
(i.e. the vector space $V(\l)$ is fixed, but the action of $U$ on it depends
analytically on $q$, such as a Verma module with dominant integral $\l$). 
The submodules contained in $V(\l)$ for generic $q$ will certainly 
survive at roots of unity,
since a highest weight vector $w$ is characterized by
$(\sum_i X_i^+) \cdot w=0$, which at roots of unity may have more, 
but not fewer
solutions than generically. In fact, highest weight modules typically develop
additional h.w. vectors at roots of unity. We can see this in the example of a
Verma module of $U_q(sl(2))$:

Let $M(j)$ be the Verma module of $U_q(sl(2))$ with highest weight $\l=j$,
i.e. $H \cdot w_j = j w_j$ and $X^+ \cdot w_j=0$.
%(the algebra is $[H,X^{\pm}] = \pm 2 X^{\pm}$ and [X^+,X^-]= [H]_q$). 
Then $M(j)$ has a basis 
$\{w_j, (X^-)^k\cdot w_j;\quad k\in \N\}$ with weights 
$j, j-2, \dots$ . For generic $q$, $M(j)$ contains another 
highest weight vector 
only if $j\in\N$, namely with weight $-j-2$; this can be seen from 
(\ref{keller_eq}).
However for $q$ a root of unity, $[H-k+1]_q=0$ if $H-k+1=M$, and
(\ref{keller_eq}) implies that there is an additional h.w. vector 
at weight $j-2k=j-2(j+1-M) = 2M-j-2$ (if this is smaller than $j$ 
and $j\in \Z$), another one
at weight $j-2M$, and so on. In fact, the weights of all the h.w. vectors 
in $M(j)$ can be obtained
from $j$ by the action of the "affine Weyl group" generated by reflections 
$\sigma_l$ with reflection centers $l M-\rho=l M-1$, for any
$l\in\Z$. An analogous statement (the "strong linkage principle")
holds in the higher rank case as
well, and will be discussed in section \ref{subsec:unitary_so5}. 
This can be used to determinine
the structure of the irreps of $\U$. 

In summary, the highest weight 
irreps at roots of unity are "usually" smaller than the irreps
with the same highest weight 
for generic $q$, and they are always finite--dimensional.

\paragraph{Tensor products}
The coproduct determines the \rep of $\U$ on a tensor product, and for $q$
as in (\ref{q_root}), one can easily see 
using a $q$ --binomial theorem that 
$\Del(X_i^{\pm})^{M_{(i)}}=(X_i^{\pm})^{M_{(i)}}\tens q_i^{M_{(i)}H_i/2} +
     q_i^{-M_{(i)}H_i/2}\tens (X_i^{\pm})^{M_{(i)}}$, cp. \cite{pasquier}.
Therefore if  $V_i$ are highest weight irreps,
then $(X_i^{\pm})^{M_{(i)}}=0$ on $V_1\tens V_2$, and similarly for any number
of factors. In this context,
it is useful to consider the various quantities as being analytic in
$h\equiv q'-q$, where $q$ is fixed to be (\ref{q_root}). Then e.g. 
$[M_{(i)}]_{q'_i}$ has a first--order zero in $h$. In particular, 
$(X_i^{\pm})^{(M_{(i)})} \equiv \frac{(X_i^{\pm})^{M_{(i)}}}{[M_{(i)}]_{q_i}}$ 
is well--defined, and the distinction between the unrestricted 
and restricted specialization mentioned in section \ref{subsec:U_q} 
becomes important. 
We will essentially work in the unrestricted specialization, 
i.e $(X_i^{\pm})^{(M_{(i)})}$ is {\em not} considered an element of  $\U$.  

Consider the tensor product $V_1\tens V_2$ of two \reps $V_1, V_2$
which are irreducible for generic $q$.
It is well known (e.g. \cite{pasquier,keller}) that 
if the $V_i$ are "large enough", $V_1\tens V_2$ does {\em not} decompose into 
the direct sum of irreps at roots of unity, but different generic irreps 
("would--be irreps") in $V_1\tens V_2$ combine 
into irreducible representations; this will be discussed in detail in 
later sections.
One should notice that this can happen because 1) all Casimirs, including
the Drinfeld Casimir $v$, approach the same value on the "would--be irreps"
which recombine as $q'\rightarrow q$, and 
2) the larger of the recombining "would--be irreps"
develops a h.w. vector, which becomes the h.w. vector of the
smaller constituent. In other words, the image of different projectors
(\ref{projectors}) becomes linearly dependent at roots of unity, and they
develop poles. Nevertheless, 

\begin{lemma}  \label{analytic_image_lemma}
The image $Im(P_{\l})$ of $P_{\l}$ (\ref{projectors}) for any given $\l$ 
is analytic even at roots of unity,
in the sense that there exists an analytic basis for it. In particular,
the dimension is the same as generically.
\end{lemma}
\begin{proof}
One can inductively define an analytic basis of $Im(P_{\l}(q'))$ for $q'$ 
near the root of unity $q$ as follows:
Suppose the $\{v_i(q')\}_{i=1}^d$ are analytic and linearly 
independent at $q'=q$, 
and satisfy $P_{\l}(q') \cdot v_i(q') = v_i(q')$. If $d$ is 
smaller than the generic dimension of $Im(P_{\l})$, take a vector 
$v_{d+1} \in Im(P_{\l}(q_0))$ for $q_0$ near $q$ which is not in the 
span of the $\{v_i(q')\}_{i=1}^d$ at $q'=q_0$.
%(the $v_i(q')$ remain independent near $q$).
Define $v_{d+1}(q') \equiv h^k P_{\l}(q')\cdot v_{d+1}$, where $k\in\Z$ 
is such that 
$v_{d+1}(q')$ is analytic and non--vanishing at $q'=q$ (this is possible 
because $P_{\l}(q')$ has only poles).
Then $v_{d+1}(q')$ satisfies 
$P_{\l}(q')\cdot v_{d+1}(q') = v_{d+1}(q')$ for $q'\neq q$,
since the $P_{\l}$ are projectors. Furthermore, $\{v_i(q')\}_{i=1}^{d+1}$
are linearly independent except possibly for isolated values of $q'$, 
and if they are linearly dependent at $q'=q$, one can redefine
$\tilde{v}_{d+1}(q') = \frac 1{h^k}(\sum v_{d+1}(q')-a_i v_i(q'))$, so that 
the new $\{v_i\}_{i=1}^{d+1}$ span the same space at $q'\neq q$,
are analytic {\em and} linearly independent at $q'=q$.
This is always possible, because the determinant defined by the
$\{v_i(q')\}_{i=1}^{d+1}$ is analytic, but not identically zero.
\end{proof}

Notice that it is essential that the $P_{\l}$ have only poles at $q'=q$,
and no essential singularities.
Furthermore, if a vector $w$ is not in $\oplus_{\l} Im(P_{\l})$ 
at the root of unity, then it is clear that
$P_{\l}(q') \cdot w$ will indeed have a pole at $q'=q$ for some 
$P_{\l}$.

\paragraph{$\R$ at roots of unity}
Finally we need to know whether $\R$ makes sense at roots of unity.
This can be answered using the explicit formulas for $\R$ given in 
\cite{kirill_resh,tolstoi,lev_soib}, refining (\ref{R_formula}). It turns out 
that $\R$ is built out of $\tilde{\R}_{\b_i}$, i.e. universal $\R$'s 
of the $U_q(sl(2))$ subalgebras corresponding to all roots.
Looking at (\ref{R_sl(2)}), the term $\frac 1{[l]_q!} ((X^+)^l\tens (X^-)^l)$ 
becomes ill--defined at roots of unity. So
strictly speaking $\R$ does not exist as element of $\U \tens\U$,
but its action on \reps $V_1 \tens V_2$ is well--defined 
at roots of unity {\em provided}  $(X_i^{\pm})^{M_{(i)}}\cdot V_i$ vanishes 
on all \reps $V_i$. In particular, $\R$ is well--defined 
if all $V_i$ are irreps, and then all the formulas for $\R$ hold 
by analyticity.
The same is true for its many--argument cousin $\R_{12...l}$.

It should be obvious by now that we are dealing with a structure which is
very different from the usual \rep theory of Lie groups and algebras.
The most remarkable objects however are the indecomposable \reps
which have barely been mentioned. They will be studied in later sections.
But first, we make a digression and consider quantum spaces.

\chapter{Quantum Spaces associated to Quantum Groups}

The classical Lie groups $SL(N)$, $SO(N)$ and $Sp(N)$ act naturally on 
$N$--dimensional vector spaces $\M$ ("vector representation"), 
preserving certain objects such as
volume--elements or bilinear forms. There exists
a perfect analog for quantum groups, introduced by Faddeev, Reshetikhin and 
Takhtadjan \cite{FRT}. In the spirit on noncommutative geometry, one does not
consider the spaces themselves, but the algebras of functions $Fun(\M)$ 
on them, which upon quantization turn into noncommutative algebras 
$Fun(\M_q)$.
Because it is customary in the literature, we will use the dual formulation 
of quantum groups in this chapter, namely $Fun(G_q)$ as explained 
in section \ref{subsec:dual_hopf}.

\section{Definitions and Examples}

\subsection{Actions and Coactions}
\label{subsec:act_coact}

So far, with "representation" we always meant a 
{\em left action} of $\U$ on a vector space $V$. In this chapter,
we will be more explicit, and instead of writing $x\cdot v$ we will
write $x\tr v$ for $v\in V$. 
A {\em left action} of an algebra $\Alg$ on a vector space $V$ is
defined by 
\beq
(x y) \tr v = x \tr (y \tr v),\quad 1 \tr v = v
\eq
for $x\in\Alg$, and $V$ is called a left $\U$-module. 
If the representation space is not only a vector space but also 
an algebra $\F$
and $\Alg$ is a Hopf algebra (such as $U_q(g)$),
we can in addition ask that this action preserve the algebra structure
of $\F$, i.e. \quad $x \tr (a b) =
(x_{(1)} \tr a) (x_{(2)} \tr b)$ and
$x \tr 1 = 1 \eps(x)$ for all $a,b \in \F$ and  $x\in \Alg$.
$\F$ is then called a left $\Alg$-module algebra.

Similarly, a {\em right action} of $\Alg$ 
on a vector space $V'$ is
defined by 
\beq
v' \tl (xy) = (v'\tl x)\tl y,\quad v'\tl q = v',
\eq
and $V'$ is called a right $\Alg$-module; correspondingly one defines
right $\Alg$--module algebras.

Just like the comultiplication is the dual operation to
multiplication, {\em right {\rm or} left coactions} are dual to left or
right actions, respectively. A left coaction
of a coalgebra $\C$ (i.e., $\C$ is equipped with a coproduct) 
on a vector space $V$ is defined as a linear map
\beq
\Del_{\C} : \; V \rightarrow \C \tens V:\quad
v \mapsto \Del_{\C}(v) \equiv v^{(1)'} \tens v^{(2)},
\eq
such that
\beq
(\id \tens \Del_{\C})\Del_{\C} = (\Del \tens \id)\Del_{\C},\quad
(\eps \tens \id )\Del_{\C} = \id.
\eq
The prime on the first factor marks a left coaction.
If $\C$ is a Hopf algebra coacting on an algebra $\F$, we say that
$\F$ {\em is a right} $\C${\em -comodule algebra} if 
$\Del_{\C}(a\cdot b) = \Del_{\C}(a) \cdot \Del_{\C}(b)$ 
and $\Del_{\C}(1) = 1 \tens 1$,
for all $a,b \in \F$. Similarly one defines right comodule algebras.

Now if the coalgebra $\C$ is dual to an algebra $\Alg$ in the sense of 
(\ref{hopf_dual}), then a left coaction of $\C$ on $V$ induces a right action 
of $\Alg$ on $V$ and vice versa, via
\beq
v \tl x \equiv <x, v^{(1)'}> v^{(2)}, \label{act_coact}
\eq
and right coactions induce left actions. More on these structures can be
found in \cite{majid,schupp_th}.

For our purpose, we will consider left coactions of $Fun(G_q)$ on
left comodule algebras $Fun(\M_q)$, which according to the above
corresponds to right actions of $\U=U_q(g)$ on $Fun(\M_q)$.
Notice that for quantum groups, a left $\U$--module algebra $\F$
can always be
transformed into a right $\U$--module algebra and vice versa 
using the (linear!)
Cartan--Weyl involution: $a\tl x \equiv \theta(x) \tr a$
for $a \in \F$ and $x\in\U$. Alternatively, one could use the antipode
instead of $\theta$, but this is a priori not compatible 
with the algebra structure of $\F$.

\subsection{Quantum Spaces and Calculus as Comodule Algebras}
   \label{subsec:q_spaces}

First a word on the conventions. We have seen in section 
\ref{subsec:properties}
that $Fun(G_{q^{-1}})$ is dual to $U_{q}(g)$. However most of the 
literature on quantum spaces uses $Fun(G_q)$, and therefore we
will do the same in this section. We may later have to 
replace $q$ by $q^{-1}$ when we make contact with $U_q(g)$. 

Recall that $Fun(G_q)$ is the algebra generated by matrix elements $A^i_j$ 
with relations (\ref{RTT})
\beq
\hat R^{ik}_{mn} A^m_j A^n_l = A^i_n A^k_m \hat R^{nm}_{jl}, 
\eq
where $\hat R^{ik}_{mn} = R^{ki}_{mn}$  and
$R^{ik}_{mn}$ is the $N$--dimensional vector \rep of $\R$. 
This is nothing but the statement that the $\hat R$ -- matrix commutes
with the action of $\U$, in the dual picture. 
The explicit form of $\hat R$  depends on the group  
and is given e.g. in \cite{FRT}. 
Unless we are dealing with $Fun(GL_q(N))$, this has to
be supplemented by additional relations corresponding to invariant 
bilinear forms or determinants (otherwise $\U$ and $Fun(G_q)$ are not 
dual, cp. \cite{majid}). 

\paragraph{Quantum Euclidean group and space}  
We only consider the case of $Fun(SO_q(N))$ and its real forms 
in detail\footnote{Here, the series $B_n$ and $D_n$ can be treated 
simultaneously.}.
In that case, the tensor product of 2 vector \reps contains a trivial
\rep corresponding to the invariant bilinear form. This can be seen
from the $\hat R$ -- matrix, which by virtue of 
section \ref{subsec:centralizer}
decomposes into 3 projectors \cite{FRT}
$\hat R^{ij}_{kl} = (q P^+ - q^{-1} P^- +q^{1-N} P^0)^{ij}_{kl}$.
The metric $g_{ij}$ is then determined by
$(P^0)^{ij}_{kl} =\frac {q^2-1}{(q^N-1)(q^{2-N}+1)} g^{ij} g_{kl}$,
where $g_{ik}g^{kj} = \delta_i^j$. Explicitely, 
\beq
g_{ij} = \d_{i,j'} q^{-\rho_i},  \label{metric}
\eq
where $j'=N+1-i$ and
$\rho_i$ are the values of the Weyl vector $\tilde\rho$ in the
vector representation. For $SO_q(N)$ with $N$ odd, 
$\rho_i = (N/2-1, N/2-2, ... 1/2,0,-1/2, ..., 1-N/2)$.
Furthermore, $D^i_j \equiv \d_{i,j} q^{-2\rho_i} = g^{ik}g_{jk}$ 
generates the square of the antipode (see section \ref{subsec:properties};
notice the replacement $q\rightarrow q^{-1}$ pointed out in 
\ref{subsec:dual_hopf}).
The last equality follows from Proposition  \ref{g_prop}.

In the language of coactions, invariance of $g_{ij}$ becomes
\beq
g_{ij}A^i_k A^j_l = g_{kl},  \label{inv_metric}
\eq
which must be imposed on $Fun(SO_q(N))$.
In section \ref{sec:weyl}, we will find a very 
interesting interpretation 
of $g_{ij}$, which will show various consistency conditions
between $g_{ij}$ and the $\hat R$--matrix. They can be used to show
that (\ref{metric}) is consistent, in particular
that the lhs is central in $Fun(SO_q(N))$. We refer to 
\cite{FRT} or \cite{OZ} for more details. 

Similarly, the tensor product of $N$ vector \reps contains a trivial \rep
corresponding to the totally antisymmetric tensor, 
\beq
A^{i_1}_{j_1} .... A^{i_N}_{j_N} \eps_q^{j_1 .... j_N} = 
  \eps_q^{i_1 .... i_N},  \label{det_1}
\eq
and $\eps_q$ also satisfies certain consistency conditions.
Both $g_{ij}$ and $\eps_q^{i_1 .... i_N}$ 
depend analytically on $q$ and reduce to the classical expressions as 
$q\rightarrow 1$. 

Now (the algebra of functions on) {\em quantum Euclidean space} $Fun(E^N_q)$ 
\cite{FRT} is generated by $x^i$  with commutation relations
\beq
(P^-)^{ij}_{kl} x^k x^l =0.
\eq
The center is generated by 1 and $r^2 = g_{ij} x^i x^j$.
One can go further and define algebras of differential
forms, derivatives, and so on, see \cite{WZ,OZ,zumino_dg}. 
The algebra of differential forms is defined by 
$(P^+)^{ij}_{kl} dx^k dx^l =0$ and  $g_{ij} dx^i dx^j = 0$, i.e.
\beq
dx^i dx^j = -q \hat R^{ij}_{kl} dx^k dx^l.
\eq
The epsilon--tensor is then determined by the unique top - (N-) form
\beq
dx^{i_1} ... dx^{i_N} = \eps_q^{i_1 ... i_N} dx^1 ... dx^N \equiv 
\eps_q^{i_1 ... i_N} \dN x.
\eq
One can introduce derivatives which satisfy
\beq
(P^-)^{ij}_{kl} \dl^k \dl^l =0,
\eq
\beq
\dl^i x^j = g^{ij} + q (\hat R^{-1})^{ij}_{kl} x^k \dl^l,
\eq
and 
\beq
\dl^i dx^j = q^{-1}\hat R^{ij}_{kl} dx^k \dl^l, \quad  x^i dx^j = 
q \hat  R^{ij}_{kl} dx^k x^l.
\label{xdx_CR}
\eq
All this is consistent, and represents one possible choice. 
For more details, see e.g. \cite{OZ}.

It can be checked that all the above relations are preserved 
under the coaction of $Fun(SO_q(N))$
\berr   \label{left_coaction}
\Delta(x^i) &=& A^i_j \tens x^j \equiv x^i_{(1)} \tens x^i_{(2)}, \nn \\
\Delta(dx^i) &=& A^i_j \tens dx^j 
\err
etc., in Sweedler - notation. 

Finally, the {\em quantum sphere} $\Sq$ is generated by $t^i = x^i/r$ 
where $r$ is central, so  $g_{ij} t^i t^j = 1$. 

So far, we have not specified any reality structure,
i.e. all the above spaces are complex. To define real quantum spaces,
we have to impose a star--structure on $Fun(SO_q(N))$ and $Fun(E^N_q)$,
i.e. an antilinear involution on these algebras. 
Again, one has to distinguish 
the cases of $q\in \reals$ and $|q|=1$. In this chapter, we will consider 
the  Euclidean case, which corresponds to $q\in \reals$. 
Later we will consider
the Anti--de Sitter case, for $|q|=1$.

So  from now on $q\in\reals$. Then there is a star--structure 
\beq
\overline{A^i_j} = g^{jm} A^l_m g_{li}  \label{conj_A}
\eq
extended as antilinear involution, which corresponds to
$Fun(SO_q(N,\reals))$ or \newline $Fun(SO_q(N,\reals))$\footnote{These 
are $C^*$ algebras \cite{podles}.}.  
The antipode can then be written as
\beq
 S(A^i_j) = \overline{A^j_i}.
\eq
On quantum Euclidean space, the corresponding involution is 
$\overline{x^i} = x^j g_{ji}$  \cite{FRT}, which
compatible with the left coaction of $Fun((S)O_q(N))$, i.e.
$\obar{\Del(x^i)} = \Del(\obar{x^i})$. Even though the metric (\ref{metric})
looks  unusual because we are working in a weight basis, this is indeed a
Euclidean space. The extension of this 
involution to the differentials and derivatives is quite complicated
\cite{OZ}, but this will not be necessary for our purpose. 

Since $\obar{r^2} = r^2$, this also induces an involution 
on the quantum sphere $\Sq$, which becomes the 
{\em Euclidean quantum sphere}\footnote{another $C^*$ algebra.}.

In this chapter, we will often write
$SO_q(N)\equiv Fun(SO_q(N,\reals))$ for this real ("compact") version of
$Fun(SO_q(N))$, abusing an earlier convention in the dual picture. Similarly,  
we will write $O_q(N)$ if the determinant 
condition (\ref{det_1}) is not imposed for the sake of generality. 

\section{Integration on Quantum Euclidean Space and Sphere}

\subsection{Introduction}

As a first application of this formalism, we will define invariant integrals
of functions or forms over $q$ - deformed Euclidean space 
and spheres in $N$ dimensions. 

In the simplest case of the quantum plane, such an integral was first 
introduced by Wess and Zumino \cite{WZ}; see also \cite{CZ}. 
In the case of quantum  Euclidean space, 
the Gaussian integration method was proposed by a number of authors  
\cite{fiore,KM}. However, 
it is tedious to calculate except in the simplest cases and
its properties have never been investigated thoroughly; in particular, 
we point out that determining the class of integrable functions is a rather 
subtle issue.

In this chapter, we will give a different definition based on spherical 
integration in $N$ dimensions and investigate its properties in 
detail \cite{integral}.
Although this idea has already appeared in the literature \cite{weich}, 
it has not been developed very far. 
We first show that there is a unique invariant
integral over the quantum Euclidean sphere, and prove that it is positive 
definite and satisfies a cyclic property involving the $D$ --matrix of 
$SO_q(N)$. The integral over quantum Euclidean space is then defined by radial
integration, both for functions and $N$ 
forms. One naturally obtains a large class of integrable functions.
It turns out not to be 
determined uniquely by rotation and translation invariance (=Stokes theorem)
alone; it is unique after e.g. imposing a general scaling law.
It is positive definite as well and thus allows to define a Hilbertspace of 
square - integrable functions, and satisfies the same cyclic property.
The cyclic property also holds for the integral of $N$ and $N-1$ --forms over 
spheres, which leads to a simple, truly noncommutative proof 
of Stokes theorem on Euclidean space with and without spherical boundary 
terms, as well as on the sphere. 
These proofs only work for $q\neq 1$,
nevertheless they reduce to the classical Stokes theorem 
for $q\rightarrow 1$.
This shows the power of noncommutative geometry.

Although only the case of quantum Euclidean space is considered here, 
the general 
approach is clearly applicable to other reality structures as well.
In particular, we will later consider the case of quantum 

Anti--de Sitter space,
which is nothing but the quantum sphere $S_q^4$ with a suitable 
reality structure.  As expected, an integral can be obtained 
from the Euclidean case by analytic continuation.
We hope that this will eventually find applications e.g. 
to define actions for 
field theories on such noncommutative spaces.

The conventions are as in the previous section with $q\in\reals$ 
except in some proofs.

\subsection{Integral on the Quantum Sphere $\Sq$}

We first define a (complex - valued) integral $<f(t)>_t$ of a function $f(t)$ 
over $\Sq$.
It should certainly be invariant under $O_q(N)$, 
which means 
\beq
A^{i_1}_{j_1} ... A^{i_n}_{j_n} <t^{j_1} ... t^{j_n}>_t = 
<t^{i_1} ... t^{i_n}>_t .
\eq
Of course, it has to satisfy
\beq
g_{i_l i_{l+1}} <t^{i_1} ... t^{i_n}>_t  =  
<t^{i_1} ...t^{i_{l-1}}t^{i_{l+2}} ... t^{i_n}>_t \quad {\rm and} \quad 
(P^-)^{i_l i_{l+1}}_{j_l j_{l+1}} <t^{j_1} ... t^{j_n}>_t = 0
\eq
We require one more property, namely that $I^{i_1 ... i_n} \equiv  
<t^{i_1} ... t^{i_n}>_t$ is analytic in $(q-1)$, i.e. its Laurent series 
in $(q-1)$ has no negative terms (we can then assume that the classical limit 
$q=1$ is nonzero). These properties in fact determine the spherical integral 
uniquely: for $n$ odd, one should define $<t^{i_1} ... t^{i_n}>_t =0$, and

\begin{prop}\label{spher_int}
For even $n$, there exists  (up to normalization) one and only one tensor 
$I^{i_1 ... i_n} = I^{i_1 ... i_n}(q)$ analytic in $(q-1)$ which is invariant 
under $O_q(N)$ 
\beq
A^{i_1}_{j_1} ... A^{i_n}_{j_n} I^{j_1 ... j_n} = I^{i_1 ... i_n} 
\label{I_inv}
\eq
and symmetric,
\beq
(P^-)^{i_l i_{l+1}}_{j_l j_{l+1}} I^{j_1 ... j_{n}} = 0
\label{I_symm}
\eq
for any l. It can be normalized such that 
\beq
g_{i_l i_{l+1}} I^{i_1 ... i_{n}} = I^{i_1 ... i_{l-1} i_{l+2} ... i_n}
\label{I_norm}
\eq
for any l. $I^{i j} \propto g^{i j}$.

An explicit form is e.g. 
$I^{i_1 ... i_{n}} = \l_n (\Delta^{n/2} x^{i_1} ... x^{i_{n}})$, where 
$\Delta=g_{ij} \dl^i \dl^j$ is the Laplacian 
(in either of the 2 possible calculi),
and $\l_n$ is analytic in $(q-1)$. 
For $q=1$, they reduce to  the  classical symmetric invariant tensors.
\end{prop}
%%%%
\proof
The proof is by induction on $n$. For $n=2$, $g^{i j}$ is in fact the only
invariant symmetric (and analytic) such tensor.

Assume the statement is true for $n$, and suppose $I_{n+2}$ and $ I'_{n+2}$
satisfy the above conditions. Using the uniqueness of $I_{n}$, we have 
(in shorthand - notation)
\berr
g_{12} I_{n+2}        &=& f(q-1) I_{n}  \\
g_{12} I'_{n+2} &=& f'(q-1) I_{n}
\err
where the $f(q-1)$ are analytic,
because the left - hand sides are invariant, symmetric and analytic.
Then $J_{n+2} = f' I_{n+2} - f I'_{n+2}$ is 
symmetric, analytic, and satisfies $g_{12} J_{n+2} = 0$. It remains to show
that $J=0$.

Since $J$ is analytic, we can write
\beq 
J^{i_1 ... i_{n}} = \sum_{k=n_0}^{\infty} (q-1)^k J_{(k)}^{i_1 ... i_n}.
\eq
$(q-1)^{-n_0} J^{i_1 ... i_n}$ has all the properties of $J$ 
and has a well-defined, nonzero limit as $q \rightarrow 1$; so we may assume
$J_{(0)} \neq 0$.
                 
Now consider invariance,
\beq
J^{i_1 ... i_n} = A^{i_1}_{j_1} ... A^{i_n}_{j_n} J^{j_1 ... j_n}.
\label{J_inv}
\eq
This equation is valid for all $q$, and we can take the limit $q
\rightarrow 1$.
Then $A^i_j$ generate the commutative algebra of 
functions
on the classical Lie group $O(N)$, and $J$ becomes $J_{(0)}$, which is just
a classical
tensor. Now $(P^-)^{i_l i_{l+1}}_{j_l j_{l+1}} J^{j_1 ... j_n} = 0$ 
implies that $J_{(0)}$ is symmetric
for neighboring indices, and therefore it is completely symmetric. 
With $g_{12} J = 0$, this implies that $J_{(0)}$ is totally traceless
(classically!).
But there exists no
totally symmetric traceless invariant tensor for $O(N)$.
%(This can be seen as follows: The only totally symmetric invariant tensor
%$J_{(0)}^{i_1...i_{n}}$ is of the
%form ${\cal S}(g^{i_1 i_2} ... g^{i_{n-1}i_{n}})$ (cp. \cite{weyl}) where
%${\cal S}$ is the (classical) symmetrizer; but then
%$g_{12} J_{(0)} \neq 0$ since it contains some positive
%but no negative terms). 
This proves uniqueness. In particular,  $I^{i_1 ... i_{n}} = 
\l_n (\Delta^{n/2} x^{i_1} ... x^{i_{n}})$ obviously satisfies the 
assumptions of the proposition; it is analytic, because in evaluating the 
Laplacians, only the metric and the $\hat R$ - matrix are involved, 
which are both analytic. Statement 
(\ref{I_norm}) now follows because $x^2$ is central. 
\endproof

Such invariant tensors have also been considered in \cite{fiore}
(where they are called S), as well as the explicit form involving the
Laplacian. Our contribution is a self - contained proof of their uniqueness.
So $<t^{i_1} ... t^{i_n}>_t \equiv I^{i_1 ... i_n}$ for even $n$ (and 0 for
odd $n$) defines the unique invariant integral over $\Sq$, 
which thus coincides with the definition given in \cite{weich}.

From now on we only consider $N\geq 3$ since for $N=1,2$, Euclidean space 
is undeformed. The following lemma is the origin of the cyclic properties 
of invariant tensors. For quantum groups, the square of the antipode 
is usually not 1. For $(S)O_q(N)$, it is
generated by the $D$ - matrix: $S^2 A^i_j = D^i_l A^l_k (D^{-1})^k_j$ where
$D^i_l = g^{ik} g_{lk}$ (note that $D$ also defines the quantum trace). Then 

\begin{lemma}     \label{inv_lemma}
For any invariant tensor  $ J^{i_1 ... i_n} = 
A^{i_1}_{j_1} ... A^{i_n}_{j_n} J^{j_1 ... j_n}$, 
$D^{i_1}_{l_1} J^{i_2 ... l_1}$ is invariant too:
\beq  \label{D_cycl}    
A^{i_1}_{j_1} ... A^{i_n}_{j_n} D^{j_1}_{l_1} J^{j_2 ... l_1} = 
D^{i_1}_{l_1} J^{i_2 ... l_1}
\eq 
\end{lemma}

\proof
From the above, (\ref{D_cycl}) amounts to 
\beq
(S^{-2} A^{i_1}_{j_1}) A^{i_2}_{j_2} ... A^{i_n}_{j_n} J^{j_2 ... j_n j_1} 
= J^{i_2 ... i_n i_1}.
\eq
Multiplying with $S^{-1}A^{i_0}_{i_1}$ from the left
and using $S^{-1}(ab) = (S^{-1}b)(S^{-1}a)$ and \newline
$(S^{-1}A^{i_1}_{j_1}) A^{i_0}_{i_1} = \delta^{i_0}_{j_1}$, this 
becomes
\beq
A^{i_2}_{j_2} ... A^{i_n}_{j_n} J^{j_2 ... j_n i_0} 
= S^{-1}A^{i_0}_{i_1} J^{i_2 ... i_n i_1}.
\eq
Now multiplying with $A^{l_0}_{i_0}$ from the right, we get
\beq
A^{i_2}_{j_2} ... A^{i_n}_{j_n} A^{l_0}_{i_0} J^{j_2 ... j_n i_0} 
= \delta^{l_0}_{i_1} J^{i_2 ... i_n i_1}.
\eq
But the (lhs) is just $J^{i_2 ... i_n l_0}$ by invariance and thus equal to 
the (rhs).
\endproof

We can now show a number of properties of the integral over the sphere:

\begin{theorem} \label{spher_thm}
\berr
    \overline{<f(t)>_t}     \quad   &=& <\overline{f(t)}>_t    
        \label{reality_t} \\
   <\overline{f(t)} f(t)>_t \quad  &\geq& 0     \label{pos_def_t} \\
   <f(t) g(t)>_t           \quad  &=&     <g(t) f(Dt)>_t     \label{cyclic_t}
\err
where $(Dt)^i = D^i_j t^j$. The last statement follows from
\beq
I^{i_1 ... i_n} = D^{i_1}_{j_1} I^{i_2 ... i_n j_1}.     \label{cyclic_I}
\eq
\end{theorem}

\proof
For (\ref{reality_t}), we have to show that 
$I^{j_n ... j_1} g_{j_n i_n} ... g_{j_1 i_1} = I^{i_1 ... i_n}$. 
Using the uniqueness of $I$, it is enough to show that 
$I^{j_n ... j_1} g_{j_n i_n} ... g_{j_1 i_1}$ is 
invariant, symmetric and normalized as $I$.
So first,
\berr
A^{i_1}_{j_1} ... A^{i_n}_{j_n} 
     \(I^{k_n ... k_1} g_{k_n j_n} ... g_{k_1 j_1}\)
&=& g_{l_1 i_1} ... g_{l_n i_n} \overline{A^{l_n}_{k_n} ... A^{l_1}_{k_1}}
     I^{k_n ... k_1} \nonumber \\
&=& \overline{A^{l_n}_{k_n} ... A^{l_1}_{k_1}I^{k_n ... k_1}}
                g_{l_1 i_1} ... g_{l_n i_n}  \nonumber \\
&=& \(I^{l_n ... l_1} g_{l_n i_n} ... g_{l_1 i_1}\).
\err
We have used that $I$ is real (since $g^{ij}$ and $\hat R$ are real), and 
$A^{i_1}_{j_1} g_{k_1 j_1} = g_{l_1 i_1}\overline{A^{l_1}_{k_1}}$.
The symmetry condition (\ref{I_symm}) follows from standard 
compatibility conditions between $\hat R$ and $g^{ij}$,
and the fact that $\hat R$ is symmetric.
The correct normalization 
can be seen easily using $g^{ij} = g_{ij}$ for $q$ - Euclidean space.

To show positive definiteness (\ref{pos_def_t}), we use the observation 
made by \cite{FRT} that
\beq
t^i \rightarrow A^i_j u^j
\eq
with $u^j = u_1\delta^j_1 + u_N\delta^j_N$ is an embedding $\Sq
\rightarrow Fun(O_q(N))$ for $u_1 u_N = (q^{(N-2)/2} + q^{(2-N)/2})^{-1}$,
since $(P^-)^{ij}_{kl} u^k u^l = 0$ and $g_{ij} u^i u^j = 1$. In fact, 
this embedding also respects the star - structure if one chooses 
$u_N = u_1 q^{1-N/2}$ and real. 
Now one can write the integral over $\Sq$ in terms of the Haar - measure 
on the compact quantum group $O_q(N,\reals)$ \cite{woron,podles}. Namely,
\beq
<t^{i_1} ... t^{i_n}>_t  
  = <A^{i_1}_{j_1} ... A^{i_n}_{j_n}>_A u^{j_1} ... u^{j_n}
\equiv <A^{\und{i}}_{\und{j}}>_A u^{\und{j}},
\eq
(in short notation)
since the Haar - measure $<>_A$ is left (and right) - invariant 
%\beq    \label{Haar_inv}
$<A^{\und{i}}_{\und{j}}>_A = A^{\und{i}}_{\und{k}} <A^{\und{k}}_{\und{j}}>_A =
<A^{\und{i}}_{\und{k}}>_A A^{\und{k}}_{\und{j}}$
%\eq
and analytic, and the
normalization condition is satisfied as well.
Then 
$<\overline{t^{\und{i}}} t^{\und{j}}>_t =
   <\overline{A^{\und{i}}_{\und{k}}} A^{\und{j}}_{\und{r}}>_A
u^{\und{k}} u^{\und{r}}$
and for $f(t) = \sum f_{\und{i}} t^{\und{i}}$ etc., 
\berr
<\overline{f(t)} g(t)>_t &=& \overline{f_{\und{i}}} g_{\und{j}}
<\overline{A^{\und{i}}_{\und{k}}}  A^{\und{j}}_{\und{r}}>_A 
  u^{\und{k}}  u^{\und{r}}
          = <\overline{\(f_{\und{i}}A^{\und{i}}_{\und{k}}u^{\und{k}}\)} 
           \(g_{\und{j}}A^{\und{j}}_{\und{r}}u^{\und{r}}\)>_A \nonumber \\
 &=& <\overline{f(Au)} g(Au)>_A.
\err
This shows that the integral over $\Sq$ is positive definite, because the 
Haar - measure over compact quantum groups is positive definite 
\cite{woron}, cp. \cite{kornw}.

Finally we show the cyclic property (\ref{cyclic_I}). (\ref{cyclic_t}) then 
follows immediately.
For $n=2$, the statement is obvious: $g^{ij} = D^i_k g^{jk}$.\\
%(For $n=4$ in 4 dimensions, it can be checked explicitely.)\\
Again using a shorthand - notation, define 
\beq
J^{12...n} = D_1 I^{23 ... n1}.
\eq
Using the previous proposition, we only have to show that $J$ is symmetric,
invariant, analytic and properly normalized. Analyticity is obvious. 
The normalization follows immediately by induction, using the property 
shown in proposition (\ref{spher_int}).
Invariance of $J$ follows from the above lemma. 
It remains to show that $J$ is symmetric, and the only nontrivial part of
that is $(P^-)_{12} J^{12...n} = 0$.
Define 
\beq
\tilde J^{12...n} = (P^-)_{12} J^{12...n},
\eq
so $\tilde J$ is invariant, antisymmetric and
traceless in the first two indices $(12)$, symmetric in the remaining indices
(we will say that such a tensor has the ISAT property), and analytic.
It is shown below that there is no such $\tilde J$ for $q=1$ (and $N \geq 3$).
Then as in proposition (\ref{spher_int}), the leading term of the expansion of 
$\tilde J$ in $(q-1)$ is classical and therefore
vanishes, which proves that $\tilde J=0$ for any $q$.

So from now on $q=1$. We show by induction that $\tilde J =0$.
This is true for $n=2$: there is no invariant antisymmetric traceless
tensor with 2 indices (for $N \geq 3$). 
Now assume the statement is true for $n$ even, and
that $\tilde J^{12...(n+2)}$ has the ISAT property.
Define 
\beq
K^{12...n} = g_{(n+1),(n+2)} \tilde J^{12...(n+2)}.
\eq
$K$ has the ISAT property, so by the induction assumption
\beq
K=0   \label{K}.
\eq
Define
\beq
M^{145...(n+2)} = g_{23} \tilde J^{12...(n+2)} = 
              {\cal S}_{14} M^{145...(n+2)} + {\cal A}_{14} M^{145...(n+2)}
\eq
where ${\cal S}$ and ${\cal A}$ are the classical symmetrizer and 
antisymmetrizer.
Again by the induction assumption, ${\cal A}_{14} M^{145...(n+2)} = 0$
(it satisfies the ISAT property). This shows that $M$ is symmetric 
in the first two indices $(1,4)$. Together with the definition of $M$, 
this implies
that $M$ is totally symmetric. Further, $g_{14} M^{145...(n+2)} = 
g_{14} g_{23} \tilde J^{12...(n+2)} = 0$ because $\tilde J$ is 
antisymmetric in $(1,2)$. But then $M$ is totally traceless, and as in
proposition (\ref{spher_int}) this implies $M=0$. 
Together with (\ref{K}) and the ISAT 
property of $\tilde J$, it follows that $\tilde J$ is totally traceless.
So $\tilde J$ corresponds to a certain Young tableaux, describing a 
larger - than - one dimensional 
irreducible representation of $O(N)$. 
However, $\tilde J$ being 
invariant means that it is a trivial one - dimensional representation. 
This is a contradiction and proves $\tilde J=0$.

\endproof

Property (\ref{pos_def_t})
(which is also implied by results in \cite{fiore}, once the uniqueness
of the invariant tensors is established)
%\footnote{as was pointed out to me by G. Fiore,
%positivity is also implied by results in \cite{fiore}}
in particular means that one can now define the 
Hilbertspace of square - integrable functions on $\Sq$. The same will be true 
for the integral on the entire Quantum Euclidean space. 

The cyclic property (\ref{cyclic_t}) is a strong constraint on 
$I^{i_1 ... i_n}$ and could actually be used to 
calculate it recursively, besides its obvious interest in its own. 
An immediate consequence of (\ref{cyclic_t}) is 
$<f(Dt)>_t = <f(t)>_t$, which also follows from rotation invariance of the
integral, because $D$ is essentially the representation of the (exponential 
of the) Weyl vector of ${\cal U}_q(SO(N))$.

Notice that although it may not look like, (\ref{cyclic_t}) is consistent
with conjugation: even though the $D$ - matrix is real, we have
\beq
\overline{\overline{f}(Dt)} = f(D^{-1}t).
\eq
To see this, take $f(t)=t^i$; then the (lhs) becomes
\berr \overline{D(\overline{t^i})} &=& \overline{D(t^j g_{ji})} = 
\overline{D^j_k t^k g_{ji}} =\\ &=& D^j_k t^l g_{lk} g_{ji} = 
  t^l g_{jl} g_{ji} = (D^{-1})^i_l t^l
\err
using the cyclic property of $g$ and $D^i_l = g_{ik} g_{lk}$,
which is the (rhs) of the above.

\subsection{Integral over Quantum Euclidean Space}

It is now easy to define an integral over quantum Euclidean space. 
Since the invariant length $r^2= g_{ij} x^i  x^j$ is central, we can use its
square root $r$ as well, and write any function
on quantum Euclidean space in the form $f(x^i) = f(t^i,r)$. 
We then define its integral
to be 
\beq
<f(x)>_x = <<f(t,r)>_t(r) \cdot r^{N-1}>_r,    \label{integral_x}
\eq
where $<f(t,r)>_t(r)$ is a classical, analytic function in $r$, 
and $<g(r)>_r$ is some 
linear functional in $r$, to be determined by requiring Stokes theorem.
It is essential that this radial integral $<g(r)>_r$ is really a functional 
of the  
{\em analytic continuation of g(r)} to a function on the (positive) real line.
Only then one obtains a large class of integrable functions,
and this concept of integration over the entire space agrees 
with the classical one. 

It will turn out that Stokes theorem e.g. in the form $<\dl_i f(x)>_x = 0$  
holds if and only if 
the radial integral satisfies the scaling property
\beq
<g(qr)>_r = q^{-1} <g(r)>_r.   \label{scaling_r}
\eq
This can be shown directly; we will instead give a more elegant proof later.
This scaling property is obviously satisfied by an arbitrary
superposition of Jackson - sums,
\beq
<f(r)>_r = \int_1^q dr_0 \mu(r_0)\sum_{n=-\infty}^{\infty} f(q^nr_0) q^{n}
                                        \label{jackson_int}
\eq
with arbitrary (positive) "weight" function $\mu(r) >0$. 
The normalization can be fixed such that e.g. $<e^{-r^2}>_r$ gives 
the classical
result. If $\mu(r)$ is a 
delta - function,
this is simply a Jackson - sum; for $\mu(r) = 1$, one obtains
the classical radial integration
\beq
<f(r) r^{N-1}>_r = \int_1^q dr_0 \sum_{n=-\infty}^{\infty}
 q^n  (q^n r_0)^{N-1} 
f(q^nr_0) = \int_0^{\infty} dr r^{N-1}f(r). \label{class_int}
\eq
This is the unique choice of $\mu(r)$ for which the scaling property 
(\ref{scaling_r}) holds for any positive real number, not just for powers
of $q$. We define $f(x^i)$ to be integrable (with respect to $\mu(r)$)
if the corresponding radial integral in (\ref{integral_x}) is finite.
We therefore obtain generally inequivalent integrals for 
different choices of $\mu(r)$, all of which satisfy Stokes theorem.

Let us try to compare the above definitions with the Gaussian approach.
In that case, one does not resort to a classical 
integral, and determining the class of integrable functions 
seems to be rather subtle. The Gaussian integration procedure
is based on the observation that the integral of functions of the type
(polynomial)$\cdot$(Gaussian) is uniquely determined by Stokes theorem 
(and therefore agrees with our definition for any normalized
$\mu(r))$;
one would then like to extend it to more general functions by a limiting
process. Lacking a natural topology on the space of functions (i.e. 
formal power -- series), this limiting process is however quite problematic.
One way to see this is because there are actually many different 
inequivalent integrals labeled by $\mu(r)$,
such a limiting process can only be unique on the (presumably small)
class of functions on which the integral is independent of $\mu(r)$.
Furthermore even classically, although one can calculate e.g. 
$\int \frac 1 {r^2+1} e^{-r^2}$ by expanding it "properly" (i.e. using 
pointwise or $L^2$ convergence) in terms of
Hermite functions, if one tries to expand it formally e.g. in terms of
$\{r^n e^{-r^2} \}$, one obtains a divergent sum of integrals. 
Thus the result may depend on the choice of basis and limiting procedure.
It is not clear to the author how to properly integrate  
functions other  than (polynomial)$\cdot$(Gaussian) in the
Gaussian sense, which would be very desirable, 
because that approach may be applied to some quantum spaces 
which do not have a central length element \cite{KM}.

The properties of the integral over $\Sq$  generalize immediately 
to the Euclidean case, for any positive $\mu(r)$:
\begin{theorem}
\berr
 \overline{<f(x)>_x}    \quad  &=& <\overline{f(x)}>_x  \label{reality_x} \\
 <\overline{f(x)} f(x)>_x \quad &\geq& 0    \label{pos_def_x} \\
 <f(x) g(x)>_x             \quad  &=&  <g(x) f(Dx)>_x,     \label{cyclic_x} 
\err
and 
\beq
<f(qx)>_x \quad   = q^{-N} <f(x)>_x     \label{scaling_x}
\eq
if and only if (\ref{scaling_r}) holds.
\end{theorem}

\proof
Immediately from theorem (\ref{spher_thm}), (\ref{scaling_r}) and
(\ref{integral_x}), using $Dr = r$ and $\mu(r_0)>0$.
\endproof

(\ref{reality_x}) and (\ref{scaling_x}) were already known for the 
special case of the Gaussian integral \cite{fiore}.
It was pointed out to me by G. Fiore that in this case,
positivity was also shown in \cite{fiore_thesis}.

\subsection{Integration of Forms}
\label{sec:forms_euclid}

It turns out to be very useful to consider not only integrals over functions, 
but also over forms, just like classically. As was mentionned before, 
there exists a unique
$N$ - form $dx^{i_1} ... dx^{i_N} = \eps_q^{i_1 ... i_N} \dN x$, and we define
\beq    \label{int_x_form}
\int_x \dN x f(x) = <f(x)>_x,
\eq
i.e. we first commute $\dN x$ to the left, and then take 
the integral over the function on the right.
Then the two statements of Stokes theorem $<\dl_i f(x)>_x = 0$ and 
$\int_x d\omega_{N-1} = 0$ are equivalent. 

The following observation by Bruno Zumino \cite{zumino_priv} 
will be very useful: 
there is a one - form
\beq
\om = \frac {q^2}{(q+1)r^2} d(r^2) = q\frac 1r dr = dr \frac 1{r} 
\eq
where $r dx^i = q dx^i r$, 
which generates the calculus on quantum Euclidean space by 
\beq
[\om, f]_{\pm} = (1-q) df
\eq
for any form $f$ with the appropriate grading. It satisfies 
\beq
d\om = \om^2 = 0.
\eq
We define the integral of a $N$ - form over the sphere $r\cdot\Sq$ 
with radius $r$ by

\beq
\int\limits_{r\cdot\Sq} \dN x f(x) = \om r^N <f(x)>_t = dr r^{N-1} <f(x)>_t,  
            \label{int_S_Nform}
\eq
which is a one - form in $r$, as classically. It satisfies
\beq
\int\limits_{r\cdot\Sq} q^N\dN x f(qx) = \int\limits_{qr\cdot\Sq} \dN x f(x)  
         \label{scaling_S_Nform}
\eq
where $(dr f(r))(qr) = q dr f(qr)$.
Now defining $\int_r dr g(r) = <g(r)>_r$, 
(\ref{int_x_form}) can be written as
\beq
\int_x \dN x f(x) = \int_r (\int\limits_{r\cdot\Sq} \dN x f(x)).
\eq
The scaling property (\ref{scaling_r}), i.e. 
$\int_x \dN x f(qx) = q^{-N}\int_x \dN x f(x)$
holds if and only if the radial integrals satisfies
\beq   \label{scaling_r_forms}
\int_r dr f(qr) = q^{-1} \int_r dr f(r).
\eq
We can also define the integral 
of a $(N-1)$ form $\a_{N-1}(x)$ over the sphere with radius $r$:
\beq   \label{int_sphere_N-1}
\int\limits_{r\cdot\Sq} \a_{N-1} 
= \om^{-1} (\int\limits_{r\cdot\Sq} \om \a_{N-1}).
\eq
The $\om^{-1}$ simply cancels the explicit $\om$ in (\ref{int_S_Nform}),
and it reduces to the correct classical limit for $q=1$. 

The epsilon - tensor satisfies  the cylic property:

\begin{prop}
\beq   \label{cycl_eps}
\eps_q^{i_1 ... i_N} = (-1)^{N-1} D^{i_i}_{j_1} \eps_q^{i_2 ... i_N j_1}.
\eq
\end{prop}

\proof
Define 
\beq
\kappa^{12...N} = (-1)^{N-1} D^1 \eps_q^{23...N1}
\eq
in shorthand - notation again. Lemma (\ref{inv_lemma}) shows that 
$\kappa$ is invariant. $\kappa^{12...N}$ is traceless and 
($q$ -) antisymmetric in (23...N). 
Now $g_{12} \kappa^{12...N}=0$ because
there exists no invariant, totally antisymmetric traceless tensor 
with $(N-2)$ indices for $q=1$, 
so by analyticity there is none for arbitrary $q$. Similarly from 
the theory of irreducible representations of $SO(N)$ \cite{weyl}, 
${P^+}_{12} \kappa^{12...N} = 0$
where ${P^+}$ is the $q$ - symmetrizer, $1=P^+ + P^- + P^0$. 
Therefore $\kappa^{12...N}$ is totally
antisymmetric and traceless (for neighboring indices), invariant 
and analytic. 
But there exists only one such tensor up to normalization
(which can be proved similarly), so 
$\kappa^{12...N} = f(q) \eps_q^{12...N}$. It remains to show $f(q) =1$. 
By repeating the above, one gets 
$\eps_q^{12...N} = (f(q))^N (\det D) \eps_q^{12...N}$ 
(here $12...N$ stands for the {\em numbers} 1,2,...,$N$), 
and since $\det D = 1$, it follows $f(q) = 1$ (times a $N$-th root of unity, 
which is fixed by the classical limit).
\endproof

Now consider a $k$ - form 
$\a_k(x) = dx^{i_1} .... dx^{i_k} f_{i_1 ... i_k} (x)$ 
and a $(N-k)$ - form $\b_{N-k}(x)$.
Then the following cyclic property for the integral over forms holds:
                           %%%%%%%%%%%%%%%%%%%%%%%%%%%%%%%%%%%%%%%
\begin{theorem} \label{cyclic_forms}

\beq   \label{cyclic_forms_any}
\int\limits_{r\cdot\Sq} \a_k(x) \b_{N-k}(x) = 
        (-1)^{k(N-k)} \int\limits_{q^{-k}r\cdot\Sq} \b_{N-k}(x) \a_k(q^N Dx)
\eq
where 
$\a_k(q^N Dx) = (q^N Ddx)^{i_1} .... (q^N Ddx)^{i_k}f_{i_1 ... i_k}(q^N Dx)$.

In particular, when $\a_k$ and $\b_{N-k}$ are forms on $\Sq$, 
i.e. they involve only $dx^i \frac 1r$ and $t^i$,
then 
\beq    \label{cyclic_forms_spher}
\int\limits_{\Sq} \a_k(t) \b_{N-k}(t) 
= (-1)^{k(N-k)}\int\limits_{\Sq} \b_{N-k}(t) \a_k(Dt).
\eq
On Euclidean space, 
\beq    \label{cyclic_forms_x}
\int_x \a_k(x) \b_{N-k}(x) = (-1)^{k(N-k)}\int_x \b_{N-k}(x) \a_k(q^N Dx)
\eq
if and only if (\ref{scaling_r_forms}) holds.
\end{theorem}
Notice that on the sphere, $\dN x f(t) = f(t)\dN x$.

\proof
We only have to show that 
\beq   \label{cycl_spher_f}
\int\limits_{r\cdot\Sq} f(x) \dN x g(x) = 
   \int\limits_{r\cdot\Sq} \dN x g(x) f(q^N Dx)
\eq 
and
\beq   \label{cycl_spher_diff}
\int\limits_{r\cdot\Sq} dx^i \b_{N-1}(x) = 
  (-1)^{N-1} \int\limits_{q^{-1}r\cdot\Sq} \b_{N-1}(x) (q^N Ddx)^i.
\eq
(\ref{cycl_spher_f}) follows immediately from (\ref{cyclic_t}) and 
$x^i \dN x = \dN x q^N x^i$. 

To see (\ref{cycl_spher_diff}), we can assume that $\b_{N-1}(x) 
= dx^{i_2} ... dx^{i_N} f(x)$.
The commutation relations
$x^i dx^j = q\hat R^{ij}_{kl} dx^k x^l$
are equivalent to
\berr   \label{CR_compact}
f(q^{-1}x) dx^j &=& \R((dx^j)_{(a)} \tens f_{(1)}) (dx^j)_{(b)}
           (f(x))_{(2)} \nonumber\\
          &=& (dx^j \triangleleft \R^1) (f(x)\triangleleft \R^2)
\err
where $\R = \R^1 \tens \R^2$ is the universal $\R$ for $SO_q(N)$, 
using its quasitriangular property
and $\R(A^j_k \tens A^i_l) = \hat R^{ij}_{kl}$.
$f\triangleleft Y = <Y,f_{(1)}> f_{(2)}$ is the right action induced by 
the left coaction (\ref{left_coaction}) of an element 
$Y \in {\cal U}_q(SO(N))$.
Now invariance of the integral implies 
\beq  \label{int_inv}
(dx^j \triangleleft \R^1) <f(x)\triangleleft \R^2>_t = dx^j <f(x)>_t,
\eq
because $\R^1 \tens\eps(\R^2) = 1$. 
Using this, (\ref{CR_compact}), (\ref{scaling_S_Nform}) and 
(\ref{int_S_Nform}), the (rhs) of (\ref{cycl_spher_diff}) becomes 
\berr
(-1)^{N-1} \int\limits_{q^{-1}r\cdot\Sq} \b_{N-1}(x) q^N D^i_j dx^j 
     &=& (-1)^{N-1} D^i_j \int\limits_{r\cdot\Sq} dx^{i_2} ... dx^{i_N} 
           f(q^{-1}x) dx^j \nonumber\\
     &=& (-1)^{N-1} D^i_j \eps^{i_2 ... i_N j} \om r^N <f(x)>_t  \nonumber\\
     &=& \eps^{i i_2 ... i_N} \om r^N <f(x)>_t \nonumber\\
     &=& \int\limits_{r\cdot\Sq} dx^i \b_{N-1}(x),
\err
using (\ref{cycl_eps}). This shows (\ref{cycl_spher_diff}), and
(\ref{cyclic_forms_spher}) follows immediately. (\ref{cyclic_forms_x}) 
then follows from (\ref{scaling_r_forms}).

\endproof

Another way to show (\ref{cycl_spher_diff}) following an idea of 
Branislav Jurco \cite{jurco} is to use
\beq
\int\limits_{r\cdot\Sq} (\a_k \triangleleft SY)\b_{N-k} 
= \int\limits_{r\cdot\Sq}  \a_k (\b_{N-k} \triangleleft Y)
\eq
to move the action of $\R^2$ in (\ref{CR_compact}) to the left picking up 
$\R^1 S \R^2$, which generates 
the inverse square of the antipode and thus corresponds to 
the $D^{-1}$ - matrix. This approach however cannot
show (\ref{cyclic_t}) or (\ref{cyclic_x}),
because the commutation relations of functions are more complicated.

(\ref{cyclic_forms_any}) shows in particular that the definition 
(\ref{int_sphere_N-1}) is natural,
i.e. it essentially does not matter on which side one multiplies with $\om$.

Now we immediately obtain {\em Stokes theorem} for the integral over 
quantum Euclidean space, if and only if (\ref{scaling_r_forms}) holds.
Noticing that $\om(q^NDx) = \om(x)$, (\ref{cyclic_forms_x}) implies
\berr
\int_x d\a_{N-1}(x) &=& \frac{1}{1-q} \int_x [\om, \a_{N-1}]_{\pm} 
               \nonumber \\
       &\propto& \int_x \om \a_{N-1} - (-1)^{N-1} \a_{N-1}\om   \nonumber \\
       &=& \int_x (-1)^{N-1} \a_{N-1}\om - (-1)^{N-1} \a_{N-1}\om = 0
\err
On the sphere, we get as easily
\berr
\int\limits_{\Sq} d\a_{N-2}(t) &\propto& \int\limits_{\Sq} 
                   [\om, \a_{N-2}]_{\pm}   \nonumber \\
      &=& \om^{-1} \int\limits_{\Sq} \om 
               (\om \a_{N-2} - (-1)^{N-2} \a_{N-2}\om)) 
     =0
\err
using (\ref{cyclic_forms_spher}) and $\om^2 = 0$. 

It is remarkable that these simple proofs only work for $q\neq 1$, 
nevertheless the statements reduce to the classical Stokes theorem 
for $q \rightarrow 1$. 
This shows the power of the $q$ - deformation technique.

One can actually obtain a version of Stokes theorem 
with spherical boundary terms. 
Define
\beq
\int\limits_{q^k r_0}^{q^l r_0} \om f(r) = 
\int\limits_{q^k r_0}^{q^l r_0} dr \frac 1r f(r) = 
 (q-1) \sum_{n=k}^{l-1} f(r_0 q^n),  \label{int_r_finite}
\eq
which reduces to the correct classical limit, because the 
(rhs) is a Riemann sum. 
Define
\beq
\int\limits_{q^k r_0\cdot\Sq}^{q^l r_0\cdot\Sq} \a_N(x) 
= \int\limits_{q^k r_0}^{q^l r_0} 
(\int\limits_{r \cdot\Sq}\a_N(x)), 
\eq
For $l \rightarrow \infty$ and  $k \rightarrow -\infty$, this becomes 
an integral over Euclidean space as defined before.
Then
\berr
\int\limits_{q^k r_0\cdot\Sq}^{q^l r_0\cdot\Sq} d\a_{N-1} 
    &=& \frac{1}{1-q}\int\limits_{q^k r_0}^{q^l r_0} 
    (\int\limits_{r\cdot\Sq}\om \a_{N-1} -(-1)^{N-1} \a_{N-1}\om) \nonumber\\
    &=& \frac{1}{1-q} \int\limits_{q^k r_0}^{q^l r_0}       
       \big(\int\limits_{r \cdot\Sq}\om \a_{N-1} 
           - \int\limits_{qr \cdot\Sq}\om\a_{N-1} )     \nonumber\\
    &=& \int\limits_{q^l r_0 \cdot\Sq}\a_{N-1} - 
        \int\limits_{q^k r_0 \cdot\Sq}\a_{N-1}.
\err
In the last line, (\ref{int_S_Nform}), (\ref{int_sphere_N-1}) and 
(\ref{int_r_finite}) was used.

%\section{Acknowledgements}
%It is a pleasure to thank Prof. Bruno Zumino for many useful
%discussions and encouragement. I also wish to thank Chris Chryssomalakos, 
%Chong -Sun Chu, Pei - Ming Ho, Branislav Jurco and Bogdan Morariu for 
%very useful discussions, and Gaetano Fiore for pointing out his results
%to me. 

\section{Quantum Anti--de Sitter Space}  \label{sec:ads_space}

Let us first review the classical Anti--de Sitter space (AdS space),
which is a 4--dimensional manifold
with constant curvature and signature \newline
$(+,-,-,-)$. 
It can be embedded as a hyperboloid into a 5--dimensional flat space 
with signature $(+,+,-,-,-)$, by
\beq
z_0^2+z_4^2-z_1^2-z_2^2-z_3^2 = R^2,  \label{ads_metric}
\eq
where $R$ will be called the "radius" of the AdS space. 
Similarly we will consider the 2--dimensional version, defined by
$z_0^2+z_2^2-z_1^2 = R^2$ (of course there is also a 3--dimensional case).
The symmetry group (isometry group) of this space is $SO(2,3)$ 
resp. $SO(2,1)$, which plays the role of the Poincar\'e group. 
In fact, the Poincare group can be obtained from $SO(2,3)$ by a contraction,
see e.g. \cite{lukierski}.
 
This space has some rather peculiar features: First, its time--like 
geodesics are finite and closed. In particular, time "translations"
are a $U(1)$ subgroup of $SO(2,3)$. The space--like geodesics are 
unbounded. Furthermore the causal structure is somewhat 
complicated, but we will not worry about these issues here. 
With the goal in mind to eventually formulate a quantum field theory on a
quantized version of "some" Minkowski--type spacetime, there
are several reason why we choose to work with this space and not e.g. with
de Sitter space (corresponding to $SO(1,4)$) or flat Minkowski space.
First, $SO(2,3)$ has unitary {\em positive--energy} \reps corresponding
to all elementary particles, as opposed to $SO(1,4)$ \cite{fronsdal}, and it
allows supersymmetric extensions \cite{zumino}.
Second, the seemingly simpler case of flat Minkowski space is actually
mathematically more difficult, because the classical Poincare group
is not semi--simple, and the theory of quantum Poincar\'e groups is
not as well developed as in the case of semi--simple quantum groups. 
But the main justification comes a posteriori, namely from the existence
of {\em finite--dimensional} unitary \reps of $SO_q(2,3)$ for any
spin at roots of unity, and some very encouraging results 
towards a formulation of gauge
theories (=theories of massless particles, strictly speaking) in this
framework, which will be presented below. 

The heavy emphasis on group theory seems justified
as a powerful guideline through the vast area of noncommutative geometry.

\subsection{Definition and Basic Properties}  \label{subsec:ads_space_def}

Quantum Anti--de Sitter space (q-AdS space) will be defined 
as a real form of the
complex quantum sphere $S_q^4$ defined above, with an (co)action of 
$SO_q(2,3)$ which is a real form of $Fun(SO_q(5))$ resp. $U_q(so(5))$.
Therefore the algebra of the coordinates $t^i \equiv x^i/r$ is
\berr
(P^-)^{ij}_{kl} t^k t^l &=& 0, \\
t\cdot t \equiv g_{kl} t^k t^l &=& 1.
\err
For $|q|=1$, consider the reality structure
\beq
\obar{t^i} = -(-1)^{E_i} t^j g_{ji} \label{ads_reality}
\eq
extended as an antilinear algebra--automorphism. Here 
$E_i=(1,0,0,0,-1)$ for $i=1,2,...,5$ 
(resp. $E_i = (1,0,-1)$ in the 2--dimensional case),
which will turn out to be the eigenvalues of energy in the 
vector representation. It is easy to check that indeed 
$\obar{t\cdot t} = t\cdot t$.
Correspondingly on $Fun(SO_q(2,3))$, one can consider the reality structure
\beq
\overline{A^i_j} = (-1)^{E_i+E_j}g^{jm} A^l_m g_{li},  \label{conj_ads}
\eq
extended as an antilinear algebra--automorphism.
The fact that $\obar{(..)}$ does not reverse 
the order is not a problem, since we will not consider the $t^i$ as operators,
only as
"coordinate functions" which will mainly be used in integrals,
e.g. to write down Lagrangians. In fact, in quantum field 
theory the coordinates are not considered as operators on a 
Hilbert space. Thus this reality structure on
q-AdS space has mainly illustrative character; some reality properties of the
integral below however will be
used to show hermiticity of interaction Lagrangians (if one would consider 
the $t^i$ as operators on a space of functions on q-AdS space,
the adjoint could be calculated from a positive--definite inner product, and
would {\em not} be given by this reality structure).
Observables like energy etc.  
do exist in our approach, in particular elements in the Cartan
subalgebra of $U_q(SO(2,3))$
which has a suitable reality structure. This is one of the reasons 
why we prefer to work with $\U$ instead of $Fun(g)$. 

To introduce proper units, define
\berr
y^i &\equiv& t^i R, \\
y\cdot y=y^i y^j g_{ij} &=& R^2  \label{ads_y}
\err
for a constant\footnote{$R$ is different from
$r$, which has nontrivial commutation relations with forms.} 
$R \in \reals_{>0}$. 

So from now on $|q|=1$.
It is easy to see that (\ref{ads_reality}) indeed corresponds to
Anti--de Sitter space for $q=1$: consider
$R^2=y\cdot y =y^i y^j g_{ij} = y^1y^5+y^2y^4+ y^3y^3+y^4y^2+y^5y^1$,
and introduce real variables $z^i$ by $y_1=\frac{z^0+iz^4}{\sqrt{2}}$,
$y_5=\frac{z^0-iz^4}{\sqrt{2}}$, $y_2=i\frac{z^1+iz^3}{\sqrt{2}}$,
$y_4=i\frac{z^1-iz^3}{\sqrt{2}}$, $y^3=iz^2$.
Plugging this into (\ref{ads_y}) gives the classical AdS space. 
This also shows that $E_i=(1,0,0,0,-1)$ is indeed the energy 
(in suitable units),
and similarly for the 2--dimensional version.

There are other possible reality structures which could define
an AdS space for $|q|=1$, such as $\bobar{t^i} =- t^i$ and 
$\bobar{A^i_j} = A^i_j$ extended as an antilinear involution.
This is however not compatible with the identification of the energy 
in $U_q(so(2,3))$ which is acting on it.
It will nevertheless be useful in some calculations involving the integral.

\paragraph{Integration.}
One can define an integral $<t^{i_1} .. t^{i_n}>_t$ on q-AdS space by analytic 
continuation in $q$ from the integral over the Euclidean sphere 
$S_q^4$ (This clearly corresponds to the Wick rotation in QFT). 
It trivially satisfies the same algebraic properties as in the Euclidean case,
and is compatible with both reality structures on AdS space:
\begin{lemma} For $|q|=1$, 
\beq
\obar{<t^{i_1} .... t^{i_n}>_t} = <\obar{t^{i_1} .... t^{i_n}}>_t = 
  <t^{i_n} ... t^{i_1}>_t
\label{obar_I_phase}
\eq
\end{lemma}
\begin{proof}
Define $J^{i_n ... i_1}(q) \equiv I^{i_1 ... i_n}(q^{-1})$; then for $|q|=1$,
$J^{i_n ... i_1} = (I^{i_1 ... i_n})^{\ast}$, since $q$ is the only 
complex quantity in the explicit formula in Proposition \ref{spher_int}. 
Applying the above $\bobar{(\;)}$ to
the statement of invariance (\ref{I_inv}), one gets 
$A^{i_n}_{j_n} ... A^{i_1}_{j_1} J^{j_1 ... j_n} = 
J^{i_1 ... i_n}$. Now from a slightly generalized Proposition \ref{spher_int}
where (\ref{I_inv}) and (\ref{I_symm}) are required only for $|q|=1$,
it follows that $J^{i_n ... i_1}(q)=I^{i_1 ... i_n}(q)$,
since $J$ and $I$ are analytic in $q$. 
Alternatively, one can 
consider the anti--algebra automorphism 
$\rho(A^i_j) =A^i_j, \quad \rho(q)=q^{-1}$,
where $q$ is treated as a formal variable.

Now (\ref{obar_I_phase}) follows from (\ref{reality_t}).
\end{proof}

At first sight, it may not look sensible to define an integral of
polynomials on a noncompact space. However we are really interested
in the case of roots of unity, where the analog of "square--integrable 
functions" are indeed obtained as (quotients of ) polynomials, as 
explained in the following sections. The normalization 
has to be refined somewhat at roots of unity, and at this point, 
we make no statement on positivity.

\subsection{Commutation Relations and Length Scale}  \label{subsec:scale}

Let us write down the algebra of coordinate functions
on q-AdS space explicitely.
This can be obtained from the Euclidean case \cite{FRT}.
In the 2--dimensional case one finds 
\beq
q y^3 y^1-q^{-1} y^1 y^3 = (q^{1/2}-q^{-1/2}) R^2,
\label{CR_2d}
\eq
where $y^2$ is eliminated by the constraint $y\cdot y=R^2$.
In 4 dimensions we find
\berr
y^i y^{i+k} &=& q y^{i+k} y^i \qquad\mbox{if $k>0$ and $2i+k \neq 6$},\nn\\
q y^5 y^1-q^{-1} y^1 y^5 &=& \frac{q^{1/2}-q^{-1/2}}{q-1+q^{-1}} R^2\nn\\
q y^4 y^2-q^{-1} y^2 y^4 &=& (1-q^2) y^5 y^1 + 
                               q\frac{q^{1/2}-q^{-1/2}}{q-1+q^{-1}} R^2\nn\\
                         &=& (q^{-2}-1) y^1 y^5 + 
                             q^{-1}\frac{q^{1/2}-q^{-1/2}}{q-1+q^{-1}} R^2
\label{CR_4d}
\err
where $y^3$ is eliminated.
The important point here is that these relations are inhomogeneous,
and therefore contain an {\em intrinsic length scale} 
\beq
L_0 \equiv \sqrt{|q^{1/2}-q^{-1/2}|} R;
\label{scale}
\eq
notice that  $(q-1+q^{-1}) \approx 1$,
having in mind that $q$ should be  very close to 1.
Since $|q|=1$, $q=e^{2\pi i h}$ with $h$ 
a small number (in fact $h=\frac{1}{m}$, as we will see below).
Then $L_0 \approx \sqrt{2\pi h} R$.
Also, notice that $L_0$ is much bigger that $|q-q^{-1}|R$ which one
might have expected naively (and which will show up later).
In flat Euclidean quantum space
for example, the commutation relations are homogeneous, and no length
scale appears.

To make these commutation relations more transparent, 
one can approximate them by 
$[y^5,y^1] = i L_0^2$ and $[y^4,y^2] = i L_0^2$. As in Quantum 
Mechanics, this means that the {\em geometry is classical for scales}
$\gg L_0$, {\em and non--classical for scales} $< L_0$.
Strictly speaking, this is only heuristic since the reality structure 
on the coordinates is
not a standard star structure. However it is clear that there really
{\em is} such a scale, and in the compact (Euclidean) version, 
the argument is indeed rigorous.

This very satisfactory, and the way it should be if this is to find 
applications in high energy physics. Being extremely optimistic, one
is tempted to identify $L_0$ with the Planck scale, where one expects
the classical behaviour of space--time to break down. Of course, there
is no justification for this so far. It means that $q$ has to be {\em very} 
close to one. These considerations are continued in 
section \ref{subsec:many_4d}.

\chapter{The Anti--de Sitter Group and its Unitary Representations}

\section{The Classical Case}

\subsection{$SO(2,3)$ and $SO(2,1)$}  \label{subsec:so_23_c}

The classical AdS group is $SO(2,3)$ resp. $U(so(2,3))$, which is a
real form of $U(so(5,\compl))$
and plays the role of the Poincar\'e group.

\begin{figure}
 \epsfxsize=5in
%  \vspace{-2.5in}
\vspace{-2.1in} 
  \hspace{3.5in}
  \epsfbox{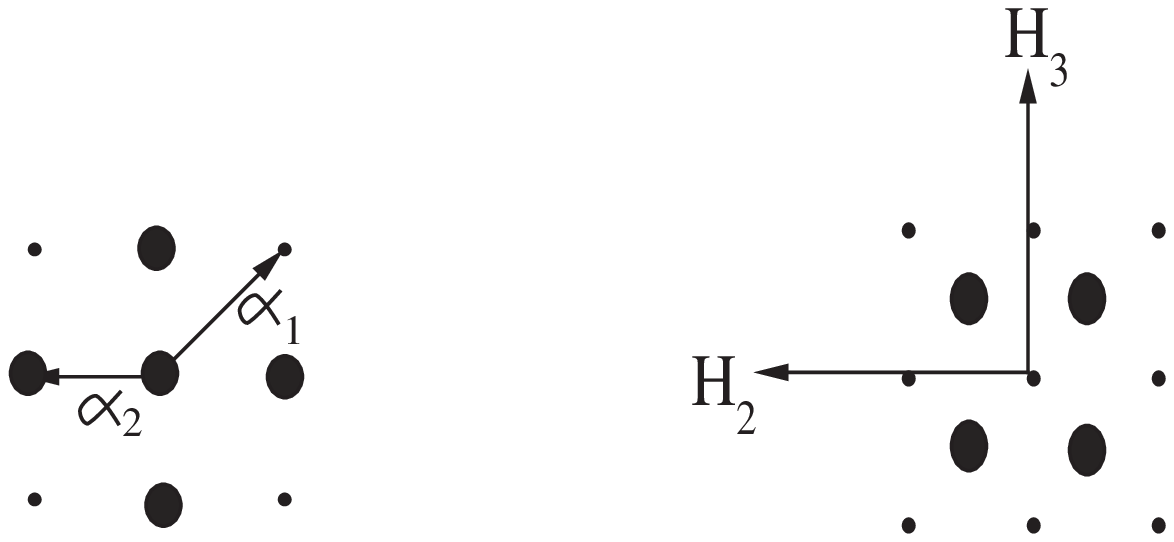}
   \vspace{-2.5in}
\caption{Vector and spinor \reps of $SO(2,3)$}
\label{fig:so23reps}
\end{figure}

The Cartan matrix for its rank 2 Lie algebra $B_2$ is
\beq
A_{ij} = \(\begin{array}{cc}    2 & -2 \\
                                -1 & 2 \end{array} \),
         \qquad (\a_i, \a_j) = \( \begin{array}{cc} 2 & -1 \\
                                                   -1 &  1  \end{array} \) ,
\eq
so $d_1 = 1$, $d_2= 1/2$, to have the standard physics normalization.
The weight diagrams of the vector \rep $V_5$
and the spinor \rep $V_4$ are  given in figure \ref{fig:so23reps} for
illustration; the adjoint $V_{10}$ is 10 --dimensional. The Weyl vector 
is $\rho = \frac 12 \sum_{\a >0} \a = \frac 32 \a_1 + 2 \a_2$.

According to (\ref{little_h}), we define
$h_1= H_1, h_2 = \frac 12 H_2$, 
$e_{\pm 1} =  X^{\pm}_1$, and 
$e_{\pm 2} = \sqrt{[\frac 12]}X^{\pm}_2$. Now one can obtain a 
Cartan--Weyl basis corresponding to all the roots,
as explained in section \ref{subsec:properties}. We choose a slightly 
different labeling here in order to have (essentially) the same 
conventions as in
\cite{lukierski}. Using the longest element of the Weyl group 
$\omega = \tau_1 \tau_2 \tau_1 \tau_2$, define
$\b_1 = \a_1, \b_2 =  \s_1 \s_2 \s_1 \a_2 = \a_2, 
\b_3 = \s_1 a_2 = \a_1 + \a_2, \b_4 = \s_1 \s_2 \a_1 = \a_1 + 2\a_2$,
see figure \ref{fig:so23reps}.
The corresponding Cartan--Weyl basis of root vectors is 
$e_3 = [e_2,e_1]$, $e_{-3} = [e_{-1},e_{-2}]$, $h_3 = h_1+h_2$
and 
$e_4 = [e_2,e_3]$, $e_{-4} = [e_{-3},e_{-2}]$, and $h_4 = h_1+2 h_2$. 
Alternatively one can use the braid--group action (\ref{braid_action}) 
which of course works for the classical case as well, i.e.
$\{e_{\pm 1}, e_{\pm 3}, e_{\pm 4}, e_{\pm 2 }\} = \{e_{\pm 1}, T_1 
e_{\pm 2}, T_1 T_2 e_{\pm 1}, T_1 T_2 T_1 e_{\pm 2}\}$.
Up to signs, this agrees with the basis used in 
\cite{lukierski}. 

The reality structure and the identification of the usual generators of the 
Poincare group in the limit $R \rightarrow \infty$ can be obtained 
by considering the algebra of generators 
leaving the metric invariant, see e.g. \cite{fronsdal,lukierski}.
It turns out that the following reality structure corresponds to $SO(2,3)$:
\beq
\obar{H_i} = H_i, \quad  \obar{X^+_1} = -X^-_1, \quad \obar{X^+_2}
= X^-_2,        \label{obar_ads}
\eq
for $i=1,2$. Then 
\beq
\obar{e_1} = -e_{-1}, \quad \obar{e_2}= e_{-2}, \quad 
\obar{e_3} = -e_{-3}, \quad \obar{e_4}= -e_{-4}.
\eq

We identify the weights of the vector \rep $(y^1,y^2,y^3,y^4,y^5)$ to be
\newline 
$(\b_3, \b_2, 0, -\b_2, -\b_3)$, see figure \ref{fig:so23reps}.
Then $\{e_{\pm 2},h_2\}$ is a compact $SU(2)$ subalgebra which acts 
only on the 
spacial variables $z^1, z^2, z^3$ in AdS space (\ref{ads_metric}). 
It corresponds to spatial rotations, and we will sometimes write 
\beq 
J_z \equiv h_2
\eq
to indicate that it can be interpreted as a component of 
angular momentum. 
Furthermore $\{e_{\pm 3}, h_{3}\}$ is a noncompact $SO(2,1)$ subalgebra 
acting on $z^0, z^2$ and $z^4$.
This is nothing but a 2 dimensional AdS group, and 
\beq
E\equiv h_3
\eq
is the energy since it generates rotations
in the $z^0, z^4$ --plane. Then $E_i = (1,0,0,0,-1)$ as in (\ref{conj_ads}), 
and $\r_i$ is as given in section \ref{subsec:q_spaces} for the Euclidean 
case. 
%This is the basis used in the literature  e.g. for the $\hat \R$ --matrix
%\cite{FRT}. 
The reality structure (\ref{obar_ads}) on $U(so(2,3))$ can now be written as
\beq
\obar{x} \equiv (-1)^E \cobar{x} (-1)^E,   \label{obar_ads_E}
\eq
where $\cobar{x}$ was defined in (\ref{cbar}) for $x\in U(so(2,3))$.

I want to give a  brief explanation for the reality structure 
of $SO(2,1)$. Let
$X = \(\begin{array}{cc}   a & b+c \\  b-c & -a \end{array}\)$, and 
$L \in SL(2,\reals)$. Then $\det(X) = -a^2-b^2+c^2$ is the quadratic form on 
2--dimensional AdS space, which is invariant under $X \rightarrow L^{-1} X L$.
Therefore $SL(2,\reals) = SO(2,1)$, at least locally. Now for
$K_1 \equiv \frac 12 \(\begin{array}{cc}   1 &0 \\  0 & -1 \end{array}\)$, 
$K_2 \equiv \frac 12 \(\begin{array}{cc}   0 & 1 \\  1 & 0 \end{array}\)$,
$K_3 \equiv \frac 12 \(\begin{array}{cc}   0 & -i \\  i & 0 \end{array}\)$
and $K_{\pm} =K_1 \pm i K_2$,
then $\{K_{\pm},K_3\}$ is a $su(2)$ Lie algebra. Furthermore 
$K_a \equiv i K_1, K_b \equiv i K_2$ are purely imaginary, and therefore  
$L=\exp(i(\a_a K_a +\a_b K_b + \a_3 K_3)) \in SL(2,\reals)$ 
for real parameters
$\a_{a,b,3}$. Now in a {\em unitary} representation of $SL(2,\reals)$ 
(which will be infinite--dimensional),
$L^{\dagger} = L^{-1}$, so $K_{a,b,3}^{\dagger} = K_{a,b,3}$. But this means
that $K_{\pm}^{\dagger} = - K_{\mp}$, and $K_3^{\dagger} = K_3$.

\subsection{Unitary Representations, Massless Particles and BRST
from a Group Theoretic Point of View}  \label{subsec:so23_reps_class}

Let us briefly discuss the classical irreducible unitary \reps of $SO(2,3)$
corresponding to elementary particles. Of course, 
they are all infinite--dimensional.

The most important unitary positive--energy 
irreducible representations are lowest--weight \reps $V_{(\l)}$
with lowest weight $\l=E_0 \b_3-s \b_2 \equiv (E_0,s)$ 
for any $E_0$ and $s$ such that $E_0 \geq s+1$, and
both integer or both half integer (i.e. $\l$ is integral, see section 
\ref{sec:rep_theory})) \cite{fronsdal}. 
Unitarity will in fact follow from the quantum case. 
Then $s$ is the spin of an elementary particle with rest energy 
$E_0$. For example, a "scalar field" has $s=0$ and $E_0 \geq 1$, see
figure 3.2; it can be realized in the space of functions
$f(y)$ on AdS space.

These \reps have only discrete weights, nevertheless they become
the usual irreps of the Poincare group in the limit
$R \rightarrow \infty$, with appropriate rescaling.

There also exist remarkable unitary irreps with non--integral weights
and all multiplicities equal to one,
namely the so--called Dirac singletons "Di" for $\l=(1,1/2)$ and "Rac" for 
$\l=(1/2,0)$ \cite{dirac}. While it is not clear if they could be 
of importance in a theory of
elementary particles, we will pursue them nevertheless.
(We will also encounter some more \reps with non--integral weights
in the root of unity case. For a more general (classical) 
discussion, see \cite{fronsdal}.)

\begin{figure}
\begin{center}
\leavevmode
\epsfysize=3in 
\epsfbox{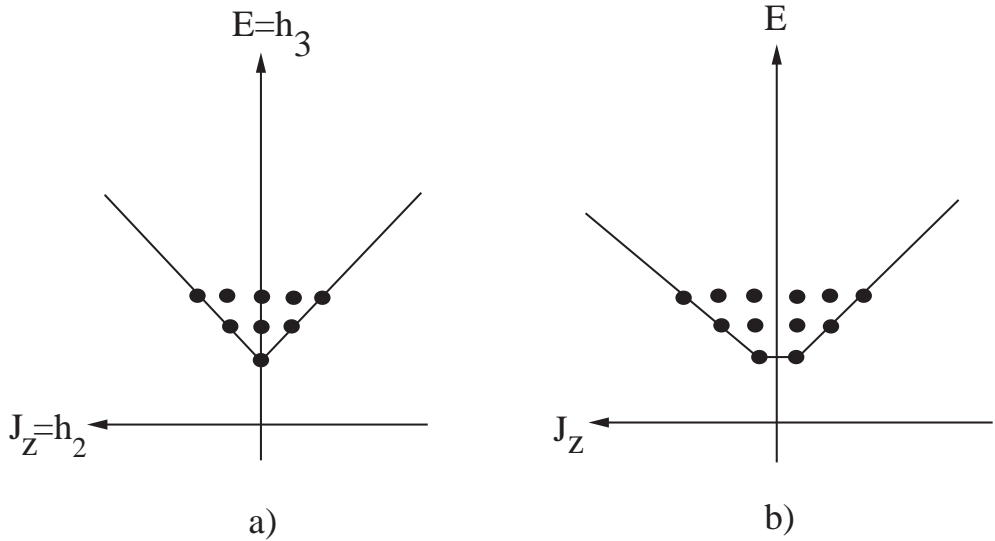}
\caption{a) Scalar field  and b) Spinor field. The vertical axis is energy,
and the horizontal axis is a component of angular momentum.}
\end{center}
\label{fields_01}
\end{figure}

The massless case should be defined as $E_0 = s+1$, cp. \cite{fronsdal}. 
In this case, the rest energy $E_0$ is the smallest possible for a unitary 
\rep of given spin. 
For $s \geq 1$, it is qualitatively different from the massive case 
with the same $s$: the lowest--weight representations, 
which are irreducible in the massive cases,
develop an invariant subspace of "pure gauge" states with lowest weight 
$(E_0+1, s-1)$. 
The \reps however do not split into the direct
sum of "pure gauges" plus the rest, i.e. they are not completely
reducible. This means that there is no complete
covariant gauge fixing, and to get rid of them and obtain a unitarizable,
irreducible representation as required in a quantum theory, 
one {\em has} to factor them out.
They are always null as we will see.

This corresponds precisely to the classical phenomenon in gauge theories,
which ensures that the massless photon, graviton etc. have only their
appropriate number of degrees of freedom. 
In general, the concept of mass
in Anti--de Sitter space is not as clear as in flat space.  Also notice 
that while
"at rest" there are actually still $2s+1$ states, the \rep 
is nevertheless reduced
by one irrep of spin $s-1$. 

The massless representations for spin 1 ("vector field", "photon") and spin 2
("graviton") are shown in figure 3.3, with their 
pure gauge subspaces. There are arrows (indicating the group action)
into the subspace, but not out of it.

\begin{figure}
\begin{center}
\leavevmode
\epsfysize=3in 
\epsfbox{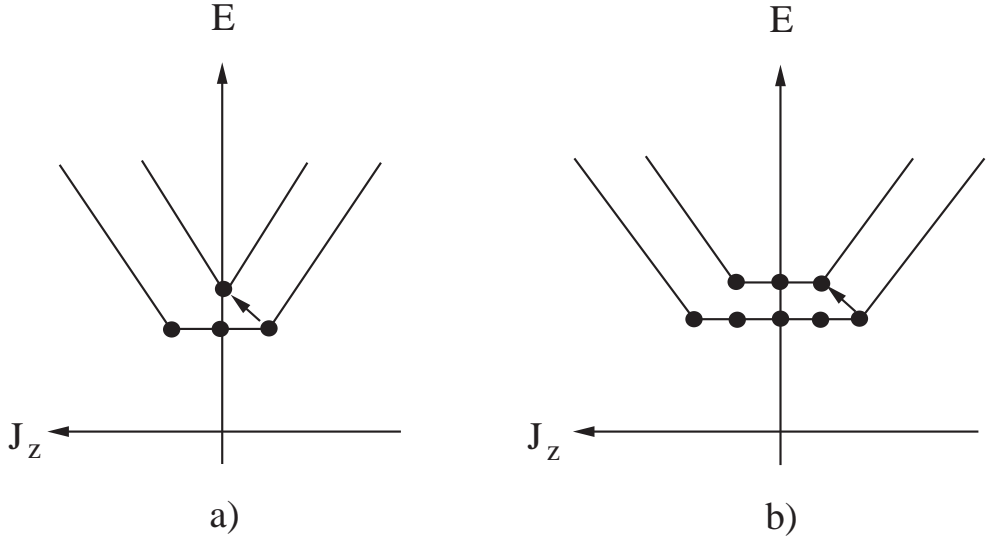}
\caption{a) Photon and b) Graviton, with their "pure gauge" subspaces.}
\end{center}
\label{fig:fields_massless}
\end{figure}

To understand the connection
with the usual formalism, let us consider the spin 1 case in more detail. 
Spin one particles
are usually described by one--forms, i.e. $A(y) = \sum A^i(y) dy^i$  
in the natural embedding of AdS space, where "automatically" 
$\sum g_{ij} y^i dy^j = 0$. 
From a group--theoretic point of view, it would be more
natural (and it is in fact unavoidable on q--AdS space) to
consider unconstrained one--forms $A = \sum A^i(x) dx^i$, 
i.e. including the "radial" component, where $x^i$ are the 
coordinates of the underlying 5--dimensional flat space.
Such a general one--form is an
element of $(\oplus_{E_0} V_{(E_0,0)}) \tens V_5$ and vice versa, 
where $(\oplus_{E_0} V_{(E_0,0)})$ is a space of functions
on AdS space spanned by 
the (unitary) scalar fields $V_{(E_0,0)}$, and $V_5$ is the 
5--dimensional vector representation. 

It is easy to see \cite{fronsdal} that as representations,
\beq
V_{(E_0,0)} \tens V_5 = V_{(E_0,1)} \oplus V_{(E_0+1,0)}
    \oplus V_{(E_0-1,0)},
\label{ads_tens_class}
\eq
see figure 3.4 . 
$V_{(E_0,1)}$ is a vector field, 
$V_{(E_0+1,0)}$ is the space of "radial" one -- forms
$A^{R}(y) dR$ which is usually not considered in the flat case,
and $V_{(E_0-1,0)}$ is what is usually called "longitudinal"
modes, which can be killed by the constraint 
$d\ast A(y) = 0$ ("Lorentz gauge") where $\ast A(y)$ is the Hodge dual of 
$A(y)$ (in 4 dimensions; notice that $d*(V_{(E_0,1)}) \equiv 0$, 
since $d\ast A$ is a scalar).
In fact, the $V_{(E_0-1,0)}$ part {\em has} to be discarded, since it
would lead to negative norm states upon canonical quantization. 

\begin{figure}
\begin{center}
\leavevmode
\epsfysize=3in
\epsfbox{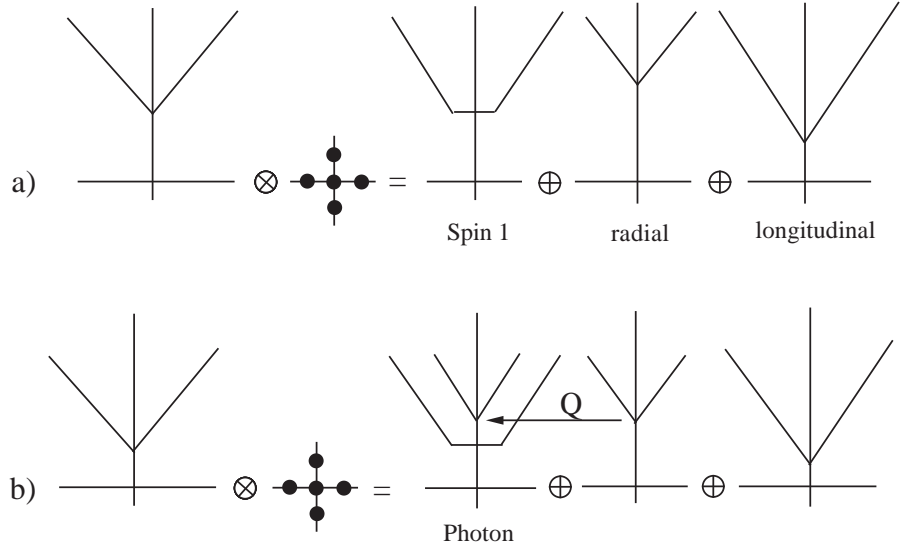}
\caption{One--forms from a group theoretic point of view: a) massive and
  b) massless case, with BRST operator $Q$}
\end{center}
\label{fig:ads_tens_class}
\end{figure}

In the massless case $E_0=2$, $V_{(E_0,1)}$ has a null subspace of pure gauges
(consisting of fields $A(y)=d\Lambda(y)$) which is isomorphic to 
$V_{(E_0+1,0)}$, and must be factored out. The essential and nontrivial point
in a gauge theory is to show that the "pure gauge" subspaces do indeed
decouple, so that they can {\em consistently} be factored out.
Generally in QFT,
this is best done using a BRST operator $Q$, which has the following
characteristic properties:
\begin{itemize}
\item [1)] The space of pure gauges is the image of $Q$ (at ghost number 0)
\item [2)] $Q$ commutes with the $S$ - matrix, the action, etc. 
\item [3)] $Q^2=0$
\label{BRST_char}
\end{itemize}
Then the physical Hilbert space can then be defined as 
the cohomology of $Q$ at ghost number 0, i.e.
\beq
{\cal H}_{phys} = {{\{Q=0\}/_{Im (Q)}}}|_{gh \# 0},
\label{BRST_cohom}
\eq
where $Im (Q)$ is the image of $Q$,
and 2) guarantees that this is consistent, i.e. $ SQ(...)$ = $Q(...)$.
In the standard formulation for photons, $\{Q=0\}$ also implies the constraint
$d\ast A=0$ (on the Hilbert space at ghost number 0), 
but this could as well be imposed by hand.

Now notice that in the AdS case (\ref{ads_tens_class}), 
the radial components of a one--form 
and the subspace of pure gauges are isomorphic, and it is tempting to
define an intertwiner $Q$ from the former to the latter.
On $V_{(E_0,1)}$ and $V_{(E_0-1,0)}$ in (\ref{ads_tens_class}), 
define $Q$ to be 0. 
Then $Q$ acting on $A$ indeed satisfies all the properties 
1) to 3) of a BRST operator,
and the radial component of $A$ plays the role of a ghost;
it is indeed a scalar, and anticommuting as a one -- form. 

Notice that we have only one ghost as opposed to 2 in the usual 
formulation, and
accordingly $\{Q=0\}$ does not constrain the longitudinal modes to vanish
(this has to be imposed in addition).
So this $Q$ does not correspond precisely to the standard BRST operator
in an abelian gauge theory\footnote{I wish to thank B. Morariu for 
discussions on this}.  Nevertheless we will  take the point of view that
the above properties 1) to 3) are the characteristic ones, and call
our $Q$ a BRST operator as well. Actually, we will relax the requirement 
$Q^2=0$ in the most general setting (see theorem \ref{struct_thm}), but it 
will hold on the sectors of representations relevant to
elementary particles.

Thus a BRST operator provides a way to define theories of 
massless elementary particles, i.e. massless unitary irreps.
I consider this to be the essential feature of a (abelian) "gauge theory",
and not some kind of "local gauge invariance" which is  unphysical anyway.

Let us try to see if and how all this works in the q--deformed case.

\section{The Quantum Anti--de Sitter Group at Roots of Unity}

The quantum Anti--de Sitter group is simply $SO_q(2,3) \equiv U_q(so(2,3))$
as explained in section \ref{sec:quantum_groups} for $|q|=1$, with the same 
reality structure (\ref{obar_ads}) as in the undeformed case, i.e.
\beq
\obar{x} = (-1)^{-E} \cobar{x} (-1)^E.
\eq
This is consistent 
with all the properties (\ref{cbar}) to (\ref{obar_R}), as explained 
in section \ref{subsec:reality}. We do not consider $q\in\reals$, because we
will mainly be interested in the roots of unity case.
The root vectors in the quantum case are defined by the braid group action 
as in section \ref{subsec:so_23_c}, now using the 
formulas (\ref{braid_action}). We obtain 
\berr
e_3 &=& q^{-1} e_2 e_1 - e_1e_2, 
       \quad e_{-3} = q e_{-1}e_{-2} - e_{-2} e_{-1}, 
        \quad h_3 = h_1 + h_2  \nonumber \\
e_4 &=& e_2 e_3 - e_3 e_2, 
       \quad e_{-4} = e_{-3} e_{-2} - e_{-2} e_{-3}, 
       \quad h_4 = h_1 + 2 h_2, \label{rootvect_q}
\err
where $h_1= H_1, h_2 = \frac 12 H_2$, 
$e_{\pm 1} =  X^{\pm}_1$ and 
$ e_{\pm 2} = \sqrt{[\frac 12]}X^{\pm}_2$.
Up to a trivial automorphism, (\ref{rootvect_q}) agrees with the basis 
used in \cite{lukierski}. 
The reality structure is 
\beq
\obar{e_1} = -e_{-1}, \quad \obar{e_2}= e_{-2}, \quad 
\obar{e_3} = -e_{-3}, \quad \obar{e_4}= -e_{-4}.
\eq
So $\{e_{\pm 2},h_2\}$ is a $SU_{q^{\frac 12}}(2)$ algebra 
(but not coalgebra), 
and the other three 
$\{e_{\pm\b}, h_{\b}\}$ are noncompact $SO_{\tilde{q}}(2,1)$ algebras. 

Now we can study q --deformed positive energy \reps such as 
vector fields. As pointed out before, the
\rep theory is completely analogous to the classical case 
if $q$ is not a root of unity, 
at least for finite--dimensional representations.
In our case as well, it is easy to see that
\beq
V_{(E_0,0)} \tens V_{5} = V_{(E_0,1)} \oplus V_{(E_0+1,0)}
    \oplus V_{(E_0-1,0)}
\label{ads_tens}
\eq
as before
for $E_0 \geq 2$, and the representation spaces are the same  as classically. 
Then everything is as in section \ref{subsec:so23_reps_class}, 
however we  will see 
below that none of these \reps is unitary unless $q$ is a root of unity.

In the following sections, we will show that for suitable roots of unity, 
there are
unitary \reps of $SO_q(2,3)$ corresponding to all the classical ones mentioned
above \cite{unitary}. They are all finite--dimensional, 
and obtained from "compact"
\reps by a simple shift in energy. Moreover a BRST operator $Q$
will arise naturally, for any spin. We start with the 2--dimensional case,
which is technically simpler.

\subsection{Unitary Representations of $SO_q(2,1)$}
   \label{subsec:unitary_2d}

In this section, we will use some results of \cite{keller} on $SU_q(2)$, 
where $2J$ equals $H$ in our notation. $SO_q(2,1)$ is defined by
\berr
[H, X^{\pm}] &=& \pm 2 X^{\pm}, \quad [X^+, X^-] = [H]_q \\
\Del(H) &=& H\tens 1 + 1 \tens H, \nonumber  \\
\Del(X^{\pm}) &=& X^{\pm} \tens q^{H/2} +          
                            q^{-H/2} \tens X^{\pm},   \nonumber \\
S(X^+) &=& -q X^+, \quad S(X^-) = -q^{-1} X^-, \quad S(H) =-H \nonumber \\
\eps(X^{\pm}) &=& \eps(H) =0 \nonumber
\err
with the reality structure 
\beq
\obar{H} = H, \quad \obar{X^+} = -X^-   \label{star_sl2}
\eq
as explained in section \ref{subsec:so_23_c}. 
Comparing with (\ref{UEA}), this corresponds to the normalization
$d=(\a,\a)/2=1$, but one can easily change to other normalizations
by rescaling $q$, as in section \ref{subsec:U_q}. 

The irreps of $U_q(su(2))$ at roots of unity are well -- 
known \cite{keller}, and we list some facts. 
As in section \ref{subsec:root_of_1}, for
\beq
q=e^{2 \pi i n/m}
\eq
with positive relatively prime integers $m,n$  
let $M=m$ if $m$ is odd, and $M=m/2$ if $m$ is even. As explained in general,
we can assume that 
\beq
(X^{\pm})^M = 0
\eq
on all irreps (this excludes cyclic representations).  Then
all finite - dimensional irreps are highest weight (h.w.) representations 
with dimension $d\leq M$. There are two types of irreps:
\begin{itemize}
\item{$V_{d,z} = \{e_m^j;\quad j= (d-1) + \frac{m}{2n}z, \quad 
                   m=j, j-2, ..., -(d-1) + \frac{m}{2n}z \}$ 
             with dimension $d$, for any 
       $1 \leq d \le M$ and $z \in \Z$, where $H e_m^j = m e_m^j$}
\item{$I^1_z$ with dimension $M$ and h.w. 
        $(M-1) +  \frac{m}{2n}z $, for
        $z \in \compl \setminus \{\Z+\frac{2n}{m}r, 1 \leq r \leq M-1\}$. }
\end{itemize}
Note that in the second type, $z \in \Z$ is allowed, in which case we 
will write $V_{M,z} \equiv I^1_z$ for convenience. We will concentrate
on the $V_{d,z}$ -- \reps from now on.
Furthermore, the fusion rules at roots of unity state that 
$V_{d,z} \tens V_{d',z'}$ decomposes into 
$\oplus_{d''} V_{d'',z+z'} \bigoplus_p I_{z+z'}^p$ where $I_z^p$ 
are the well - known reducible, but indecomposable 
representations of dimension $2M$, see  figure \ref{fig:indecomp}
and \cite{keller}. 

Let us consider the invariant inner product  
$(u,v)$ for $u,v \in V_{d,z}$, as defined in section \ref{subsec:invar_forms}, 
i.e. $\obar{x}$ is the adjoint of $x\in\U$. If $(\ ,\ )$ is 
positive--definite, 
we have a unitary representation.

\begin{prop}
The representations $V_{d,z}$ are 
unitarizable w.r.t  $SO_q(2,1)$ if and only if 
\beq
(-1)^{z+1} \sin(2 \pi n k/m) \sin(2 \pi n(d-k)/m) >0
\eq
 for all  $k=1,...,(d-1)$. 

For $d-1 < \frac{m}{2n}$, this holds precisely if $z$ is odd. 
For $d-1 \geq \frac{m}{2n}$, 
it holds for isolated values of $d$ only, i.e. if it holds for $d$, 
then it (generally) does 
{\em not} hold for $d \pm1, d \pm 2, ...$.

The representations $V_{d,z}$ are 
unitarizable w.r.t  $SU_q(2)$ if
$z$ is even and $d-1 < \frac{m}{2n}$.
 \label{unitary_2d}
\end{prop}

\begin{proof} 
Let ${e^j_m}$  be a basis of $V_{d,z}$ with h.w. $j$.
After a straightforward calculation, invariance implies
\beq
\((X^-)^k \cdot e^j_j, (X^-)^k \cdot e^j_j\) = 
  (-1)^k [k]! [j] [j-1] ... [j-k+1] 
\(e^j_j, e^j_j\)
\eq
for $k=1, ... , (d-1)$, where $[n]! = [1] [2] ... [n]$.
Therefore we can have a positive definite inner product 
$(e^j_m, e^j_n) = \d_{m,n}$  if and only if 
$a_k \equiv (-1)^k [k]! [j] [j-1] ... [j-k+1]$ is a positive number for 
all $k=1,... ,(d-1)$, in which case 
$e^j_{j-2k} = (a_k)^{-1/2} (X^-)^k \cdot e^j_j$. 

Now $a_{k}= -[k] [j-k+1] a_{k-1}$, and
\berr
-[k] [j-k+1]  &=& -[k] [d-k+\frac{m}{2n}z] = -[k] [d-k] e^{i \pi z} \\                  
                     &=& (-1)^{z+1} \sin(2 \pi n k/m) \sin(2 \pi n(d-k)/m) 
\frac{1}{sin(2 \pi n/m)^2}  \nn ,
\err
since $z$ is an integer. Then the Proposition follows.
The compact case is known \cite{keller}.
\end{proof}

In particular, all of them are finite--dimensional, and clearly if $q$ is
not a root of unity, none of the \reps are unitarizable. 

We will be particularly interested in the case of (half)integer 
representations of type $V_{d,z}$ and $n=1, m$ even, for reasons to 
be discussed below. Then $d-1 < \frac{m}{2n} = M$ always holds, and 
{\em the $V_{d,z}$ are unitarizable if and only if $z$ is odd}. These 
representations are centered around $M z$, with dimension 
$\leq M$.

Let us compare this with the classical case. For the 
Anti--de Sitter group $SO(2,1)$, $H$ is 
nothing but the energy. At $q=1$, the unitary 
irreps of $SO(2,1)$ are lowest weight representations with 
lowest weight $j > 0$ resp. highest weight representations with 
highest weight $j<0$. For any given such lowest resp. highest weight 
we can now find a {\em finite--dimensional} unitary 
representation with the same lowest resp. highest weight, 
provided $M$ is large enough (we only consider (half)integer $j$ here).  
These are unitary 
representations which for low energies look like the classical  
one--particle representations, but have an intrinsic high--energy cutoff
if $q \neq 1$, which goes to infinity as $q \rightarrow 1$.
The same will be true in the 4 --dimensional case.

\subsection{Tensor Product and Many--Particle Representations of $SO_q(2,1)$}
\label{subsec:many_2d}

So far we only considered  what could be called one--particle 
representations. Many--particle representations should be defined by some 
tensor product of 2 or more such irreps, which should be 
unitary as well and agree with the classical case at least for
low energies. 

\begin{figure}
 \epsfxsize=4in
% \vspace{-2in}
  \vspace{-1.6in} 
   \hspace{0.8in}
   \epsfbox{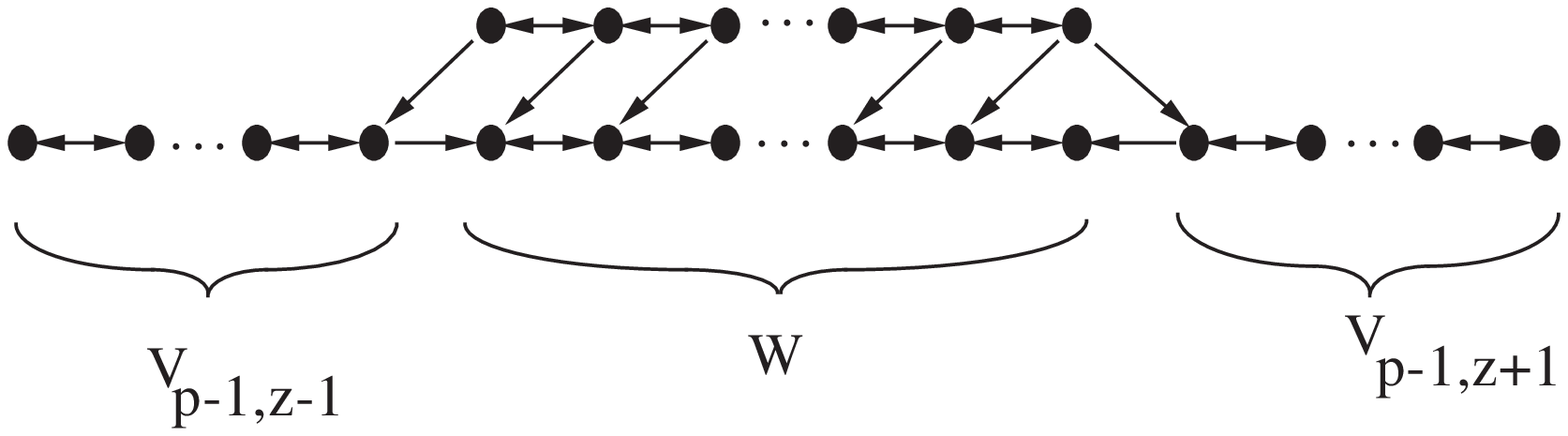}
   \vspace{-2.2in}
 \caption{Indecomposable \rep $I_z^p$}
\label{fig:indecomp}
\end{figure}

Since $\U$ is a Hopf algebra, there is a natural notion of a tensor 
product of two representations, given by the coproduct $\Del$. 
However, it is not unitary a priori. 
As mentioned above, the tensor product of two irreps of type 
$V_{d,z}$ is
\beq
V_{d,z} \tens V_{d',z'} = \oplus_{d''} V_{d'',z+z'}  
\bigoplus_{p=r,r+2,\dots}^{d+d'-M} I_{z+z'}^p  
\label{fusion}
\eq
where $r=1$ if $d+d'-M$ is odd or else $r=2$, and
$I_z^p$ is a indecomposable representation of dimenson $2M$ 
whose structure is shown in figure \ref{fig:indecomp}.
The arrows indicate the 
rising resp. lowering operators. 

In the case of $SU_q(2)$, one defines a truncated tensor 
product $\hat{\tens}$ in the context of CFT by omitting all $I_z^p$ 
representations \cite{mack_schom}. Then the remaining reps are 
unitary w.r.t. $SU_q(2)$; see \cite{mack_schom}.

This is not  the right thing to do for $SO_q(2,1)$.
Let $n=1$ and $m$ even, and consider e.g. $V_{M-1,1}\tens V_{M-1,1}$.
Both factors have lowest energy $H=2$,
and the tensor product of the two corresponding {\em classical} \reps
is the sum
of \reps with lowest weights 4,6,8,\dots . In our case, these weights
are in the $I_z^p$  representations, while the $V_{d'',z''} $ have
$H \geq M \rightarrow \infty$ and are not unitarizable.
So we have to keep the $I_z^p$'s
and throw away the $V_{d'',z''} $'s in (\ref{fusion}). A priori however,
the $I_z^p$'s are
not unitarizable, either. To get a unitary tensor product, note that
as a vector space,
\beq
I_z^p = V_{p-1, z-1} \oplus W \oplus V_{p-1,z+1}
\label{I_decomp}
\eq
where
\beq
W=V_{M-p+1,z} \oplus V_{M-p+1,z}  \label{decomp}
\eq
as vector space. Now
$(X^+)^{p-1} \cdot e_l$ is a lowest weight state where $e_l$  is the
lowest weight vector of $I_p^z$, and similarly $(X^-)^{p-1} \cdot e_h$ is a
highest weight state with $e_h$ being the
highest weight vector of $I_p^z$ (see figure \ref{fig:indecomp}).
It is therefore consistent to consider the submodule of
$I_p^z$ generated by $e_l$, and factor out its submodule generated by
$(X^+)^{p-1} \cdot e_l$; the result is an irreducible \rep
equivalent to $V_{p-1,z-1}$  realized on the left summand in
(\ref{I_decomp}).
Similarly, one could consider the submodule of
$I_p^z$ generated by $e_h$, factor out its submodule generated by
$(X^-)^{p-1} \cdot e_h$, and obtain an irreducible \rep
equivalent to $V_{p-1,z+1}$.
In short, one can just "delete" $W$ in (\ref{I_decomp}).
These  two $V$ - type \reps are unitarizable provided $n=1$ and $m$
is even, and one can either keep both
(notice the similarity with band structures in solid--state
physics), or for simplicity keep the
low--energy part only, in view of the physical application we have in mind.
We therefore  define a truncated tensor product  as
\begin{definition}
For $n=1$ and even $m$,
\beq
V_{d,z} \ttens V_{d',z'} \equiv \bigoplus_{\tilde{d}=r, r+2, \dots}^{d+d'-M}
V_{\tilde{d},z+z'-1}
\label{fusion_t}
\eq
\end{definition}

This can be stated as follows:
Notice that any representation
naturally decomposes as a vector space into sums of $V_{d,z}$'s,
cp. (\ref{decomp}); the
definition of $\ttens$ simply means that only the smallest value of
$z$ in this decomposition is kept, which is the submodule of irreps
with lowest weights $\leq \frac m{2n} (z+z'-1)$.
%(Incidentally, $z$ is the eigenvalue of $D_3$ in the {\em classical}
%$su(2)$ -- algebra $\{D^{\pm} = \frac{(X^{\pm})^M}{[M]!}, 2D_3 = [D^+,
%D^-]\}$, where $\frac{(X^{\pm})^M}{[M]!}$ is understood by some
%limes procedure).
With this in mind, it is obvious that $\ttens$ is associative: both in
$(V_1\ttens V_2) \ttens V_3$ and in $V_1 \ttens (V_2 \ttens V_3)$,
the result is simply the $V$'s with minimal $z$, which is the
{\em same} space, because the ordinary tensor product is
associative and $\Del$ is coassociative. This is in contrast
with the "ordinary" truncated tensor product $\hat{\tens}$
\cite{mack_schom}. Of course, one could give a similar definition for
negative--energy representations. 

This will be generalized to the 4--dimensional case, and in a later section,
we will give a conjecture on a elegant definition of a completely
reducible tensor product using a BRST operator.

$V_{d,z} \ttens V_{d',z'}$ is unitarizable
if all the $V$ 's on the rhs of (\ref{fusion_t}) are unitarizable.
This is certainly true if $n=1$ and $m$ is even. In all other cases,
there are no terms on the rhs of  (\ref{fusion_t}) if the factors
on the lhs are unitarizable, since no $I_z^p$ --type \reps
are generated (they are too large).
This is the reason why we concentrate on this case,
and furthermore on
$z=z'=1$ which corresponds to low--energy representations.
Then $\ttens$ defines a two--particle Hilbert space with the
correct classical limit. So

\begin{prop}
$\ttens$ is associative, and
$V_{d,1} \ttens V_{d',1}$ is unitarizable.
\end{prop}

How the inner product can be induced from the single--particle Hilbert 
spaces will be explained in section \ref{sec:inner_prod_g0}.

Before discussing $SO_q(2,3)$, we will consider the compact case.

\subsection{Unitary Representations of $SO_q(5)$}
   \label{subsec:unitary_so5}

Again $q=e^{2\pi i n/m}$. As explained in sections 
\ref{subsec:invar_forms} and \ref{subsec:root_of_1}, the irreducible h.w. 
\reps $L(\l)$ with highest weight $\l$ can be obtained from the corresponding 
Verma module $M(\l)$ by factoring out its maximal submodule. 
The latter is precisely the null spaces w.r.t. its invariant inner product,
and this  is what we have to determine first.

The following discussion until the paragraph before 
Definition \ref{cpct_def} is
technical and may be skipped upon first reading.
As in section \ref{sec:rep_theory}, 
$Q = \sum \Z \a_i$ is the root lattice, $Q^+ = \sum \Z_+ \a_i$, and
\beq
\l \succ \mu \quad \mbox{if} \quad \l - \mu \in Q^+.
\eq
For $\eta \in Q$, denote \cite{deconc_kac}
\beq
\Par(\eta) = \{\underline{k} \in \Z_+^N; \quad \sum k_i \beta_i = \eta\}.
\eq 
Let $M(\l)_{\eta}$ be the weight space with weight $\l - \eta$ in $M(\l)$. 
Then its dimension is given by 
$|\Par(\eta)|$. If $M(\l)$ contains a h.w. vector with weight
$\sigma$, then the multiplicity of the weight space 
$\(M(\l) / M(\sigma) \)_{\eta}$ is given by 
$|\Par(\eta)| - |\Par(\eta +\sigma- \l)|$,
and so on. 

The character of a representation $V(\l)$ with maximal weight $\l$ 
is the function on weight space defined
by
\beq
\ch V(\l) = e^{\l} \sum_{\eta \in Q^+} \dim{V(\l)_{\eta}} e^{-\eta},
\eq
where again $V(\l)_{\eta}$ is the weight space of $V(\l)$ at weight $\l-\eta$,
and $e^{\l-\eta}(\mu) \equiv e^{(\l-\eta,\mu)}$.
The characters of inequivalent highest weight irreps are linearly independent;
remember that they are all finite--dimensional at roots of unity.
The sum makes sense even for Verma modules and agrees with the classical 
result, 
\beq
\ch M(\l) = e^{\l} \sum_{\eta\in Q^+} |\Par(\eta)| e^{-\eta},
\label{char_verma}
\eq
see \cite{jantzen_m}.

In general, the structure of Verma modules
is complicated and it is
not always enough to know all highest weight vectors, cp. \cite{jantzen_m}.
The proper technical tool to describe the structure of a Verma module is 
its {\em composition series}, or Jordan--H\"older series.
For any module $M$ with a maximal weight, 
consider a sequence of nested submodules 
$ ... \subset W_2 \subset W_1 \subset W_0 = M$ such that 
$W_k / W_{k+1}$ is irreducible, and thus $W_k / W_{k+1}\cong L(\mu_k)$
for some $\mu_k$; this is called a Jordan--H\"older series (it is infinite
for roots of unity, but this is not a problem for the following arguments).
It can be constructed inductively by fixing a
maximal submodule of the $W_k$ (e.g. by factoring out inductively all but one
highest weight submodules of $W_k$, the sum of which is a 
possible $W_{k+1}$).
There are many ways to construct a Jordan--H\"older series, 
but for all of them we obviously have 
$\ch M = \sum \ch (W_k / W_{k+1}) = \sum \ch L(\mu_k)$. 
Since the characters of irreps
are linearly independent, this decomposition of $\ch M$ 
is unique, and so are the subquotients
$L(\mu_k)$. We want to determine these $L(\mu_k)$.

The main tool to find them will be a 
remarkable formula by De Concini and Kac for 
$\det(M(\l)_{\eta})$, the determinant of the invariant inner product matrix of  
$M(\l)_{\eta}$ in a P.B.W. basis, for arbitrary highest weight $\l$.
For $|q|=1$, their result is as follows \cite{deconc_kac}:
\beq
\det(M(\l)_{\eta}) = \prod_{\b \in R^+} \prod_{k \in \N} \([k]_{d_\b} 
                       \frac{q^{(\l+\r-k\b/2,\b)} - 
q^{-(\l+\r-k\b/2,\b)}}{q^{d_{\b}} - q^{-d_{\b}}} \)^{|\Par(\eta - k\b)|}    
\label{deconc}
\eq
where $R^+$ denotes the positive roots,
$d_{\b} =  (\b,\b)/2$, and $k=k_{\b}$ really.

To get some insight, notice first of all that due to  $|\Par(\eta-k\b)|$ 
in the 
exponent, the product is finite. Now for some positive root $\b$, let
$k_{\b}$ be the smallest integer such that 
$D(\l)_{k_{\b},\b} \equiv \([k_{\b}]_{d_\b} \frac{q^{(\l+\r-k_{\b}\b/2,\b)} - 
q^{-(\l+\r-k_{\b}\b/2,\b)}}{q^{d_{\b}} - q^{-d_{\b}}} \) = 0$ (assuming such a
$k_{\b}$ exists) and consider 
the weight space at weight $\l - k_{\b}\b$, i.e. $\eta_{\b} = k_{\b}\b$.
Then $|\Par(\eta_{\b} - k_{\b}\b)| =1$ and
$\det(M(\l)_{\eta_{\b}})$ is zero, so there is a h.w.
vector $w_{\b}$ with weight $\l-\eta_{\b}$ 
(assuming that there is no other with weight $\succ (\l-\eta_{\b})$). 
It generates a submodule which is again a Verma module 
(because $\U$ does not have zero divisors \cite{deconc_kac}), with 
dimension $|\Par(\eta - k_{\b}\b)|$ at weight
$\l - \eta$. This is the origin of the exponent. 
However the submodules generated by the $\w_{\b_i}$ are not independent,
i.e. they contain common h.w. vectors, and there might be other
h.w. vectors at different weights. 
Nevertheless, we will see that all the highest weights $\mu_k$ 
of the composition series of $M(\l)$ are precisely 
obtained in this way. This "strong linkage principle" will be formulated
carefully below. The corresponding statement in the classical case
is well--known \cite{jantzen_m}. While it is not a new insight 
for the quantum case either \cite{dobrev,anderson},
it seems that no explicit proof applicable to our purpose has been 
given (the results in \cite{anderson} apply only
to certain odd roots of unity, and we will see that in fact the even ones 
are most interesting here),
and we will provide one, adapting arguments in \cite{jantzen_m}.

To make the structure more transparent, 
let $\N_{\b}^T$ be the set of positive integers $k$ with $[k]_{\b} = 0$,
and $\N_{\b}^R$ the positive integers $k$ such that 
$(\l + \r -\frac k2 \b,\b) \in \frac m{2n}\Z$. 
Then
\beq
D(\l)_{k,\b} = 0 \Leftrightarrow k \in \N_{\b}^T 
                              \quad\mbox{or} \quad k\in \N_{\b}^R.
\eq
The second condition is 
$k = 2 \frac{(\l+\rho, \b)}{(\b, \b)} + \frac m{2n} \frac 2{(\b,\b)} \Z$,
which means that 
\beq
\l-k \b = \sigma_{\b,l}(\l)
\eq
 where
$\sigma_{\b,l}(\l)$
is the reflection of $\l$ by a plane perpendicular to $\b$ through 
$-\rho+\frac m{4nd_{\b}} l \b$, for some integer $l$. 
For general $l$, $\sigma_{\b,l}(\l) \notin \l+Q$; 
but $k$ should be an integer,
so it is natural to define the {\em (modified) affine Weyl group} 
$\W_{\l}$ of reflections 
in weight space to be generated by those $\sigma_{\b_i,l_i}$ which map $\l$
into $\l+Q$, cp. section \ref{subsec:root_of_1}. 
For $q=e^{2\pi i n/m}$,
two such allowed reflection planes $\perp \b_i$  will differ 
by multiples of $\frac 12 M_{(i)}\b_i$; in the case of $SO_q(5)$,
$M_{(2,3)}=m$ and
$M_{(1,4)}= m$ resp. $m/2$ if $m$ is odd resp. even. Thus $\W_{\l}$
is generated by all reflections by these planes.
Alternatively, it is generated by the usual Weyl group
with a suitable reflection center, and translations by $M_{(i)}\b_i$,
which correspond to $\N_{\b_i}^T$.

%Among all reflections $\sigma_{\b_i,l_i} \in \W_{\l}$ which take a weight 
%$\mu$ 
%into a smaller weight, denote the one "closest to $\mu$" by $\sigma^-_i$;
%that is, $\mu \succ \sigma^-_i(\mu) \succeq \sigma_{\b_i,l_i}(\mu)$ whenever
%$\mu \succ\sigma_{\b_i,l_i}(\mu)$ 
%(we will drop the obvious dependence on $\mu$). In particular, 
%$\sigma^-_i(\l) = \l -k_{\b_i} \b_i$. 

Now the {\em strong linkage principle} states the following:
\begin{theorem}
$L(\mu)$ is a composition factor of the Verma module $M(\l)$ 
if and only if $\mu$ is {\em strongly linked} to $\l$, 
i.e. if there is a descendant sequence of weights
related by the affine Weyl group as
\beq
\l \succ \l_i = \sigma_{\b_i,l_i}(\l)\succ \dots \succ 
 \l_{kj...i} = \sigma_{\b_k,l_k}(\l_{j...i}) = \mu
\label{W_link}
\eq 
\label{h_w_theorem}
\end{theorem}

\begin{proof}
The main tool to show this is the
formula (\ref{deconc}). To make use of it, 
deform $\l$ to $\l'=\l + h\rho$ 
for\footnote{on complex weights, see \cite{deconc_kac} below
Prop. 1.9.} $h \in \compl$, and $q$ to $q' = q e^{i\pi h}$, so that 
$D(\l')_{k_{\b},\b} \neq 0$. 
Consider the inner product matrix
$(a,b)_{\l'}$ for $a,b$ being P.B.W. basis vectors of $M(\l')$;
here $h$ is treated as a formal variable, i.e. no complex conjugation
is implied by the "sesquilinear" form (this is customary in the mathematical 
literature). Then $(a,b)_{\l'}$ is hermitian
if $h\in\reals$, and (\ref{deconc}) holds for any $h\in \compl$. 

Although the $M(\l')$ strictly speaking depend on $\l'$,
we can identify them for different $h$  via the P.B.W. basis.
In this sense,
the action of $X_i^{\pm}$ is  analytic in $h$ since it only depends on the 
commutation relations of the $X^{\pm}_{\b}$, cp. \cite{deconc_kac},
and so is $(a,b)_{\l'}$. According to a theorem
(\cite{kato}, chapter 2, theorem 1.10) for analytic matrices
which are normal for real $h$,
its eigenvalues $e_{\a}$ are analytic, and there exist
analytic projectors $P_{e_{\a}}$
on the eigenspaces $V_{e_{\a}}$ which span the entire 
vectorspace (except possibly at isolated points where some eigenvalues 
coincide; for $h\in\reals$ however, the generic eigenspaces
are orthogonal and therefore remain independent even at such points).
These projectors provide an analytic 
basis of eigenvectors of $(a,b)_{\l'}$ near $\l$.
We can now define
\beq
V_k \equiv \bigoplus_{e_{\a} \propto h^k} V_{e_{\a}},
\eq
i.e. the sum of the eigenspaces with eigenvalues 
$e_{\a}$ with a zero of order $k$ (precisely) at $h=0$.
Of course, $V_k \perp V_{k'}$ for $k \neq k'$. The $V_k$ span the entire 
space, they have an analytic basis as discussed, and have the following 
properties:

\begin{lemma}
\label{order_lemma}
\begin{itemize}

  \item [1)] $(v_k,v)_{\l'} = o(h^k)$ for  
             $v_k\in V_k$ and {\em any} (analytic) $v \in M(\l')$.  
  \item [2)] $X_i^{\pm} v_k ={\displaystyle \sum_{l\geq k}} a_l v_l +
             {\displaystyle \sum_{l =1}^k}  h^l b_l v_{k-l}$ 
            for $v_l\in V_l$ and  $a_l, b_l$ analytic. In particular at $h=0$,
            \beq
              M^k \equiv \oplus_{n \geq k} V_n
            \eq
             is invariant.
%  \item [3)] At $h=0$, 
%             no h.w. state of the form $\U^-V_k$
%             has a finite component in $V_k$.
\end{itemize}
\end{lemma}

\begin{proof} 
\begin{itemize}
  \item [1)] Decomposing $v$ according to $\oplus_l V_l$, only the (analytic)
      component in $V_k$ contributes in $(v_k,v)$, with a factor $h^k$
      by the definition of $V_k$ ($o(h^k)$ means at least $k$ factors of $h$).
  \item [2)] Decompose $X_i^{\pm} v_k = \sum_{e_{\a}} a_{e_{\a}} v_{e_{\a}}$
      with analytic coefficients $a_{e_{\a}}$, 
      corresponding to the eigenvalue $e_{\a}$.  
      For any $v_{e_{\a}}$ appearing on the rhs, consider 
      $(v_{e_{\a}}, X_i^{\pm} v_k) = a_{e_{\a}} (v_{e_{\a}},v_{e_{\a}}) = 
      c\ a_{e_{\a}} e_{\a} $ with $c \neq 0$ at $h=0$ 
      ($v_{e_{\a}}$ might not be normalized).
      But the lhs is $(X_i^{\mp} v_{e_{\a}}, v_k) = o(h^k)$ as shown above.
      Therefore $a_{e_{\a}} e_{\a} =o(h^k)$, which implies 2). 
\end{itemize} \end{proof}

In particular, the quotient
$M(\l)/_{M^1}$ is irreducible and isomorphic to $L(\l)$. 
(The sequence of submodules
$ ... \subset M^2 \subset M^1 \subset M(\l)$ is similar to the Jantzen 
filtration \cite{jantzen_m}.)

By the definition of $V_n$ resp. $M^k$, we have
\beq
\mbox{ord} (\det(M(\l)_{\eta})) = \sum_{k\geq 1} \dim M^k_{\l-\eta}
\eq
where $M^k_{\l-\eta}$ is the weight space of $M^k$ at weight $\l - \eta$,
and $\mbox{ord} (\det(M(\l)_{\eta}))$ is the order of the 
zero of $\det(M(\l)_{\eta})$
as a function of $h$, i.e. the maximal power of $h$ it contains.
Now from (\ref{deconc}) and the above definition of $\N_{\b}^{T,R}$, 
it follows
\berr
\sum_{k\geq 1} \ch M^k &=& e^{\l} \sum_{\eta\in Q^+} (\sum_{k\geq 1}
       \dim M^k_{\l-\eta}) e^{-\eta}  \nn\\
  &=& e^{\l} \sum_{\eta\in Q^+} \mbox{ord}(\det(M(\l)_{\eta})) e^{-\eta}\nn\\
  &=& \sum_{\b\in \R^+} (\sum_{n\in \N_{\b}^T} + \sum_{n\in \N_{\b}^R})
          e^{\l} \sum_{\eta\in Q^+} |\Par(\eta-n\b)| e^{-\eta} \nn\\
  &=& \sum_{\b\in \R^+} (\sum_{n\in \N_{\b}^T} + \sum_{n\in \N_{\b}^R})
          \ch M(\l-n\b) \nn \label{char_calc}\\
\err
where we used (\ref{char_verma}). 

Now we can now prove  (\ref{h_w_theorem}) inductively.
Both the left and the right side of
$(\ref{char_calc})$ can be decomposed into a sum of characters of
highest weight irreps, according to their Jordan--H\"older series. 
These characters are linearly independent.
Suppose that $L(\l-\eta)$ is a composition factor of $M(\l)$. Then the 
corresponding character is contained in the lhs of 
$(\ref{char_calc})$, since $M(\l)/_{M^1}$ is irreducible.
Therefore it is also contained in one of the $\ch M(\l-n\b)$
on the rhs. Therefore $L(\l-\eta)$ is a composition factor  
of one of these 
$M(\l-n\b)$, and by the induction assumption we obtain that 
$\mu \equiv \l-\eta$ is strongly linked to $\l$ as in (\ref{W_link}).

Conversely, assume that $\mu$ satisfies (\ref{W_link}). By the induction
assumption, there exists a $n\in \N_{\b}^T \cup \N_{\b}^R$ such that 
$L(\mu)$ is a subquotient of $M(\l-n\b)$. Then (\ref{char_calc}) shows
that $L(\mu)$ is a subquotient of $M(\l)$.
\end{proof}

Obviously this applies to other quantum groups as well.
In particular, we see again that for 
$q=e^{2\pi i n/m}$, all $(X^-_i)^{M_{(i)}} w_{\l}$ are 
h.w. vectors, and factored out in an irrep. 

\begin{figure}
\epsfxsize=5in
%  \vspace{-2in}
 \vspace{-1.6in}
  \hspace{2.6in}
\epsfbox{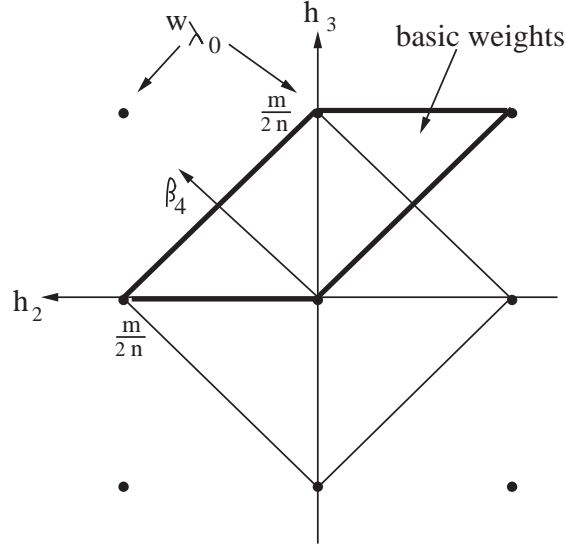}
  \vspace{-1.9in}
\caption{Envelope of compact representations, basic weights and the 
lattice of $w_{\l_0}$}
%\end{center}
\label{fig:basic}
\end{figure}

With these tools, we are now ready to study irreps and determine which ones
are unitarizable, i.e.
for which the inner product is positive definite. As mentioned before,
there exist remarkable nontrivial one--dimensional 
\reps $w_{\l_0}$ with weights $\l_0 = \sum \frac m{2n} k_i \a_i$
for integers $k_i$. By tensoring
any \rep with $w_{\l_0}$, one obtains another \rep with identical structure, 
but all weights shifted by $\l_0$. We will see below that by such a shift, 
\reps which are 
unitarizable w.r.t. $SO_q(2,3)$ are in one--to--one correspondence 
with \reps which are unitarizable w.r.t. $SO_q(5)$.
It is therefore enough to consider highest weights in the following 
domain:

\begin{definition}
A weight $\l = E_0 \b_3 + s \b_2$ 
is called {\em basic} if 
\beq
0 \leq (\l, \b_3) = E_0 < \frac m{2n}, \quad 0 
  \leq (\l, \b_4) = (E_0+s) < \frac m{2n}.
  \label{basic}
\eq
In particular, $\l \succ 0$. It is {\em compact} if in addition 
it is integral (i.e. $(\l, \b_i) \in \Z d_i$),  
\beq
s \geq 0 \quad \mbox{and} \quad (\l, \b_1) \geq 0.
  \label{cpct}
\eq
An irrep with compact h.w. will be called  compact.
\label{cpct_def}
\end{definition}
The region of basic weights is drawn in figure \ref{fig:basic},
together with the lattice of $w_{\l_0}$'s. The compact \reps are
centered around 0, and the (quantum) Weyl group \cite{kirill_resh} 
acts on them, as classically (it is easy to see that the action of
the quantum Weyl group resp. braid group on the compact \reps 
is well defined at roots of unity as well).

A \rep with basic highest weight can be unitarizable w.r.t. $SO_q(5)$ 
(with conjugation $\cobar{(..)}$)
only if all the $SU(2)$'s are unitarizable. For compact 
$\l$, all the $SU_q(2)$'s are indeed unitarizable according to 
section \ref{subsec:many_2d},
using $M_{(2,3)} = m$ and $M_{(1,4)} = m$ resp. $m/2$ if $m$ is odd 
resp. even. This alone
however is not enough to show that they are
unitarizable w.r.t. to the full group.

Although it may seem surprising, there are actually unitary 
representations with nonintegral basic highest weight, namely 
for 
\beq
\l=\frac{m-1}2 \b_3 \quad \mbox{and } \quad 
\l=(\frac m2 -1)\b_3 + \frac 12 \b_2
\label{nonint_cpct}
\eq
for $n=1$ and $m$ even. It follows from theorem (\ref{h_w_theorem}) that
there is a h.w. vector at $\l-2\b_3$ resp. $\l-\b_3$, and all the
multiplicities turn out to be one in the irreps. 
Thus all $SU_q(2)$ modules in
$\b_1, \b_4$ direction have maximal length $M_{(1)} = m/2$,
from which it follows that they are unitarizable. 
The structure is that of shifted Dirac singletons which were 
already studied in \cite{dobrev}, and we will come back to them.

It appears that all other irreps must have integral highest weight 
in order to be unitarizable w.r.t. $SO_q(5)$.
Furthermore, if the highest weight 
is not compact, some of the $SU_q(2)$'s will not be unitarizable.
On the other hand, all irreps with compact highest weight are indeed
unitarizable: 

\begin{theorem}
The structure of the irreps $V(\l)$ with compact highest weight $\l$
is the same as classically except in the cases
\begin{itemize}
  \item [a)] $\l = (m/2-1-s)\b_3 + s \b_2$ for $s\geq 1$ and $\frac m{2n}$ 
        integer, where one additional highest weight state at weight 
       $\l - \b_4$ appears and no others, and 
  \item [b)] $\l=\frac{m-1}2 \b_3$ and 
       $\l=(\frac m2 -1)\b_3 + \frac 12 \b_2$  for $n=1$ and $m$ odd, 
       where one additional highest weight state at weight $\l-2\b_3$ 
        resp. $\l-\b_3$ appears and no others,
\end{itemize}
which are factored out in the irrep.
They are unitarizable w.r.t.  $SO_q(5)$ (with conjugation $\theta^{\ast}$).

The irreps with nonintegral highest weights 
(\ref{nonint_cpct}) as discussed above are unitarizable as well.
\label{cpct_thm}
\end{theorem} 

\begin{proof}
The statements on the structure follow easily from theorem
\ref{h_w_theorem}.

To show that these irreps are unitarizable, consider the
compact representation with highest weight $\l$ before factoring out 
the additional h.w. state,
so that the space is the same as classically. 
For $q=1$, they are known to be unitarizable, so the 
inner product 
is positive definite. Consider the eigenvalues of the inner product matrix
of $(\ , \ )_q$
as $q$ goes from 1 to $e^{2 \pi i n/m}$ along the unit circle.
The only way an eigenvalue could become negative is that it is zero
for some $q'\neq q$. This can only happen
if $q'$ is a root of unity, $q'=e^{2 i \pi n'/m'}$ with  $n'/m' < n/m$.
But then the "non--classical" reflection planes of $\W_{\l}$ 
are further away 
from the  origin and are relevant only in the case
$\l=\frac{m-1}2 \b_3$ for $n=1$ and $m$ odd;
but as pointed out above, no additional eigenvector 
appears in this case for $q' \neq q$.

Thus the eigenvalues might only become zero at $q$.
This happens precisely if a new h.w. vector appears,
i.e. in the cases listed. 
Since there is no null vector in the remaining irrep, all its
eigenvalues are positive by continuity.
\end{proof}

So far all results were stated for h.w. modules; of course the 
analogous statements for lowest weight modules are true as well.
%which can be seen e.g. using the algebra automorphism
%$X^+_i \rightarrow X^-_i$, $H \rightarrow -H$, $q \rightarrow q^{-1}$. 
All the $V(\l)$ in the above theorem have lowest weight $-\l$.

\subsection{Unitary Representations of $SO_q(2,3)$}
   \label{subsec:unitary_so23}

In this section, we will finally see that there are finite--dimensional, 
unitary positive--energy irreps of $SO_q(2,3)$ corresponding to all the 
classical unitary \reps discussed in section \ref{subsec:so23_reps_class},
for suitable roots of unity $q$.
At low energies, their structure is the same as classically 
including the appearance of "pure gauge" 
subspaces in the massless case, for spin $\geq 1$. 
Again, these "pure gauge states" 
can be factored out to obtain the physical, unitary representations. 
At high energies, there is an intrinsic cutoff.

These lowest weight representations
can be obtained from the compact ones by a shift,
as indicated in section \ref{subsec:root_of_1}: 
if $V(\l)$ is a compact h.w. representation, then 
\beq 
V_{(\mu)}\ \equiv V(\l) \tens \w
\eq
with $\w \equiv w_{\l_0}, \l_0 = \frac m{2n} \b_3$ 
has lowest weight $\mu =  -\l + \l_0 \equiv E_0 \b_3 -s \b_2 \equiv (E_0, s)$.
It is a positive--energy representation, i.e. the eigenvalues of 
$h_3$ are positive. 

\begin{figure}
 \epsfxsize=5in
%   \vspace{-2in}
\vspace{-1.2in} 
  \hspace{0.5in}
\epsfbox{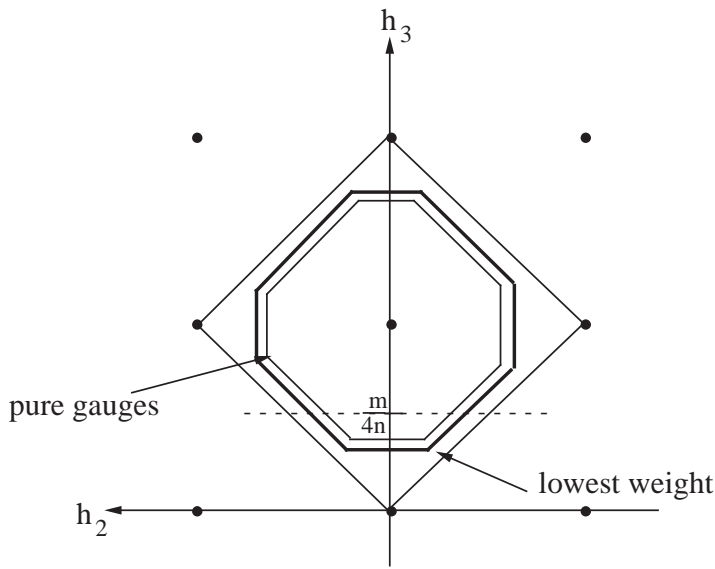}
   \vspace{-2.1in}
\caption{Physical \rep with subspace of pure gauges
  (for $\frac m{2n}$ integer), schematically.
          For $h_3 \leq \frac m{4n}$,
          the structure is the same as for $q=1$.}
\label{fig:gauge}
\end{figure}

For $\frac m{2n}$ integer, $V_{(\mu)}$ corresponds precisely to the
classical positive--energy representation
with the same lowest weight. Again, $E_0$ is the rest energy
and $s$ the spin. For $h_3 \leq m/4n$, the structure  is the same 
as classically, see figure \ref{fig:gauge}.
The irreps with nonintegral highest weights (\ref{nonint_cpct})  
correspond upon this shift to the Dirac singleton 
\reps "Rac" with lowest weight $\mu=(1/2,0)$ and "Di" with $\mu=(1,1/2)$,
as discussed in \cite{dobrev}. 

If $\frac m{2n}$ is not integer, the weights of shifted compact \reps
are not integral. For $n=1$ and $m$ odd, the irreps in 
b) of theorem (\ref{cpct_thm}) now correspond to the singletons, 
again in argeement with \cite{dobrev}. We will see however that this case
does not lead  to an interesting tensor product. 

For $\frac m{2n}$ integer, the cases
$\mu = (s+1, s)$ for $s \geq 1$ will be called "massless" for the same reasons
as in section \ref{subsec:so23_reps_class}.
$E_0$ is the smallest possible rest energy for 
a unitarizable \rep with given $s$ (see below), and  an additional 
lowest weight state with $E_0'=E_0+1$ and $s' = s-1$ appears as classically, 
which generates a null--subspace of "pure gauge" states. 
But now, all these representations are finite--dimensional.

This motivates the following

\begin{definition}
A lowest weight irrep $V_{(\mu)}$ with lowest weight 
$\mu =  (E_0,s) \equiv E_0 \b_3 - s \b_2$ (resp. $\mu$ itself)
is called {\em physical} if it is unitarizable w.r.t. $SO_q(2,3)$
(with conjugation as in (\ref{obar_ads}) ff.).

For $n=1$, $V_{(\mu)}$ is called {\em Di} if $\mu=(1,1/2)$ and {\em Rac} if
$\mu=(1/2,0)$.

For $\frac m{2n}$ integer, $V_{(\mu)}$
is called {\em massless} if $\mu=(s+1,s)$ for $s \in \frac 12 \Z$ 
and $s \geq 1$.
\label{physical}
\end{definition}

\begin{theorem}
A lowest weight irrep $V_{(\mu)}$ is physical precisely if
the shifted irrep with lowest weight
$\mu - \frac m{2n} \b_3$ is unitarizable w.r.t. $SO_q(5)$.

All $V_{(\mu)}$ where  
$-(\mu - \frac m{2n} \b_3)$ is compact
are physical, in particular the massless irreps, as well as
the singletons Di and Rac. For $h_3 \leq \frac m{4n}$,
they are obtained from a (lowest weight)
Verma module by factoring out the 
submodule with lowest weight $(E_0, -(s+1))$ only, except for the massless 
case, where an additional lowest weight state 
with weight $(E_0+1, s-1)$ appears, and for the Di resp. Rac, where an 
additional lowest weight state with weight $(E_0+1,s)$ resp. $(E_0+2,s)$ 
appears.
This is the same as classically, see figure \ref{fig:gauge}.
\label{noncpct_thm}
\end{theorem} 
For the singletons, this was already shown in \cite{dobrev}.

\begin{proof}
As mentioned before, we can write every vector in such a \rep uniquely 
as $a \cdot \w$, where $a$ belongs to a unitarizable
irrep of $SO_q(5)$. Consider the inner product
\beq
\<a\cdot \w, b\cdot\w \> \equiv (a,b),
\eq
where $(a,b)$ is the hermitian inner product on the {\em compact} (shifted) 
representation. Then
\berr
\<a\cdot\w, e_1 (b\cdot\w)\> &=& \<a\cdot\w,(e_1 \tens q^{h_1/2} + q^{-h_1/2}      
                    \tens e_1)b\tens\w \>  \nonumber \\
   &=& q^{h_1/2}|_{\w} (a, e_1 b) = i (a, e_1 b) 
\err
using $h_1|_{\w} = \frac m{2n}$. Similarly, 
\berr
\< e_{-1} (a\cdot\w), b\cdot\w\> &=& \< (e_{-1} \tens q^{h_1/2} +q^{-h_1/2} 
              \tens e_{-1})a\tens\w, b\tens\w\> \nonumber \\
   &=&  q^{-h_1/2}|_{\w} (e_{-1}a, b) = -i (e_{-1} a, b)
\err
because $\<\;,\;\>$ is antilinar in the first argument and linear 
in the second.
Therefore
\beq
\<a\cdot\w, e_1 (b\cdot\w)\>  = -\< e_{-1} (a\cdot\w), b\cdot\w\>.
\eq
Similarly $\<a\cdot\w, e_2 (b\cdot\w)\>  = \< e_{-2} (a\cdot\w), b\cdot\w\>$.
This shows that $\<\ ,\ \>$ is hermitian w.r.t. $\obar{x}$, 
and positive definite
because $(\ ,\ )$ is positive definite by definition. 
Theorem \ref{cpct_thm} now completes the proof.
\end{proof}

As a consistency check, one can see again from section \ref{subsec:unitary_2d}
that all the $SO_{\tilde{q}}(2,1)$ 
resp. $SU_{\tilde{q}}(2)$ subgroups are unitarizable in these representations, 
but this 
is not enough to show unitarizability for the full group. 
Note that as $m \rightarrow \infty$ for $n=1$, 
one obtains the classical one--particle \reps for given $s, E_0$.
We have therefore also proved the unitarizability at $q=1$
for (half)integer spin, which appears to be non--trivial in itself 
\cite{fronsdal}. Furthermore, {\em all \reps obtained 
from the above by shifting
$E_0$ or $s$ by a multiple of $\frac mn$ are unitarizable as well}. 
One obtains in weight space 
a cell--like structure of representations which  are unitarizable w.r.t. 
$SO_q(2,3)$ resp. $SO_q(5)$. 

\subsection{Tensor Product and Many--Particle Representations of $SO_q(2,3)$}
\label{subsec:many_4d}

Finally we want to consider many--particle representations, i.e. find  a
tensor product such that the tensor product of unitary \reps 
is unitarizable, 
as in section \ref{subsec:many_2d}. The idea is the same as there, 
the tensor product of 2 such \reps
will be a direct sum of representations, and we only keep appropriate  
physical lowest--weight subspaces. 
To make this more precise, consider two physical irreps 
$V_{(\mu)}$ and $V_{(\mu')}$ as in Definition (\ref{physical}). 
For a basis $\{u_{\l'}\}$ of physical lowest 
weight states in $V_{(\mu)} \tens V_{(\mu')}$, consider the linear 
span $<\U u_{\l'}>$ of its lowest--weight submodules, and 
let $Q_{\mu,\mu'}$ be the quotient of it after factoring out 
all proper submodules of the $\U u_{\l'}$. 
Let $\{u_{\l''}\}$ be a basis of lowest 
weight states of $Q_{\mu,\mu'}$.
%(notice that 
%the massless $\{u_{\l'}\}$ get factored out
%in $Q_{\mu,\mu'}$, and are not contained any more in $\{u_{\l''}\}$).
Then $Q_{\mu,\mu'} = \oplus V_{(\l'')}$ where $V_{(\l'')}$ are the
corresponding (physical) irreducible lowest weight modules, 
i.e. $Q_{\mu,\mu'}$ is completely reducible.
Therefore the following definition makes sense:

\begin{definition} 
\label{trunc_tensor}
In the above situation, let $\{u_{\l''}\}$ be a basis of physical
lowest--weight states of $Q_{\mu,\mu'}$, and 
$V_{(\l'')}$ be the corresponding physical lowest weight irreps. 
Then define  
\beq
V_{(\mu)} \ttens V_{(\mu')} \equiv \bigoplus_{\l''} V_{(\l'')}
\eq
\end{definition} 
Notice that if $\frac m{2n}$ is not integer, then the physical states
have non--integral weights, and  the full tensor product of two
physical irreps $V_{(\mu)} \tens V_{(\mu')}$ does not contain any 
physical lowest weights. Therefore 
$V_{(\mu)} \ttens V_{(\mu')}$ is zero, and there seems to be no reasonnable 
way to get around this.

Again as in section \ref{subsec:many_2d}, one might also include 
a second "band" of high--energy states. Now

\begin{theorem} 
If all weights $\mu, \mu', \dots$ involved are integral, then
$\ttens$ is associative, and 
$V_{(\mu)} \ttens V_{(\mu')}$ is unitarizable w.r.t. $SO_q(2,3)$.
\end{theorem}
\begin{proof}
First, notice that the $\l''$ are all integral and none of them gives rise
to a  massless \rep or a singleton. So by the strong linkage principle, 
none of the $\U u_{\l'}$ can contain a physical 
lowest--weight submodule.
Also, lowest weight states for generic $q$ cannot disappear 
at roots of unity. 
Therefore $Q_{\mu,\mu'}$ contains all the physical 
lowest--weight states of the
full tensor product. Furthermore, no physical lowest--weight states
are contained in (discarded states)$\tens$ (any states).
Then associativity follows from associativity of the full $\tens$, 
and the stucture is the same as classically 
for energies $h_3 \leq \frac m{4n}$ 
(observe that $\tens$ contains no massless representations, so
classically inequivalent physical \reps cannot recombine into 
indecomposable ones).
\end{proof}

In particular, none of the low--energy
states have been discarded. Therefore our definition is physically sensible,
and the case of $q=e^{2\pi i/m}$ with $m$ even appears to be most 
interesting physically.

The highest ("cosmological") energies available in this "low--energy band" 
are of order $E_{max}=\frac 1{hR}$ in appropriate units, 
where $R$ is the radius of AdS space, and $h=1/m$.
This is by a factor $\frac 1{\sqrt{h}}$ larger than the energy scale 
$L_0 \approx \frac 1{\sqrt{h} R}$ where
the geometry becomes noncommutative, see section \ref{subsec:scale}.
From a QFT point of view, the latter should be the interesting scale.
Thus the hierarchy from the curvature radius $R$ to the geometrical scale
$L_0$ is the same as between $L_0$ and $1/E_{max}$.
This hierarchy has to be large in order to have a large number of 
physical states available.
In any case, this shows that such a theory (imaginary, so far)
could in principle accomodate large systems, with an interesting
relation between  "cosmological"  scales and a geometrical scale
(even though we do not advertise this as a cosmologically interesting model).

\subsection{Massless Particles, Indecomposable Representations and BRST 
  for $SO_q(2,3)$}
     \label{subsec:so23_reps_quant}

Let $n=1$ and $m=2M$, i.e. $q=e^{i \pi/M}$. 
$V_{(E_0,0)}$ will again be called "scalar field",
$V_{(E_0,1)}$ "vector field", and so on. 
$V_{(E_0,s)}$ has a subspace of pure gauges in the
"massless" case, otherwise they are irreducible.
They are unitarizable w.r.t. $SO_q(2,3)$ after factoring out 
the pure gauge states. 

Consider again vectorfields as one--forms.
Instead of studying (\ref{ads_tens}) at roots of unity, 
we can study the tensor product of compact representations, using a shift.
In particular for $\l = -(E_0-M)\b_3$, consider $V(\l) \tens V_5$ at
roots of unity. For $E_0 > 2$, one can easily see that
\beq
V(\l) \tens V_5 =  V(\l+\b_2) \oplus V(\l-\b_3)
    \oplus V(\l+\b_3),
\label{ads_tens_cpct}
\eq
which is the same as for generic $q$. This follows by analycity from the
generic case (this is why we consider the shifted representations),
since the \reps on the rhs remain irreducible at $q=e^{i \pi/M}$
according to section \ref{subsec:unitary_so23} (notice e.g.
that the Drinfeld--casimir is different on these representations).

In the (shifted) massless case $E_0 = 2$ however, the generic highest weight
\rep $V(\l+\b_2)$ 
has an invariant subspace as in Theorem \ref{cpct_thm}. In fact, $V(\l+\b_2)$ 
and $V(\l-\b_3)$ (the "radial" part according to section 
\ref{subsec:so23_reps_class}) 
are combined in one reducible, but indecomposable 
representation, similar as the \reps $I_z^p$ encountered in section 
\ref{subsec:unitary_2d}; see figure 3.8 for the
noncompact case. 
This kind of phenomenon at roots of unity is well -- known, cp. the general
discussion in section \ref{subsec:root_of_1}. It will be analyzed in general 
in the next chapter,
but we can understand it more directly here. 

\begin{figure}
\begin{center}
\leavevmode
\epsfysize=3in 
\epsfbox{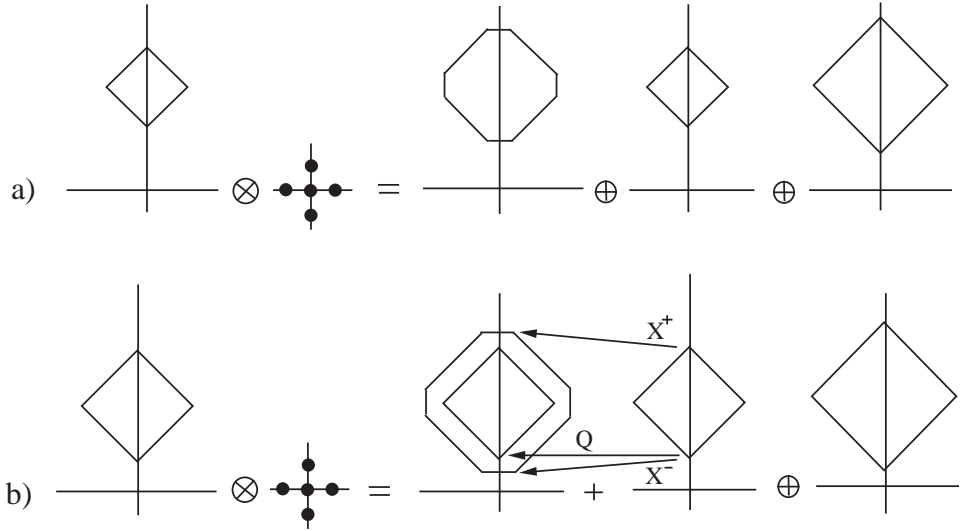}
\caption{$V_{(\l)} \tens V_5$ for the a) massive and 
b) massless case. The $+$
in b) means sum as vector spaces, but not as representations.}
\end{center}
\label{fig:ads_tens_q}
\end{figure}

As explained in \ref{subsec:root_of_1},
$\R$ is well--defined for $q=e^{i \pi/M}$ if acting on irreps with
$(X_i^{\pm})^{M_{(i)}} = 0$ such as the compact $V(\l)$ or $V_{(E_0,s)}$.
Therefore  the invariant sesquilinear form (\ref{inner_R}) 
on $V(\l) \tens V_{5}$,
\beq
(v_1\tens w_1, v_2\tens w_2)_{\R} =
(v_1\tens w_1, \R v_2\tens w_2)_{\tens}
\label{inner_prod_vf}
\eq
is well--defined and nondegenerate for $q=e^{i \pi/M}$
(but not positive definite in general).

Now it is easy to see that for $q=e^{i \pi/M}$ and $\l=(M-2)\b_3$,
$V(\l+\b_2)$ and $V(\l-\b_3)$ 
on the rhs of (\ref{ads_tens_cpct})
are combined into one indecomposable representation:
Let $w_{\l+\b_2}$ resp. $w_{\l-\b_3}$ be their h.w. vectors 
for generic $q$. We know from Theorem \ref{cpct_thm}
that for $q=e^{i \pi/M}$, $V(\l+\b_2)$ contains a descendant 
h.w. vector $\varphi_{\l-\b_3}$
at weight $\l-\b_3$, which means that  $\varphi_{\l-\b_3}$ is 
orthogonal to {\em any} descendant of $w_{\l+\b_2}$, for any invariant 
sesquilinear form (while it could happen that this h.w. vector is zero
in a given representation, it can be seen easily that this is not 
the case here). But since our $(\;,\;)_{\R}$ is
nondegenerate, there must be some vector 
$\chi \notin V(\l+\b_2)$ which is not orthogonal to $\varphi_{\l-\b_3}$.
Since $w_{\l+\b_2}$ is analytic at $q=e^{i \pi/M}$
and $w_{\l-\b_3}$ is orthogonal to $V(\l+\b_2)$ for $|q|=1$, 
this is only possible if 
$w_{\l-\b_3}\rightarrow \varphi_{\l-\b_3} \in \U^- w_{\l+\b_2}$ 
as  $q \rightarrow e^{i \pi/M}$, i.e. the generically independent h.w. modules 
$V(\l-\b_3)$ and $V(\l+\b_2)$ become dependent, so that one has to include
different states such as $\chi$ and its descendants 
to span the tensor product at roots of unity.
Furthermore, notice that 
$\chi$ cannot be a h.w. vector because $(\varphi_{\l-\b_3},\chi)_{\R} \neq 0$, 
so $X^+ \chi \in V(\l+\b_2)$, and
the structure is really indecomposable.
This is the reason for the appearance of indecomposable representations.

\paragraph{BRST operator.}
At first, this may look  complicated. However, the main point is
that this is actually very nice from the BRST point of view:
In fact, the BRST operator $Q$ which
was defined "by hand" in the generic case is now
an element of the center of $\U$, and maps $\chi$ into 
$\varphi_{\l-\b_3}$. This is exactly what one would like in QFT.
It is not so surprising, since $X^+\chi \in V(\l+\b_2)$, so 
some $Q \approx \sum X^- X^+$ should do the job. We will show that 
\beq
Q \equiv (v^{2M} - v^{-2M})
\label{brst}
\eq
is indeed a BRST operator. One could take higher powers of $v$ as well.

To see this, consider first the characteristic equation (\ref{char_eq}) 
of $v$ in the \rep $V(\l) \tens V_5$ for generic $q$: 
\beq
(v-q^{-c_{\l+\b_2}})(v-q^{- c_{\l-\b_3}})(v-q^{-c_{\l+\b_3}}) =0 . 
\label{char_eq_vf}
\eq
For compact representations, $c_{\l} =(\l,\l+\rho) \in \frac 12 \Z$. 
As $q \rightarrow e^{i \pi/M}$, 
the first two factors above coincide, and
$(v^{4M}-1)^2$ contains all the factors in (\ref{char_eq_vf}) separately
(and more). So by continuity it follows that $(v^{4M}-1)^2 = 0$, which implies
\beq
Q^2=0 \quad \mbox{for}\quad q=e^{i\pi/M}  \label{Q2_0}
\eq
on $V(\l) \tens V_5$, since $v$ is invertible.

It remains to show that $Q$ maps some $\chi$ into $\varphi_{\l-\b_3}$
for $\l = (M-2)\b_3$. We have seen that as
$h\equiv (q-e^{i\pi/M}) \rightarrow 0$, 
$w_{\l-\b_3}$ becomes $\varphi_{\l-\b_3} \in \U^- w_{\l+\b_2}$, so
there is a (fixed) $u^- \in \U$ such that 
$u^-w_{\l+\b_2} = w_{\l-\b_3} + h\tilde{\chi}(h)$, where 
$\tilde{\chi}(h) \in V(\l)\tens V_5$ is analytic in $h$. 
Applying $Q$ to that equation for generic $q$, we get
\beq
Q(\frac{u^-w_{\l+\b_2}-w_{\l-\b_3}}{h}) = Q\tilde{\chi}(h),
\eq
wich using (\ref{v_h_w}) becomes
\berr
-4M(c_{\l+\b_2} u^-w_{\l+\b_2} - c_{\l-\b_3} w_{\l-\b_3}) + o(h) &=&
-4M(c_{\l+\b_2}-c_{\l-\b_3})w_{\l-\b_3} + o(h) \nn\\
  &=& Q(\tilde{\chi}(h)).
\err
Now $w_{\l-\b_3}=\varphi_{\l-\b_3}+o(h)$, and therefore for $h=0$,
\beq
\varphi_{\l-\b_3} = Q \chi
\label{imQ_ads}
\eq
where $\chi = c^{-1}\tilde{\chi}(0)$ and 
$c=-4M(c_{\l+\b_2}-c_{\l-\b_3}) \neq 0$ using Lemma \ref{casimir_ord}.
Thus 2) and 3) of the properties of a BRST operator in section \ref{BRST_char} 
hold for our $Q$,
and 1) would certainly be satisfied in any covariant theory based on
$SO_q(2,3)$. States on which $Q$ does not vanish will be called "ghosts".
It is obvious that all this generalizes to higher spins, and
works at roots of unity only.

\chapter{Operator Algebra and BRST Structure 
    at Roots of Unity}
\label{chap:brst}

In this chapter, we study the tensor product of irreducible \reps
for any quantum group at roots of unity using a BRST operator $Q$,
generalizing results of the last section.
$Q$ allows us to define in a very simple way
a new algebra of \reps of $\U$ similar as in QFT, which has a 
"ghost--free" subalgebra with involution. For the AdS group, this generalizes 
the physical many--particle Hilbert space introduced above, and
can be used to define correlators. Finally we give a conjecture 
on complete reducibility, generalizing the standard truncated tensor
product used in CFT \cite{mack_schom}. The problem of a
symmetrization postulate is also briefly discussed.

\section{Indecomposable Representations and BRST Operator}
\label{sec:tensor_brst}

Let $V_i$ be compact irreps for generic $q$, i.e. with with dominant integral
h.w. $\mu_i$, which remain irreducible at the root of
unity
\beq
q_0 = e^{2\pi n/m}.
\eq
By a simple shift, this also covers e.g. the unitary \reps
of $SO_q(2,3)$, but excludes the massless \reps for the AdS case
(they can be built by tensor products).
Then for generic $q$,
\beq
V \equiv V_1 \tens ... \tens V_l =\oplus V(\l_l)
\label{decomp_V}
\eq
with irreps $V(\l_l)$ because it is completely reducible.
The space $V$ is the same for any $q$, and the \rep
of $\U$ depends analytically on 
\beq
h\equiv q-q_0.
\eq
The projectors $P_{\l_l}$ (\ref{projectors}) may have a pole at $h=0$, 
but no worse singularity.
If $q$ is a phase, consider the involution 
$\cobar{x} = \theta(x^{\ast})$ as in (\ref{cbar_tens}).

Now for  $q_0=e^{2\pi i n/m}$ with 
$M=m$ if $m$ is odd and $M=m/2$ if $m$ is even,
the BRST operator $Q$ is defined as in section \ref{subsec:so23_reps_quant}, 
\beq
Q \equiv (v^{dM} - v^{-dM})
\eq
where $d$ is an integer depending on the group, such that
$q_0^{2dM c_{\l}}=1$ for any integral weight $\l$; notice that $c_{\l}$ is 
a rational number.

Of course $Q$ can be considered for any $q$.
For $|q|=1$, it satisfies
\beq
\cobar{Q} = -Q,
\label{Q_conj}
\eq
since $\cobar{v} = v^{-1}$, see (\ref{cobar_v}).

If $V$ is completely reducible at $q_0$, then
$Q$ vanishes at $h=0$ by construction, using (\ref{v_h_w}).
In general, it follows that
all the eigenvalues of $Q$ are zero at $h=0$. This implies that
$Q^N=0$ for large enough $N$ (depending on the representation),
however $Q$ will not be zero on the indecomposable representations.
It is not clear whether $Q^2= 0$ in general, but this is not essential.

As in the previous section, our essential tools will be 
the sesquilinear forms
$(a, b)_{\tens}$ and $(a, b)_{\R}$
on $V$ defined in section \ref{subsec:invar_forms}, for
$a=a_1 \tens ... \tens a_l \in V_1 \tens ... \tens V_l$ 
and similarly $b \in V$. 
Furthermore, define 
\beq
(a, b)_{k} \equiv (a, Q^k b)_{\R}
\label{inner_Q}
\eq
for $k\in\N$,
which is invariant but degenerate. This will play an important role in 
analyzing the structure of $V$. If $q$ is a phase, then (\ref{Q_conj}) 
implies
\beq
Q^{\dagger} = -Q,
\eq
where $\ ^{\dagger}$ is 
the operator adjoint w.r.t. any of these invariant inner products.

Now define vectorspaces $G_k \subset V$ as follows:
\beq
G_k \equiv \{a_k\in V; \quad Q^{k+1} a_k=0 \mbox{ at } h=0 \}
\eq
(again, $V$ does not depend on $q$, only the representation does).
At $h=0$, $Q^N=0$ 
for sufficiently large $N$ because of (\ref{char_eq}).
Since $Q$ is a Casimir, the $G_k$ are invariant under the action of $\U$ 
at $h=0$, and
$G_0 \subset G_1 ... \subset G_k \subset ... \subset G_N = V$ for some $N$.
This is a "filtration"; it is well known that the
tensor product at roots of unity has the structure 
of a filtration \cite{CH_P}.
We can now define the quotients
\beq
\G_k = G_k/G_{k-1}
\eq
which are representations again (for $h=0$), and as vectorspace,
\beq
V \cong \oplus_k \G_k
\eq
(but not as representation). The $\G_k$ with $k>0$ will be called "ghosts".
Roughly speaking, $\G_k$ essentially contains 
the $a_k$ with $Q^k a_k \neq 0$, but $Q^{k+1} a_k =0$.
From the definition, $Q$ maps $\G_k$ into $\G_{k-1}$ at $h=0$, 
and it is injective (i.e. it never vanishes).
In particular, $Q^k: \G_k \rightarrow G_0$ is injective as well, 
and $Q(G_0) =0$. Therefore at $h=0$, $(a,b)_k = 0$  
if either $a$ or $b$ $\in G_{k-1}$,
which means that $(\;,\;)_k$ is a well--defined sesquilinear form on $\G_k$
or more generally on $V/G_{k-1} \cong \oplus_{l \geq k} \G_l$.

From now we will work with these spaces for $q=q_0$.
Consider $v$ acting on $V$ in its Jordan normal form, 
$v=D+N$ where $D$ is a diagonal matrix
and $N$ nilpotent, and $DN = ND$. Since $v$ is invertible,
its (generalized) eigenvalues are
nonzero, and $D$ is invertible. Then
$v^n = D^n + n D^{n-1} N + N^2(...)$ is the Jordan normal form of $v^n$,
so $v$ is diagonalizable
if and only if $v^n$ is. In particular for  $a \in G_0$,
$Qa = (v^{dM}-v^{-dM})a=0$, therefore $(v^{2dM}-1)a=0$ since 
$v$ is invertible.
This means that $a$ is an eigenvector of $v^{2dM}$, and therefore
$a$ is contained in the sum of the {\em proper} eigenspaces of $v$,
as we have just seen. Conversely of course, any eigenvector of $v$
will be annihilated by $Q$, since the eigenvalues
and the characteristic equation of $v$ are analytic. 
Therefore we see again that $Q$ is nilpotent on $V$.
In summary,
\begin{lemma} For $h=0$,
\beq
Q: \G_k \rightarrow \G_{k-1} \quad \mbox{is injective},
\eq
and $G_0$ 
is the sum of the proper eigenspaces of $v$.
\label{Q_lemma}
\end{lemma}

Now one can define 
\beq
\H_k = \G_k /(Q\G_{k+1}). 
\label{H_cohom}
\eq
generalizing (\ref{BRST_cohom}); we do not require that $Q^2=0$.
Since $Q\G_{k+1}$ is invariant,  $\H_k$ is a \rep of $\U$ as well.
An element  $h_k \in \H_k$ can be represented by an element of
$\G_k$  (not uniquely of course), which will still be denoted as $h_k$.
Using our sesquilinear forms, one can choose them in a particular way: 

\begin{theorem}
Consider $V = V_1 \tens ... \tens V_l$ as above for $h=0$, with
$V \cong \oplus_{k=0}^{k_0} \G_k$
as vector space (but not as representation). 
Then one can choose $H_l \subset \G_l$ 
such that  $(\; , \;)_k$ is nondegenerate on $H_k$, and 
\beq
(H_k, H_{k+l})_k =0 \quad  \mbox{ for } l \geq 1.
\label{H_orth}
\eq
Furthermore,
\beq
\G_k = H_k \oplus Q H_{k+1} \oplus ... \oplus Q^{k_0-k} H_{k_0}
\label{G_decomp}
\eq
as vector space.
Therefore $\H_k \cong H_k$, and since $(H_k, Q(...))_k=0$, 
$V/G_{k-1}$ is the direct sum of $H_k$ and its 
orthogonal complement w.r.t $(\;,\;)_k$, as vectorspace. 
\label{struct_thm}
\end{theorem}
Notice that this is not trivial, since the $(\;,\;)_k$ are
not positive definite in general. Furthermore, if $Q \neq 0$, then the 
\rep is indecomposable.
%The usual truncated tensor product used in CFT \cite{mack_schom} coincides
%with that part of $G_0$ which is irreducible,
%i.e. which does not contain any submodule of the form $Q(...)$, so that 
%$H_0 = \H_0$.

\begin{proof}
Let $k_0$ be the maximal $k$ such that $G_k \neq 0$. Then for any  
$h_{k_0} \in H_{k_0} \equiv \G_{k_0}$, $Q^{k_0} h_k \neq 0$, 
and since $(\;,\;)_0 = (\;,\;)_{\R}$
is nondegenerate, there exists some $a$ such that 
$(a,Q^{k_0} h_{k_0})_0 = (a, h_{k_0})_{k_0} \neq 0$.
Furthermore from the definition of $(\;,\;)_k$ it follows that 
$(\H_k,G_l)_k=0$ if $l<k$. 
Therefore $(\;,\;)_{k_0}$ is nondegenerate on $H_{k_0}$. This implies as usual
that any linear form on $H_{k_0}$ can be written as $(h_{k_0}, . )_{k_0}$
with a suitable $h_{k_0} \in H_{k_0}$, i.e. there is a isomorphism
from the dual space $H_{k_0}^{\ast}$ to $H_{k_0}$. 

In particular, any $a_0 \in \G_0$ defines a linear form on $H_{k_0}$ by 
$h_{k_0} \rightarrow (a_0, h_{k_0})_0$. Therefore there is a unique element 
$i(a_0) \in H_{k_0}$ which satisfies 
\beq
(a_0-Q^{k_0-k} i(a_0), H_{k_0})_k = 0.
\eq
Now define $W^{k_0}_0 \subset G_0$ by 
\beq
W^{k_0}_0 \equiv  (id-Q^{k_0} i) G_0;
\eq
then by definition $(W^{k_0}_0, H_{k_0})_0 =0$, and since 
$a_0 = (a_0-Q^{k_0} i(a_0)) + Q^{k_0} i(a_0) 
\in W^{k_0}_0 \oplus Q^{k_0} H_{k_0}$ for any $a_0 \in G_0$,  we have
\beq
G_0 = W^{k_0}_0 \oplus Q^{k_0} H_{k_0} \quad \mbox{with } 
 (W^{k_0}_0, H_{k_0})_0 =0,
\eq
and $W^{k_0}_0$ is determined uniquely by this requirement.

Now we can define $W^{k_0}_k \subset \G_k$ for any $k$
to be the space which is mapped into $W^{k_0}_0$ by $Q^k$, 
\beq
W^{k_0}_k = (Q^k)^{-1} W^{k_0}_0
\eq
(as set). 
For any $a_k \in \G_k$, decompose $Q^k a_k = w^{k_0}_0 + Q^{k_0} h_{k_0}$
accordingly. Then $w^{k_0}_0=Q^k(a_k-Q^{k_0-k}h_{k_0})$,
and we have $a_k = w^{k_0}_k + Q^{k_0-k}h_{k_0}$ for 
$w^{k_0}_k \equiv a_k-Q^{k_0-k}h_{k_0}$.
This means that
\beq
\G_k = W^{k_0}_k \oplus Q^{k_0-k} H_{k_0}, \quad \mbox{ and } \quad
 (W^{k_0}_k, H_{k_0})_k =0
\label{decomp_k}
\eq
using the definition of $(\;,\;)_k$. 
This decomposition is unique since $Q^k: \G_k \rightarrow G_0$ is 
injective.

Now for $\G_{k_0-1}$, define 
\beq
H_{k_0-1} \equiv W^{k_0}_{k_0-1}.
\eq
Then obviously $\H_{k_0-1} \cong H_{k_0-1}$, and 
$(H_{k_0-1}, H_{k_0})_{k_0-1} =0$ using (\ref{decomp_k}). 
Thus we have shown (\ref{G_decomp})
and (\ref{H_orth}) for $k=k_0-1$. Of course, $(H_{k_0-1}, Q(...))_{k_0-1}=0$.

Now we can repeat this construction: since 
$(H_{k_0-1},H_{k_0})_{k_0-1}=0$ and $Q^{k_0-1} h_{k_0-1} \neq 0$
for any $h_{k_0-1} \in H_{k_0-1}$, it follows that
$(\;,\;)_{k_0-1}$ is nondegenerate on $H_{k_0-1}$. Again, this means that 
for every $a_0 \in W^{k_0}_0$ there exists some 
$h_{k_0-1} = i(a_0) \in H_{k_0-1}$
such that $(a_0-Q^{k_0-1} i(a_0), H_{k_0-1})_0 = 0$. Define
$W^{k_0-1}_0 \equiv  (id-Q^{k_0-1} i) W^{k_0}_0$; then 
\beq
W^{k_0}_0 = W^{k_0-1}_0  \oplus Q^{k_0-1} H_{k_0-1}
\eq
and $(W^{k_0-1}_0, H_{k_0-1})_0 =0$. Therefore 
\beq
G_0 = W^{k_0-1}_0 \oplus Q^{k_0-1} H_{k_0-1} \oplus  
    Q^{k_0} H_{k_0},
\eq
and we already know $(W^{k_0-1}_0, H_{k_0})_0 =0$. 
This implies that for  $W^{k_0-1}_k \equiv (Q^k)^{-1} W^{k_0-1}_0$,
$W^{k_0}_k = W^{k_0-1}_k  \oplus Q^{k_0-1-k} H_{k_0-1}$ and 
$(W^{k_0-1}_k, H_{k_0-1})_k =0$.
Finally with
\beq
H_{k_0-2} \equiv W^{k_0-1}_{k_0-2},
\eq
(\ref{G_decomp})
and (\ref{H_orth}) follows. Repeating this argument, we arrive at the 
decomposition as stated.
\end{proof}

\subsection{A Conjecture on BRST and Complete Reducibility}

There are many indications that the following extension of Theorem 
\ref{struct_thm} holds, in the same context:
\begin{conjecture}
\begin{itemize}
\item[1)] $\H_0$ is completely reducible, i.e.
  it is the  direct sum of irreducible representations.
\item[2)]  $\H_0$ defines an associative,  completely reducible modified
  tensor product of irreducible representations.
\end{itemize}
\label{conjecture}
\end{conjecture}
Furthermore, it appears that $G_0$ is the sum of the analytic images of the
projectors $P_{\l}$ at $h=0$, see Lemma \ref{analytic_image_lemma},
and $G_0\cap Q(...)$ is the subspace where they are linearly dependent.
This could be used to show 2) from 1).

In the context of the AdS group, 2) would generalize the definition of the 
tensor product $\tilde{\tens}$ in sections \ref{subsec:many_2d} 
and \ref{subsec:many_4d}, and include the "high--energy" bands as well as 
irreps which are unitary w.r.t. the "compact" reality structure. 
This would also generalize the standard truncated tensor product
used in the context of CFT, and provide a common framework for
$\ttens$ and $\hat{\tens}$.

\section{An Inner Product on $G_0$}
\label{sec:inner_prod_g0}

In this section, we will define a hermitian inner product on $G_0$. 
To do this, we first show how to find an element in $\U$ which
implements $\sqrt{v}$ (or many other
functions of $v$) analytically on $G_0$ for $q=q_0$.

As already noted, $v$ becomes degenerate on different \reps at $h=0$,
because $q^{-c_{\l}}$ does. For a weight $\l$, define the 
$\l$ --{\em group} $g_{\l}$ of $V$
to be the set of highest weights  in the generic decomposition 
(\ref{decomp_V}) of $V$ for which $v$ is degenerate,
\beq
g_{\l} \equiv \{\l_l;\quad q_0^{c_{\l_l}} = q_0^{c_{\l}} \}.
\eq
Then define
\beq
P_{g_{\l}} \equiv \frac{\prod_{g_{\l'} \neq g_{\l}}(v-q^{-c_{\l'}})}
            {\prod_{g_{\l'} \neq g_{\l}}(q^{-c_{\l}}-q^{-c_{\l'}})},
\eq
where the products go over all possible $\l$ --groups once. 
As opposed to $P_{\l}$, 
this is analytic at $q_0$. It is not a projector in $V$,
but it {\em is} a projector for $h=0$ in $G_0$, since then $v$ is 
diagonalizable on $G_0$ as pointed out in Lemma \ref{Q_lemma};
it is simply the projection on the eigenspaces of $v$ corresponding
to different $\l$ --groups.

Now define
\beq
\sqrt{v} \equiv \sum_{g_{\l}} P_{g_{\l}} q^{-\frac 12 c_{\l}},                                         
\label{sqrt_v}
\eq
which is an element of $\U$ and analytic at $q_0$, 
where $\l$ is some element of $g_{\l}$.
It depends on the choice of $\l\in g_{\l}$, which simply corresponds to 
choosing a different branch; pick any of them. 
It is easy to see that on $G_0$,
\beq
(\sqrt{v})^2 = v \quad \mbox{ for } h=0,
\eq
and its inverse on $G_0$ is given by
\beq
\sqrt{v^{-1}} \equiv \sum_{g_{\l}} P_{g_{\l}} q^{\frac 12 c_{\l}}.
\label{sqrt_v_inv}
\eq
This follows because the $P_{g_{\l}}$ are projectors on $G_0$ as shown above.

Using this, we can finally show the following:
\begin{prop}
Define
\beq
(a,b)_H \equiv c_V (a,\sqrt{v}\cdot b)_{\R},
\label{inner_H}
\eq
where $c_V = q^{\frac 12(c_1 + ... + c_l)}$ and $c_i$ is the quadratic 
Casimir on the $V_i$. Then for $a,b \in G_0$, $(a,b)_H$ 
is {\em hermitian} at $h=0$, i.e. it is an inner product 
on $G_0$ for $q=q_0$. 
\end{prop}
\begin{proof}
To see this, notice first that 
\beq
\Del_{(l)}v = (\R_{l...21}\R_{12...l})^{-1} (v\tens ... \tens v).
\label{del_v_l}
\eq
This follows inductively by applying $(\Del_{(l-1)}\tens \id)$ to 
(\ref{v_coprod}):
\berr
\Del_{(l)}(v) &=&  \((\Del_{(l-1)}\tens \id)\R_{21}(\Del_{(l-1)}\tens \id)
                                  \R_{12}\)^{-1} (\Del_{(l-1)}v\tens v) \nn\\
   &=& \((\Del_{(l-1)}\tens \id)\R_{21} (\R_{12...(l-1)}^{-1}\tens 1) 
                              \R_{12...l}\)^{-1} (\Del_{(l-1)}v\tens v) \nn\\
   &=& \((\R_{12...(l-1)}^{-1}\tens 1) ((\Del'_{(l-1)}\tens 1)\R_{21}) 
                              \R_{12...l}\)^{-1} (\Del_{(l-1)}v\tens v)\nn\\
   &=& \((\R_{12...(l-1)}^{-1}\tens 1) (\R_{(l-1) ... 21}^{-1}\tens 1) 
                   \R_{l...21}\R_{12...l}\)^{-1} (\Del_{(l-1)}v\tens v) \nn\\ 
   &=& (\R_{l...21}\R_{12...l})^{-1} 
   (\R_{(l-1) ... 21} \R_{12...(l-1)} \tens 1) (\Del_{(l-1)}v\tens v) \nn\\
   &=& (\R_{l...21}\R_{12...l})^{-1} (v\tens ...\tens v) 
\err   
using (\ref{R_l_a}), (\ref{R_l_b}) and Lemma \ref{R_l_lemma}.
%For $|q|=1$, this can be written as 
%\beq
%\Del_{(l)}(v) = \(\R_{12..l}^{-1}
%     (\cobar{()}\tens ... \tens \cobar{()})\R_{12...l}\) (v\tens .. \tens v)
%\eq
Furthermore $\cobar{\sqrt{v}} = \sqrt{v^{-1}}$. Then
\berr
c_V(a,b)_H^{\ast} &=& \(a,\R_{12...l} \Del_{(l)}(\sqrt{v}) 
                                                  b\)_{\tens}^{\ast} \nn\\
 &=& \(\R_{12...l} \Del_{(l)}(\sqrt{v}) b,a\)_{\tens} \nn\\
 &=& \(\Del'_{(l)}(\sqrt{v})\R_{12...l} b,a\)_{\tens} \nn\\
% &=& \(b,((\cobar{()}\tens ...\tens\cobar{()})\R_{12...l}) 
%                                \Del_{(l)}(\sqrt{v^{-1}}) a\)_{\tens} \nn\\
 &=& \(b, \R^{-1}_{l..21} \Del_{(l)}(\sqrt{v^{-1}}) a\)_{\tens} \nn\\
 &=& \(b, \R_{12...l} \R_{12...l}^{-1} \R^{-1}_{l...21} 
                                  v^{-1}\cdot \sqrt{v}\cdot a\)_{\tens} \nn\\
 &=& (b, (v^{-1}\tens .. \tens v^{-1}) \sqrt{v} \cdot a)_{\R} \nn\\
 &=& c_V (b,a)_H
\err
for $h=0$,
using (\ref{qtr_l_3}). 
\end{proof}

%Notice that this 
%only works on $G_0$ for roots of unity, since otherwise there is no suitable 
%$\sqrt{v}\in\U$. 

The above definitions may seem a bit complicated at first. We will show
below that all this can be formulated in a very simple way, 
using an extension of $\U$ by a universal element 
which implements a Weyl reflection. But before that, we show 
how the above inner product defines a {\em Hilbert space} of 
physical many--body representations 
of the Anti--de Sitter group, with the correct 
classical limit. Then the adjoint of an operator acting on any component of a
tensor product is determined by the positive definite inner product, and
is guaranteed to have the correct classical limit, since the inner product has
(for $q\neq 1$, its adjoint will act on the entire tensor product, see below).
This finally settles any doubts 
whether the reality structure (\ref{cbar_tens}) is suitable to describe
physical many--body states. 

\subsection{Hilbert Space for the Quantum Anti--de Sitter group}
\label{subsec:hilbert}

It follows from Theorem \ref{struct_thm} 
that $(a,Q(...))_H=0$ for $a \in G_0$,
i.e. the image of $Q$ is null with respect to this inner product. Therefore 
$(\;,\;)_H$ induces an invariant inner product on $H_0$ as defined in section 
\ref{sec:tensor_brst}. Furthermore, $(\;,\;)_H$ is nondegenerate on $H_0$. 
We want to show that it induces in fact a 
{\em positive definite} inner product on the physical \reps 
\beq
V_{(\mu_1)} \ttens ... \ttens  V_{(\mu_l)}=
      \bigoplus_{\l_k} V_{(\l_k)}
\label{ads_phys_many}
\eq
defined in section \ref{subsec:many_4d}, where all $V_{(\mu_l)}$ are massive
(this is not an essential restriction). 
The terms on the rhs are  (quotients of) analytic, 
generic representations and therefore contained in (a quotient of) $G_0$.
Thus the results of the last section apply, and the 
eigenvalues of $(\;,\;)_H$ on $G_0$ are either positive or negative. 
It is clear that 
for low energies, they will be positive as in the classical limit.
We claim that they are positive on all the $V_{(\l_k)}$ above, and outline 
a proof:
%for lowest--weight states with energy $\leq \frac m{4n}=M/2$ 
%(only $\frac m{2n}$ integer is of interest here. If the "strong linkage
%principle" \cite{anderson} applies in general, 
%then this gives a complete proof):

Consider the tensor product $V$ of the infinite--dimensional (generic)
lowest--weight modules corresponding to the  massive \reps in
(\ref{ads_phys_many}). 
For fixed $\l_k$, let $G_{\l_k}$ be the sum of the lowest--weight 
submodules of $V$ with lowest weight $\l_k$. 
According to the strong linkage principle, none of them will contain physical
lowest-weight vectors, for any $q$ on the arc from 1 to $e^{2 i\pi/m}$.
This implies that the physical lowest--weight vectors of $G_{\l_k}$ 
are linearly independent of the other $G_{\l_k'}$ for $q$ 
on the arc from 1 to 
$q=e^{2 i\pi/m}$, and analytic in $q$. 

Now we can define $(\;,\;)_H$ as in (\ref{inner_H}) for states with weights
$\l_k$ corresponding to physical $G_{\l_k}$, 
where one has to use the value of the 
classical quadratic Casimir on $G_{\l_k}$. 
$\R$ is well--defined for such states,  
as can be seen from (\ref{R_formula}) and the discussion in
section \ref{subsec:root_of_1}. Therefore $(\;,\;)_H$
is hermitian and analytic for $q$ on the arc from 1 to $e^{2 i\pi/m}$. 
For $q=e^{2 i\pi/m}$, it reduces precisely to our inner product on $G_0$. 
Furthermore, all states with such weights are contained 
in $\oplus G_{\l_k}$.

Now $(\;,\;)_H$ cannot become null on the physical 
lowest--weight states of $G_{\l_k}$, where it is non---degenerate 
for $q$ on the arc from 1 to $e^{2 i\pi/m}$, 
since the $G_{\l_k}$ remain linearly 
independent, so $\sqrt{v}$ is analytic and invertible, 
and the other $G_{\l_{k'}}$ are orthogonal. Therefore
the eigenvalues of $(\;,\;)_H$ on $G_{\l_k}$ are positive, as classically.

\section{The Universal Weyl Element $w$}
\label{sec:weyl}

The proper mathematical tool to obtain this inner product 
and an involution is an element 
of an extension of $\U$ by generators of the braid group $\w_i$, introduced
in \cite{kirill_resh} and \cite{lev_soib}. The $w_i$ act on \reps of $\U$ and 
implement the braid group action (\ref{braid_action}) 
on $\U$ via $T_i(x) = w_i x w_i^{-1}$
for $x\in \U$.
All we need is the generator corresponding to the longest element
of the Weyl group, $w$. Acting on a highest weight irrep, $w$ maps
the highest weight vector 
into the lowest weight vector of the contragredient representation.
It has the following important properties \cite{kirill_resh,lev_soib}:
\berr
\Del(w) &=& \R^{-1} w\tens w = w\tens w \R^{-1}_{21} \label{del_om} \\
w^2 &=& v \vareps  \label{w_2}
\err
where $\vareps$ is a Casimir with 
\beq
\Del(\vareps) = \vareps\tens\vareps
\label{del_eps}
\eq
(for $SL_q(2)$ and $SO_q(2,3)$, $\vareps$ is $+1$ for integer "spin"
and $-1$ for half--integer "spin", see Appendix B).
(\ref{del_om}) justifies the name "universal Weyl element".
One can also find the antipode and counit of $w$.
Furthermore, for the "real" quantum groups (=those having only self--dual
representations, i.e. all except $A_n$, $D_n$ and $E_6$), 
such as $SO_q(5,\compl)$ and its real forms, the following holds
\cite{resh_1}:
\beq
w x w^{-1} = \theta S x = S^{-1} \theta x.  \label{autom_w}
\eq 
We will only consider this "real" case.
For "complex" groups, this gets corrected by an automorphism of the Dynkin
diagram. 
Actually, (\ref{autom_w}) and (\ref{w_2}) have only been proved explicitely
for the $SL_q(2)$ case in the literature, therefore we will supply proofs 
in Appendix B.

Similarly, we define 
\beq
\tw \equiv  (-1)^E q^{2\tilde{\rho}} w
\label{tw}
\eq
with the same properties except
\beq
\tw x \tw^{-1} = \ttheta S^{-1} x = S \ttheta x \label{autom_tw}.
\eq
Here $\ttheta x \equiv (-1)^E \theta(x) (-1)^E$ implements the 
reality structure on $\U$ according to section
\ref{subsec:reality}: in the case of $SO_q(2,3)$, $E$ is the energy operator, 
while the compact case corresponds to $E=0$. 

Denote the (left) action of $x\in\U$ on a \rep $V$ by
\beq
x\tr v_i = v_j \pi^j_i(x), 
\eq
for a basis $v_i$ of $V$.
We are mainly interested in (tensor products of) unitary representations.
The following will be very useful:
\begin{lemma}
If an irrep $V(\l)$ is analytic and unitary at a phase $q$
(for the compact involution (\ref{cbar}), say), then it has
a basis which is orthogonal and normalized w.r.t. 
both its symmetric bilinear form {\em and} its
invariant inner product, i.e.
\berr
\pi^i_j(\theta(x)) &=& \pi^j_i(x)\quad\mbox{and}\quad \\
\pi^i_j(\cobar{x}) &=& \pi^j_i(x)^{\ast} \equiv (\pi^j_i(x))^{\ast},
\err
in that basis.
\label{orth_lemma}
\end{lemma}
\begin{proof}
First, one can check that this holds for the fundamental \reps $V_f$
(i.e. the spinor \rep for $SO_q(5)$) \cite{resh_1}. 
We will show the general statement inductively by taking tensor products
with the fundamental representation. Suppose it holds  for 
$\pi^i_j$ on $V(\mu)$, and onsider 
$V(\mu)\tens V_f = \oplus V(\l_i)$. All the
multiplicities are known to be one in this case (this can be seen e.g. using
the Racah--Speiser algorithm). Then the Clebsch--Gordan coefficients 
$K^{ij}_m(q)$ defined by  
$v_m^{(\l_i)} = K^{ij}_m v_i\tens v_j \in V(\l_i)$  satisfy \cite{resh_1}
\berr
K^{ij}_m(q)R^{ls}_{ij} &=& 
    (-1)^{\nu_i}q^{\frac 12(c_{\l_i}-c_{\mu}-c_f)} \wtilde{K}^{sl}_m(q), 
                                                         \label{clebsch_R}\\
K^{ij}_m(q^{-1}) &=& (-1)^{\nu_i} \wtilde{K}^{ji}_m(q). \label{clebsch_q-1}
\err
Here $\wtilde{K}^{sl}_m(q)$ is the Clebsch in the reversed tensor product, 
and $(-1)^{\nu_i}$ is the same as classically (and just a convention
unless the factors are identical). The first can be seen similarly as in
section \ref{subsec:centralizer}. Furthermore, $K^{ij}_m(q)$ is real for 
$q\in\reals$.

Using this, we can write an invariant inner product
of 2 vectors in $V(\l_i)$ as in (\ref{inner_H}),
\berr
(v_m^{(\l_i)},v_n^{(\l_i)})_{\H} &=& q^{-\frac 12(c_{\l_i}-c_{\mu}-c_f)}
         K^{ij}_m(q^{-1}) K^{st}_n(q) R^{kl}_{st}
                              (v_i\tens v_j,v_k\tens v_l)_{\tens} \nn\\
    &=& \wtilde{K}^{ji}_m(q) \wtilde{K}^{lk}_n(q)
        \delta^i_k \delta^j_l \nn\\
    &=& \delta^m_n, 
\err
choosing an orthogonal basis of $V(\l_i)$, which is unitarizable 
by assumption.
But then the invariant bilinear form of these vectors is
\berr
(v_m^{(\l_i)},v_n^{(\l_i)})_{\tens}^{(bi)} &=& 
 K^{ij}_m(q)K^{kl}_m(q)(v_i,v_k)^{(bi)} (v_j,v_l)^{(bi)}\nn\\
&=& \delta^m_n,
\err
using $(v_i,v_k)^{(bi)} = \d^i_k$ on the components, 
by the induction assumption. This means that the $v_m^{(\l_i)}$ are 
orthogonal not only w.r.t. the sesquilinear inner product, but also the 
above bilinear form.
\end{proof}

In the case of $SU_q(2)$, this can easily be checked explicitely.

We will always use this basis from now on.
By inserting $(-1)^E$, one gets the same statement for the 
shifted noncompact representations, with $\theta$ replaced by $\ttheta$.
Moreover, the result also holds for the irreducible quotients of analytic 
\reps at roots of unity of a given unitarity type, such as the physical
many--particle \reps defined in section \ref{subsec:many_2d} and 
\ref{subsec:many_4d}.

Now we can get a better understanding of $\tw$. 
First, it intertwines a unitary \rep with its
contragredient (=dual) representation, defined by 
$x\tilde{\tr} a_i = a_j \pi^i_j(S^{-1}x)$ (remember that we only 
consider "real" groups):
\berr
x\tr(\tw\tr a_i) &=& (x\tw)\tr a_i = \tw\tr(\ttheta S^{-1}x)\tr a_i \nn\\
        &=& \tw\tr(a_j \pi^j_i(\ttheta S^{-1}x)) = 
              \tw\tr(a_j \pi^i_j(S^{-1}x))  \\
        &=& \tw\tr(x\tilde{\tr} a_i).
\err

Other important properties are as follows. Define
\beq
g_{ij} \equiv \pi^i_j(\tw) q^{\frac 12 c_{\l}} = (-1)^{f} g^{ij}
\label{metric_w}
\eq
where $c_{\l}$ is the classical quadratic Casimir of the representation, and
$(-1)^{f}$ is the value of $\vareps$.
The last equality follows from (\ref{w_2}), where
$g_{ij} g^{jl} = \d_i^l$.
This is nothing but the invariant tensor 
(again, for "real" groups):
\begin{prop}
\berr
\pi^i_k(x_1) \pi^j_l(x_2) g^{kl} &=& \eps(x) g^{ij} \label{g_inv1}\\
g_{ij} \pi^i_k(x_1) \pi^j_l(x_2) &=& \eps(x) g_{kl} \label{g_inv2}\\
g_{ij} &=& q^{\frac 12 c_{\l}}\pi^j_i((-1)^E w)   \label{g_transposed} \\
g_{ls} \hat R^{sk}_{ut} g^{tv} &=& (\hat R^{-1})^{kv}_{lu}. \label{R_g}
\err
\label{g_prop}
\end{prop}
(\ref{R_g}) and some more similar relations are contained in \cite{resh_1}.
\begin{proof}
(\ref{g_inv1}) follows from
\berr
\pi^i_k(x_1) \pi^j_l(x_2) \pi^k_l(\tw) &=& 
                                 \pi^j_l(x_2) \pi^i_l(x_1\tw)\nn\\
   &=& \pi^j_l(x_2)\pi^i_n(\tw) \pi^n_l(\ttheta(S^{-1}x_1)) \nn\\
   &=& \pi^j_l(x_2)\pi^i_n(\tw) \pi^l_n(S^{-1}x_1) \nn \\
   &=& \pi^j_n(x_2 S^{-1}x_1) )\pi^i_n(\tw) \nn\\
   &=& \eps(x) \d^j_n \pi^i_n(\tw),
\err
and similarly (\ref{g_inv1}). 
(\ref{g_transposed}) follows from the uniqueness of the invariant tensor
(cp. Lemma \ref{inv_lemma}),
noting that 
\berr
\pi^i_k(x_1) \pi^j_l(x_2) \pi^l_k((-1)^E w) &=&
                   \pi^j_k(x_2 q^{-2\tilde{\rho}}\tw) \pi^i_k(x_1)\nn\\
 % &=& \pi^j_k(q^{-2\tilde{\rho}}(S^2 x_2)\tw)\pi^i_k(x_1) \nn\\
  &=& \pi^j_k(q^{-2\tilde{\rho}}\tw \ttheta S x_2)\pi^i_k(x_1) \nn\\
 &=& \pi^j_t(q^{-2\tilde{\rho}}\tw) \pi^t_k(\ttheta S x_2)\pi^i_k(x_1) \nn\\
 &=& \pi^j_t(q^{-2\tilde{\rho}}\tw) \pi^k_t(S x_2)\pi^i_k(x_1) \nn\\
 &=& \pi^j_t((-1)^E w) \d^i_t \eps(x)
\err
is indeed invariant, where we used (\ref{S2_inner}).
This shows (\ref{g_transposed}) up to a constant, which is one because
\berr
\pi^i_j(\tw) \pi^j_k(\tw) &=& \pi^i_k(v \vareps) \quad\mbox{and} \nn\\
\pi^j_i((-1)^E w) \pi^k_j((-1)^E w) &=& \pi^k_i(v\vareps),
\err
which is the same. Finally,
(\ref{R_g}) is obtained by taking \reps
of 
\beq
(1\tens\tw) \R (1\tens\tw^{-1}) = (1\tens \ttheta S^{-1})\R = 
            (1\tens\ttheta)\R^{-1}.
\eq
\end{proof}

Furthermore, from (\ref{del_om}) one immediately obtains (\ref{R_l_equal}), 
(\ref{qtr_l_3}), as well as (\ref{del_v_l}).

For $|q|=1$, it is consistent to define
\beq
\obar{\tw} = \tw^{-1}
\label{om_conj}
\eq
and similarly for $w$. 
Correspondingly, we have 
\beq
\pi^i_j(\tw)^{\ast} = \pi^j_i(\tw^{-1}).
\label{om_unit_rep}
\eq
This can be checked explicitely using the known formulas for
$w$ in terms of the $w_i$ \cite{kirill_resh} 
and their action on a representation \cite{jantzen,lusztig_book}.
 
\section{An Algebra of Creation and Anihilation Operators}
\label{sec:creat_anihil_op}

In this section, we will show how to define an interesting algebra of 
creation and anihilation operators, with involution. 
This allows 
us to work with states of different particle numbers, and
to write down correlators as in QFT. It is however not clear at present 
how identical particles should be defined, so we only consider 
distinguishible particles.

Denote the states of a physical \rep $V_{(\l)}$ of $SO_q(2,3)$\footnote{
This works for other real groups as well.} by $a_i$.
We can make $V_{(\l)}$ into a
$\U$ --module algebra $\F^{(a)}$ (see section \ref{subsec:act_coact})
by defining either $a_i a_j =0$, or more somewhat more interestingly by
\beq
a_i a_j = g_{ij} a^2, \quad a_i a_j a_k=0
\label{a_algebra}
\eq
where $a^2$ is a (scalar) variable, and $g_{ij}$ as in (\ref{metric_w}).
We will use this algebra only to to show how to define an involution,
and to write the inner product on $G_0$ in a elegant way.
Notice that in the noncompact case,
$g_{ij}\neq 0$ only if one of the "factors" is a positive and one a
negative energy representation (an antiparticle wavefunction,
i.e. shifted by $-2M\b_3$ as explained in section \ref{subsec:unitary_so23}).
We are using the same letter $a_i$,
because both positive and negative energy \reps can be obtained as 
quotients of one "big" self--dual representation, namely the $\H_0$ part 
of a tensor product as in
Theorem \ref{struct_thm}, which is the direct sum of a 
positive an a negative energy representation. 
This is very nice from the QFT point of view,
and one should keep this in mind for the following.

Given many left $\U$ --module algebras $\F^{(a)}, \F^{(b)}, ...$
generated by $a_i,b_i, ...$, one can define a combined
(left) $\U$ --module algebra $\F$ using the
{\em braided tensor product} \cite{majid_braid}. 
As vector space, this is simply $\F = \F^{(a)}\tens \F^{(b)} \tens ...$,
with commutation relations
\beq
a_i b_j = (\R_2 \tr b_j)(\R_1\tr a_i) \g
\label{braiding}
\eq
for  some $\g \in \compl$; similarly for more variables.
This definition requires an (arbitrary) "ordering" $a > b > ...$ of the 
different algebras. It is consistent because of the standard
properties of $\R$. 

Finally on the vector space $\U\tens \F$, one can define a 
{\em cross product} algebra $\U\smash \F$ via
\beq
x a_i = (x_1 \tr a_i) x_2, \quad\mbox{or equivalently}\quad
       x\tr a_i = x_1 a_i Sx_2.
\label{smash}
\eq
This is an algebra, because $\U$ is a Hopf
algebra and $\F$ is a (left) $\U$ --module algebra. 

To make the connection with QFT more obvious, we define a ("Fock") vacuum
$\>$
which reduces $\U\smash\F$ to its "vacuum--representation" 
$\U\smash\F\>$ by
\beq
f(a,b,..) x\> \equiv \eps(x) f(a,b,...)\>.
\eq
This is a tensor product of representation of $\U$, and
\beq
x f(a,b,...)\> = x_1\tr f(a,b,...)x_2\> = x\tr f(a,b,...)\>.
\eq 
In particular this contains the subspace
$G_0$ where $v$ is diagonalizable,
and we can apply the results of sections \ref{sec:tensor_brst}
and \ref{sec:inner_prod_g0}.

Define
\beq
\Om = \tw \sqrt{v^{-1}} 
\label{Omega}
\eq
with $\sqrt{v^{-1}}$ as in (\ref{sqrt_v_inv}).
Acting on $G_0$, this satisfies
\beq
\Om^2 = \vareps
\label{Om_2_eps}
\eq
at roots of unity, using (\ref{w_2}) and $S(v)=v$. 
Furthermore from (\ref{om_conj}), 
\beq
\obar{\Om} = \Om^{-1}
\label{Om_conj}
\eq
if acting on $G_0$.

Now at roots of unity,
define $\H$ to be the algebra $\U\smash\F$ with the {\em additional}
relation $Q=0$, i.e.
\beq
\H \equiv (\U\smash\F)/Q.
\label{phys_alg}
\eq
It is closely related to the $\H_0$ \reps of section
\ref{sec:tensor_brst}, see also Conjecture \ref{conjecture}.

This is very similar to the definition of the physical Hilbert space
in QFT using the BRST operator (\ref{BRST_cohom}), 
formulated in terms of an algebra
instead of representations, which is essentially the same.
Since $Q$ vanishes on all generic representations,
this contains in particular the (semidirect product of $\U$ with the)
physical many--particle states in the case of $SO_q(2,3)$, as well
as the states of the usual truncated tensor product in the context of CFT.

Using $\Om$, we can essentially define an involution on 
$\H$ (resp. for $G_0$ --type \reps of $\U\smash \F$) as follows: 
for $x\in\U$, it is simply the involution 
$\obar{x}$ corresponding to the reality structure considered.
If $x$ is either $X_i^{\pm}$ or $H_i$, this can be written as
\beq
\obar{x} = \Om S(x) \Om^{-1}. 
\label{invol_U}
\eq
For a generator $a_i$ of $\F$, we define
\beq
\obar{a_i} = \Om a_i \Om^{-1},
\label{invol_F}
\eq
and extend this as an antilinear antialgebra--homomorphism on
$\H$ (resp. $\U\smash\F$). It is shown in Appendix A that
this is consistent with the algebra $\H$
provided the $a_i$ are unitary w.r.t. the reality structure on $\U$,
and
with the braiding algebra (\ref{braiding}) if $\g$ is a phase.
It is also consistent with (\ref{a_algebra}) provided 
$\obar{a^2} = (-1)^{f_a}a^2$ where $(-1)^{f_a}$ is the value of $\vareps$ 
on $a_i$; note that 
$g_{ij}^{\ast} = (-1)^{f_a} g_{ji}$, using (\ref{om_unit_rep}).
Again in the  noncompact case,
$\Om$ can act on $a_i$ only if it contains both positive and negative energy
states, such as $\H_0$ in Theorem \ref{struct_thm}.
Now
\berr
\obar{\obar{a_i}} &=& v \sqrt{v^{-1}}^2 \vareps a_i v^{-1}
                \sqrt{v^{-1}}^{-2} \vareps^{-1}\\
           &=&  a_i (-1)^{f_a}
\label{obar_2}
\err
in $\H$ (resp. on $G_0$ --type representations),
using (\ref{del_eps}) resp. ({\ref{Om_2_eps}). 
This is as good as an involution, 
and the main result 
of this section. In fact, one could define instead 
$\obar{a_i} = \eps \Om a_i \Om^{-1}$,
which really is an involution; we choose not to do this here.
Again, the operator adjoint can be calculated once there
is a positive definite inner product. We want to indicate how this could
be achieved:

Let $\F$ be as above for
braided copies $a_i, b_i,...$ of the algebra (\ref{a_algebra}).
Define an evaluation of $\H$
\beq
\<x f(a,b,...)\> \equiv \eps(x) \<f(a,b,...)\>
\eq
by first collecting the generators of the same algebra using the 
braiding relations, and defining $\<a^2\> \equiv (-1)^{f_a} g_a$ with 
$g_a\in\reals$, and 
$\<a_j\> \equiv 0$; similarly for the other variables. This is independent of 
ordering if $\g = \pm 1$, because $a^2$ is 
then central. It satisfies
\berr
\<x f(a,b,...)\> &=& \<f(a,b,...)x\> =  
                 \eps(x) \<f(a,b,...)\> \label{invar_corr}\\
\<f(a,b,...)\>^{\ast} &=& \<\obar{f(a,b,...)}\> (-1)^f.  \label{herm_corr}
\err
on $\H$ where $f=f_a+f_b+ ...$, using $\obar{a^2} = a^2 (-1)^{f_a}$.

One can now write states of $\H_0$ in the form
$f\> = f^{i_a i_b ...} a_{i_a} b_{i_b} ...\>$ and $g\> =
g^{i_a i_b ...} a_{i_a} b_{i_b} ...\>$ where $a_{i_a}$,  $b_{i_b}$, \dots are 
positive--energy (physical) states, and define an inner 
product as follows:
\beq
(f,g) \equiv \<\obar{f}g\>.
\label{inner_om} 
\eq
This  is hermitian, invariant
\beq
(x\tr f,g) = (f,\obar{x}\tr g)
\eq
and it is shown in Appendix A that it is in fact the same as 
the inner product defined in (\ref{inner_H}),
\beq
(a_i b_j ... ,a_k b_l...) \equiv  \<\obar{a_i b_j ...}a_k b_l...\> = 
  (a_i\tens b_j\tens...., a_k\tens b_l\tens...)_H ,
\label{inner_equal}
\eq
if the normalization on the rhs is chosen as $(a_i,a_j)=g_a \d^i_j$
(remember that we work in an orthogonal basis).
Invariance can be seen easily:
\berr
(x\tr f,g) &=& \left<(\obar{x_1 f Sx_2}) g\right> = 
   \left<\obar{Sx_2} (\obar{f}) \obar{x_1} g\right> = 
              \left<\obar{f} \obar{x} g\right> \nn\\
&=& (f,\obar{x}\tr g),
\err
using (\ref{invar_corr}). Hermiticity follows from (\ref{herm_corr}).
Positivity was discussed in 
section \ref{subsec:hilbert} for the case of the Anti--de Sitter group.
 
One could formulate all this without using $w$ explicitely, in the form
$\obar{a_i} = (\Om_1 \tr a_i)\Om_2$ which can be written in terms of
the universal $\R$. Needless to say, this would be much more
complicated. However it helps to understand the main point 
of this definition,
namely the $\R$ involved which "corrects" the flipping of the tensor product
in the reality structure (\ref{cbar_tens}).
In this form, a somewhat similar--looking conjugation was
introduced in \cite{mack_schom}.

\subsection{On Quantum Fields and Lagrangians}

In this section, we want to show how the above formalism could find 
application in a QFT. This can only be very vague at present, because
an important piece is still missing -- the implementation of a 
symmetrization postulate, in order to define identical particles.
We can nevertheles write down a few generic  
formulas in an ad--hoc way.

Consider a (large) number of braided copies of the algebra (\ref{a_algebra})
with generators $a_i^{\l,(n)}$, for $n=1,2,...,N$ and $\l$ going through all
possible highest weights of physical (unitary) \reps of a given spin;
remember that there exist only finitely many at roots of unity.

Then consider the following object:
\beq
\Psi(y) \equiv \frac 1{\sqrt{N}} \sum_{i,n} a_i^{\l,(n)} f^i_{\l,(n)}(y).
\label{quant_field}
\eq
Here $f^i_{\l,(n)}(y)$ is the dual \rep of $a_i^{\l,(n)}$, realized as
functions (or forms, ...) on quantum AdS space, which is a {\em right} 
$\U$ --module algebra in the dual picture of chapter 2. 
This works as in Theorem \ref{struct_thm}, where the factors $V_i$ are the
5--dimensional \reps $y^i$. Then the unitary \reps are the quotients $\H_0$,
in the space of functions on quantum AdS space. By this construction,
$\Psi(y)$ contains both positive and negative 
energy \reps as discussed earlier,
so that $\Om$ can act on it.
Thus we can assume that
\beq
u\tr \sum a_i^{\l,(n)} f^i_{\l,(n)}(y)  
 = \sum a_j^{\l,(n)}\pi^{j \l}_i(u) f^i_{\l,(n)}(y)
 = \sum a_j^{\l,(n)} (f^j_{\l,(n)}(y) \tl u).
\eq
for $u\in \U$. $\Psi(y)$  behaves very much 
like an off--shell quantum field. Similarly, we can consider
$\Psi(y_1), \Psi(y_2), ...$ with (braided) copies $y_i$ of quantum AdS space.
Then for example,
\beq
\<\Psi(y_1)\Psi(y_2)\> = 
     \pm \sum_{\l} g_{\l} g_{ij} f^i_{\l,(n)}(y_1) f^j_{\l,(n)}(y_2) +o(h)
\eq
becomes a correlator in the classical limit, depending on the choice of
the $g_{\l} \equiv g_{a^{\l}}$ for the representations involved.
In particular, $g_{\l} = \frac i{\Box_{\l} -m^2}$ 
corresponds to a 
Green's function, where $\Box_{\l}$ is the (quantum) quadratic 
Casimir (see \cite{FRT}).  More generally,
\beq
\<\Psi(y_1) ... \Psi(y_k)\>
\eq
will reproduce the sum of Wick contractions at $q=1$  
for $N \rightarrow \infty$.

Using this, one can write down e.g. an "interaction Lagrangian"
\beq
{\cal S} = \int_y \Psi(y) ... \Psi(y),
\label{S_interact}
\eq
where the integral over quantum AdS space is defined in section 
\ref{subsec:ads_space_def} (in different notation),
and many similar terms.
This is invariant under $SO_q(2,3)$, i.e. $x {\cal S} = {\cal S} x$
in $\U\smash \F$ for $x\in \U$, using the invariance of the integral.
The only thing we want do here is to show that ${\cal S}$
is in fact hermitian, with the involution defined in the previous section:
\beq
\obar{\int_y \Psi(y) ... \Psi(y)} = \int_y \Psi(y) ... \Psi(y).
\label{S_interact_herm}
\eq
If one can find an algebra replacing (\ref{a_algebra}) such that 
the  inner product (\ref{inner_om}) 
is positive definite, this implies that $e^{i{\cal S}}$
is a unitary operator on $H_0$ (in the present version, this is not the 
case because of the additional generator $a^2$). 

To see (\ref{S_interact_herm}), 
let us simplify the notation first by writing
$a_i, b_j,c_k$ for $a_i^{\l,(1)}$,
$a_j^{\l,(2)}$ and $a_k^{\l,(3)}$, respectively, and to be specific consider
\beq
\tilde{{\cal S}} = \int_y \psi_a(y) \psi_b(y) \psi_c(y)
\eq
where $\psi_a(y) \equiv \sum_{i} a_i f^i(y)$,
and similarly $\psi_b(y), \psi_c(y)$.
We claim that 
\beq
\obar{\tilde{{\cal S}}} = \int_y \psi_c(y) \psi_b(y) \psi_a(y).
\label{stuff}
\eq
It is clear that this implies (\ref{S_interact_herm}). We consider only 
scalar fields here for simplicity.

To see (\ref{stuff}), first observe that
\beq
\bobar{f^i(y)} = \pm f^i(y) 
\eq
using the (auxiliary) antilinear involution on quantum AdS space 
defined in section
\ref{subsec:ads_space_def}. This follows from $\bobar{y^i} = -y^i$ and their 
algebra,
(\ref{clebsch_R}) and (\ref{clebsch_q-1}). 
Together with (\ref{obar_I_phase}), this implies
\berr
\obar{\int_y \psi_a(y) \psi_b(y) \psi_c(y)} &=& 
      \sum \Om c_k b_j a_i\Om^{-1} \int_y f^k(y) f^j(y) f^i(y) \nn\\
  &=& \int_y \psi_c(y) \psi_b(y) \psi_a(y)
\err
as claimed, since this is a scalar and it therefore commutes with $\U$ 
and $\Om$.

One could now go ahead and define "ad hoc" correlators such as 
$\<\Psi(y_1) e^{i {\cal S}} \Psi(y_2)\>$, which in fact can be expressed
as sum of contractions as seen above, similar to the classical
Wick expansion. Moreover for $q\neq 1$, 
these contractions can be interpreted naturally
as generalized links, with interaction vertices being Clebsch--Gordan
coefficients defined by the integral, 
and crossings given by the $\hat R$ --matrices in the particular
representations. All these diagrams would be finite, since
there exist only finitely many "physical" representations, which are 
all finite--dimensional. Moreover, 
these correlators are in fact symmetric in the
classical limit, since the $\Psi(y_i)$ commute for $q=1$.
Nevertheless, this is not really satisfactory, since an
explicit symmetrization postulate is missing, as well as
a dynamical principle
determining "on--shell" states. These two open problems are probably
related.

\paragraph{Symmetrization.}
Let us briefly discuss the problem of symmetrization and 
identical particles.
For generic $q$, this has been studied in \cite{schupp}. 
At roots of unity, the situation can be expected to be somewhat different. 

In principle,  one can define projectors $(P^{S,A})^{ij}_{kl}$ which 
act on the tensor product of 2 identical representations $V(\mu)$, and 
project out the totally symmetric resp. antisymmetric representations.
This can be done using the fact that the $\hat R$ --matrix discussed in
section \ref{subsec:centralizer} has eigenvalues
$\pm q^{\frac 12(c_{\l} -2c_{\mu})}$ with the same sign as classically,
and one "just" has to correct the $q^{\frac 12(c_{\l} -2c_{\mu})}$ --factor.
The problem is that this may not give an interesting associative algebra of 
"totally (anti)symmetric" \reps for more than 2 particles, it is probably 
too restrictive in general. One may hope  that something more 
favorable happens at roots of unity on the space $\H_0$.

%is quite different, and there appears to be an "obvious" candidate
%for a symmetrization postulate. Consider a physical, unitary
%representation $V$ with elements $a_i$, and recall from section 
%\ref{subsec:centralizer} that the $\hat R$ --matrix acts on
%$V\tens V$, commutes with the action of $\U$, and has eigenvalues 
%$\pm q^{\frac 12(c_{\l} -2c_{\mu})}$. The sign is the same as in the 
%classical limit, namely $+$ for symmetric \reps and $-$ for
%antisymmetric representations. Moreover, it turns out that at least
%for \reps with integer spin, the classical Casimir $c_{\l}$ for 
%the AdS group is always an ... integer. Thus for $q=e^{i\pi/M}$
%and odd $M$, $(\hat R)^M $ has eigenvalues $+1$ for the 
%symmetric representations, and $-1$ for the antisymmetric
%^ones, on $G_0$ resp. $H_0$. This is exactly what we want, and one 
%can define an algebra
%\beq
%a_i a_j  = a_k a_l (\hat R^M)^{kl}_{ij}
%\label{symm_alg}
%\eq
%for bosons. Instead of $\hat R^M$, one could as well use
%$c_V \tau\circ(\pi\tens\pi)\R\Del(\sqrt{v})$, 
%and generalize (\ref{symm_alg}) to fermions.

%As attractive as this may look, it is not clear whether this algebra
%is interesting for our purpose. It might be too restricitive,
%i.e. the \reps obtained in this way for more than 2 particles 
%may be smaller than classically, since $(\hat  R)^M $
%is not expected in general to satisfy the Yang--Baxter equation. 
%It would be very desirable to gain a better understanding of these
%issues.

In this context, we can define an interesting algebra:
\beq
a_i \obar{a_j} = \pm \obar{a_j} a_i
\eq
with the bar as in (\ref{invol_F}), taking advantage of
$\U\smash\F$ resp. $\H$.
It is easy to check that this is compatible with the cross--product 
and our "involution" on $\H$.
Acting on the Fock--vacuum $\>$, this in fact defines totally 
symmetric resp. antisymmetric 2--particle representations, 
with a suitable definition of $\sqrt{v}$.
Again, it is not
clear if this algebra is interesting for our purpose
for more than 2 particles. It would be very desirable to get a 
better understanding of these issues.

\subsection{Nonabelian Gauge Fields from Quantum AdS Space}

In section \ref{subsec:so23_reps_quant}, we have found a BRST operator 
for spin one particles, corresponding to abelian gauge fields.
Needless to say that one would also like to consider the nonabelian case.
Nonabelian gauge fields are usually described by connections on
principal fiber bundles. Therefore one might try to do the same
in the q--deformed case, see e.g. \cite{BM,pflaum}. However 
there is a much simpler, extremely fascinating way 
to obtain such objects on quantum AdS space 
(and certain other quantum spaces). 
It is in fact much simpler than classically. We will only give a rough outline 
here.

Consider the calculus of differential forms on q--AdS space, which is the 
same as on quantum Euclidean space, except for the reality structure.
As in \ref{sec:forms_euclid}, the following observation by Bruno Zumino 
\cite{zumino_priv} will be crucial: there exists a "radial" 
one--form $\om = \frac q{\sqrt{q}^{-1}+\sqrt{q}} \frac 1{r^2} d(r^2)$ 
which generates the calculus on quantum Euclidean space, sphere 
resp. on q-AdS space by 
\beq
[\om, f]_{\pm} = (\sqrt{q}^{-1}-\sqrt{q}) df \equiv \xi df.
\label{om_generate}
\eq
Furthermore on the sphere, it is not possible for $q\neq 1$ 
to work with "tangential"
one--forms only, due to the commutation relations 
(\ref{xdx_CR}). Therefore $\om$ must be included in the calculus.

Now consider a matrix of one--forms $\Om^i_j$ (this has nothing to do with
the $\Om$ in section \ref{sec:creat_anihil_op}), 
and write it in the following way:
\beq
\Om^i_j = \om \d^i_j +\xi B^i_j .
\label{Omega_ij}
\eq
Physically, we can imagine that this comes from some spontaneous 
symmetry breaking in the radial fields, which are scalars.
One can decompose the one-forms $B^i_j$ into tangential
and radial components, e.g. using a Hodge--star operator $\ast$ which 
can be defined in a straightforward way. Then
\berr
\Om &=& \om 1 + \xi(\Phi +  A),  \quad\mbox{i.e.} \nn\\
\Om^i_j &=& \om \d^i_j + \xi (\phi^i_j \om +  A^i_{j \mu} dy^{\mu})
\err
with the condition $\om\wedge\ast (A^i_{j,\mu} dy^{\mu})=0$, 
changing notation
for the coordinates on AdS space.
Furthermore, we can imagine a reality condition like 
$(\Om^i_j)^{\dagger} = -\Om^j_i$ (without going into details here),
so that $\Om^i_j$ corresponds to some Lie algebra, and
(\ref{Omega_ij}) corresponds to $tr \phi =0$.

Now consider $(\Om^2)^i_j$, which after a simple calculation using 
(\ref{om_generate}) becomes
\berr
\Om^2 &=& \xi^2 (dB+BB),\quad\mbox{i.e. } \nn\\
\Om^i_k \Om^k_j &=& \xi^2 (dB^i_j + B^i_k B^k_j).
\err
Decomposing it into radial and tangential components, we get
\berr
\Om^2 &=& \xi^2 (dA +AA + d\Phi + \Phi A+A \Phi ) \nn\\
      &=& \xi^2 (dA+AA+ (d\phi+ [A,\phi])\om + o(h))
\err
where $q=e^{2\pi i h}$ as usual. Thus 
\beq
{\cal S} = \frac 1{\xi^2} \int tr (\Om^2 \ast \Om^2)
\eq
gives precisely the Yang--Mills
action for a gauge field $A$ coupled to a scalar in the adjoint, like
a Higgs field in some GUT models! We do not have to define curvature by hand.
This also contains massless BRST ghosts in a nonstandard form,
according to section \ref{subsec:so23_reps_quant}.
Moreover, if $\Om$ transforms like 
\beq
\Om(y) \rightarrow \g^{-1}(y)\Om(y) \g(y),
\eq
then the components of $\Om$ transform like 
\berr
A &\rightarrow& \g^{-1} A \g + \g^{-1} d\g +o(h), \quad\mbox{and}\nn\\
\phi &\rightarrow& \g^{-1} \phi \g + o(h),
\err
which are precisely the gauge transformations in a Yang--Mills Theory.
Similarly, any trace of polynomials in $\Om$ gives a gauge invariant 
Lagrangian in the limit $h\rightarrow 0$.

How exactly this fits together
with the BRST operator etc. remains to be seen.
At the very least, this shows that there is no need to define
objects like  connections on principal bundles for $q \neq 1$,
they arise naturally form the mathematical structure considered.
Without elaborating this any further here, we see once again that
$q$ --deformation is more than just a "deformation", 
it allows to do things which
cannot be done for $q=1$, and which look very interesting from the point of 
view of QFT.  

\chapter{Conclusion}

We have shown that many of the essential ingredients of quantum field
theories, or more properly quantum theories of elementary particles,
have their counterparts in an approach where the classical Poincare group  
and Minkowski--space
are replaced by the quantum Anti--de Sitter group $SO_q(2,3)$
and a corresponding AdS space, with $q$ a suitable root of unity. 
First of all, it turned out that there are indeed unitary representations 
which may describe
elementary particles of any spin, with the same low--energy structure
as classically. We have defined many--particle representations, which are 
unitary as well; this is not trivial. In the massless case, it turns out that
there is  a very natural way to define a Hilbert space using a BRST operator 
$Q$, which is an element of the center of $U_q(so(2,3))$; in the classical
case, it has to be defined by hand. Moreover, the same $Q$
works for any spin, and it seems that it can be used to define the 
many--particle Hilbert space in a similar way. We have also seen how
the structure of a nonabelian gauge theory can arise naturally from
quantum Anti--de Sitter space; again in the classical case, its
description using connections on fiber bundles is quite ad--hoc.

Moreover, everything is manifestly finite in the quantum case. This should not
be considered a technicality; it seems that if there is a complete theory
of elementary particles, it should be possible to formulate it in a completely
well--defined way, without any vague "remedies" under the
name of regularization; it should be regular by itself.
While it may be too ambitious to attempt finding such a theory,
{\em any} finite version of a 4--dimensional quantum field theory
would be very interesting in itself.

Apart from all these mathematical features, one may still wonder if it makes
any sense at all 
to assume that the spacetime we see looking out of the window 
is noncommutative, i.e. not a classical manifold. I think that this has been 
answered as well, by pointing out that 
if $q=e^{i\pi h}$ is very close to 1 (which it has to be, otherwise there is 
an obviously clash with observation), 
the coordinate algebra of quantum Anti--de 
Sitter space is classical up to corrections involving a length scale
$L_0 = \sqrt{h} R$, where $R$ is the curvature radius of AdS space. Thus one
should expect it to look like a classical manifold on length scales bigger 
than $L_0$, just like quantum mechanics behaves classically 
on scales large compared to $\hbar$. 
This shows that $h$ must indeed be a 
{\em very} small number. 

All this can be said in favour of the approach chosen. 
Nevertheless, we have not
yet succeeded in formulating a theory of elementary particles,
because we do not know at this time how to define a Hilbert space of 
identical particles. 
There may be a satisfactory solution of this
problem  in the form of a suitable algebra of creation and
anihilation operators (in which case we have presented an nice way
to define an involution, which is usually a difficult part in the 
noncommutative case), or perhaps in a completely unexpected 
way, -- or it might be that this is 
the downfall of the approach. Of course one should hope for the first
alternatives, and  in view of all the promising features found so far, 
I cannot believe that this is the end of the story.

\appendix

\chapter{Consistency of Involution, Inner Product}
\paragraph{Compatibility of (\ref{invol_F})  with  cross product.} 

Applying $\obar{(.)}$ to $x a_i = x_1\tr a_i x_2$ in $\H$, we get
\berr
\Om a_i \Om^{-1} \obar{x} &=&(\obar{x_2})\obar{a_l}\pi^l_i(x_1)^{\ast} \nn\\
  &=& \obar{x_2} \Om a_l \Om^{-1} \pi^i_l(\obar{x_1}).
\err
Multiplying from the left with $\Om^{-1}$ and from the right with $\Om$,
this becomes
\beq
a_i \Om^{-1} \obar{x} \Om \stackrel{!}{=}
      \Om^{-1}(\obar{x})_1 \Om a_l \pi^i_l((\obar{x})_2)
\eq
where $(\obar{x})_{1,2}$ is the Sweedler notation for the coproduct, 
which becomes using (\ref{autom_tw}) 
\berr 
a_i \ttheta S^{-1} \obar{x} &\stackrel{!}{=}& (\ttheta S^{-1}(\obar{x})_1)_1
          \tr a_l (\ttheta S^{-1}(\obar{x})_1)_2 \pi^i_l((\obar{x})_2) \nn\\
   &=& (\ttheta S^{-1}(\obar{x})_{12})
           \tr a_l \ttheta S^{-1}(\obar{x})_{11} \pi^i_l((\obar{x})_2) \nn\\
%   &=& a_k \pi^k_l(\ttheta S^{-1}(\obar{x})_{12})\pi^i_l((\obar{x})_2)
%                                        \ttheta S^{-1}(\obar{x})_{11} \nn\\
   &=& a_k \pi^l_k(S^{-1}(\obar{x})_{12}) \pi^i_l((\obar{x})_2)
                                         \ttheta S^{-1}(\obar{x})_{11} \nn\\
   &=& a_k \pi^i_k((\obar{x})_2S^{-1}(\obar{x})_{12})
                                       \ttheta S^{-1}(\obar{x})_{11} \nn\\
   &=& a_k \pi^i_k(\eps(\obar{x}_2)) \ttheta S^{-1}(\obar{x})_1 \nn\\
   &=& a_i \ttheta S^{-1}(\obar{x})
\err
as desired, using Lemma \ref{orth_lemma} and standard
properties of Hopf algebras. 
                                    
\paragraph{Compatibility of (\ref{invol_F}) with braiding.}

Applying $\obar{(.)}$ to \newline
$a_i b_j = \g (\R_2\tr b_j)(\R_1\tr a_i)$, we get
\beq
\Om b_j a_i \Om^{-1} \stackrel{!}{=} \g^{\ast} \Om a_k \pi^k_i(\R_1)^{\ast}
             b_l \pi^l_j(\R_2)^{\ast} \Om^{-1},
\eq
or 
\berr
b_j a_i  &\stackrel{!}{=}& \g^{\ast} a_k \pi^i_k(\obar{\R_1})
                               b_l \pi^j_l(\obar{\R_2}) \nn\\ 
    &=& \g^{\ast} a_k \pi^i_k(\R^{-1}_2) b_l \pi^j_l(\R^{-1}_1) \nn\\ 
    &=& \g^{\ast} a_k \pi^k_i(\R^{-1}_1) b_l \pi^l_j(\R^{-1}_2),
\err
using $(\theta\tens\theta) \R = \R_{21}$ and similarly for $\ttheta$
in the last line. Multiplying with $\pi^j_s(\R_b) \pi^i_t(\R_a)$, we get
\beq
b_j \pi^j_s(\R_b) a_i \pi^i_t(\R_a) \stackrel{!}{=} \g^{\ast} a_t b_s,
\eq
which is just the braiding algebra provided $\g^{\ast} = \g^{-1}$.

\paragraph{Equality of the inner products (\ref{inner_equal})}

Using $\Del(\Om) = \Del(\sqrt{v^{-1}})\R^{-1}(\tw\tens\tw)$ 
and $\<f(a)x = \< (S^{-1}x)\tr f(a)$ we can write the lhs of 
(\ref{inner_equal}) explicitely as
\berr
(a_i b_j ...,a_k b_l...) &=& 
         \<((\sqrt{v^{-1}})_1 \R^{-1}_1 \tw)\tr(...b_j a_i) 
       (\sqrt{v^{-1}})_2 \R^{-1}_2 \tw\tw^{-1}\sqrt{v}  a_k b_l...\> \nn\\
  &=& \<((\sqrt{v^{-1}})_1 \R^{-1}_1\tw)\tr(...b_j a_i) 
          (\sqrt{v^{-1}})_2 \R^{-1}_2\sqrt{v}  a_k b_l...\> \nn\\
  &=& \<(S^{-1}((\sqrt{v^{-1}})_2 \R^{-1}_2\sqrt{v})
       (\sqrt{v^{-1}})_1 \R^{-1}_1\tw)\tr(...b_j a_i) a_k b_l...\> \nn\\
  &=& \<(S^{-1}(\R^{-1}_2\sqrt{v})\eps(\sqrt{v^{-1}})\R^{-1}_1\tw)
         \tr(...b_j a_i) a_k b_l...\> \nn\\
  &=& \<(\sqrt{v} \R_2 S^2 \R_1 \tw)\tr(...b_j a_i) a_k b_l...\> \nn\\
  &=& \<(\sqrt{v} q^{-2\tilde{\rho}}v^{-1}\tw)
                           \tr(...b_j a_i) a_k b_l...\> \nn\\
  &=& \<(\sqrt{v} (-1)^E v^{-1} w)\tr(...b_j a_i) a_k b_l...\> \nn\\
  &=& \<(v^{-1}w)\tr(...b_j a_i) \sqrt{v} (-1)^E  a_k b_l...\>
 \label{inner_calc}
\err
using  (\ref{R_properties}) ff., (\ref{v_inv}), (\ref{tw}) and 
$\<(x\tr a)b\>= \<a(Sx) b\>$, $\eps(\sqrt{v^{-1}}) = 1$, 
which is easy to see.
In particular, with (\ref{g_transposed}) we can verify that 
\berr
 (a_i,a_j) \equiv  \<\obar{a_i}a_j\> &=& 
                      \< (v^{-1}w \tr a_i) \sqrt{v} (-1)^E a_j\> \nn\\
  &=&  g_a q^{\frac 12 c_a} (-1)^{E_j}(-1)^{f_a} g_{kj} \pi^k_i(w) \nn\\
  &=&  g_a q^{c_a} (-1)^{E_j+f_a} \pi^j_k((-1)^E w) \pi^k_i(w)\nn\\
  &=&  g_a \delta^i_j
\label{help_calc}
\err
is a (positive definite) inner product on the unitary \reps $a_i$;
it had to come out this way, since we always use the orthogonal basis
of Lemma \ref{orth_lemma}.

Using the first line of (\ref{help_calc}) and 
$\Del_{(n)}(v^{-1}w) = (v^{-1}w\tens...\tens v^{-1}w)\R_{(n)}$, 
we can continue (\ref{inner_calc}) as
\berr 
rhs &=& \<... (v^{-1}w)\tr b_s\pi^s_j(\R_{n-1})(v^{-1}w)\tr a_t\pi^t_i(\R_n)
            \sqrt{v} (-1)^E  a_k b_l...\> \nn\\
  &=& q^{\frac 12(c_a+c_b+...)} \pi^t_i(\R_n)\pi^s_j(\R_{n-1}) ... 
     (a_t\tens b_s\tens ..., \sqrt{v} \tr (a_k\tens b_l\tens...))_{\tens} 
\label{inner_calc_2}
\err
where $\R_{12...}$ are the components of $\R_{(n)}$ assuming there are $n$ 
factors, and using $\Del(-1)^E = (-1)^E \tens ... \tens (-1)^E$. 
Now notice that (\ref{help_calc}) and (\ref{R_l_theta}) imply 
\berr
\pi^t_i(\R_2) \pi^s_j(\R_1)(a_t\tens b_s,a_k\tens b_l)_{\tens} &=&
  g_a g_b \pi^k_i(\R_2) \pi^l_j(\R_1)  \nn\\
 &=&(a_i\tens b_j,a_t\tens b_s)_{\tens} \pi^t_k(\R_1) \pi^s_l(\R_2) \nn\\
 &=& (a_i\tens b_j,a_k\tens b_l)_{\R}
\err
and similarly for several factors, so (\ref{inner_calc_2}) is nothing but
\berr
 \quad rhs &=& c_V (a_i\tens b_j\tens...., 
               \sqrt{v}\tr(a_k\tens b_l\tens...))_{\R}  \nn\\
  &=& (a_i\tens b_j\tens...., a_k\tens b_l\tens...)_H
\err
as in (\ref{inner_H}).

\chapter{On the Weyl Element $w$}

In this appendix, we want to give some explanations to the remarkable
formulas (\ref{del_om}), (\ref{w_2}) and (\ref{autom_w}). 

As mentioned before, the braid group action (\ref{braid_action}) on $\U$ 
can in
fact be extended to a braid group action $T_i$ on any representation, 
with explicit formulas in terms of (infinite) sums of  generators of
$\U$ \cite{lusztig_book}, see also \cite{jantzen}. Then one can consider the 
element of the braid group corresponding to 
the longest element of the Weyl group, call it $T$. It was shown in 
\cite{kirill_resh,lev_soib} that these $T_i$ and therefore 
$T$ can be implemented by a
conjugation with $w_i$ resp. $w$, which are elements of an extension of $\U$,
and that $w$ satisfies (\ref{del_om}) with suitable 
definitions. Unfortunately, this requires complicated calculations.

To get some faith in these formulas, 
we want to point out first that if (\ref{del_om})
\beq
\Del(w) = \R^{-1} w\tens w
\label{del_om_left}
\eq
holds on the tensor product of 2 fundamental representations, it holds for
any representations. This can be seen inductively: if (\ref{del_om_left})
holds on $V_i\tens V_j$, then it also holds on the \reps in 
$V_1\tens V_2\tens V_3$, since the rhs of
\beq
({\id}\tens \Del)\Del(w) \stackrel{!}{=} \R^{-1}_{1,(23)} (w\tens w)
\eq 
acting on $V_1\tens(V_2\tens V_3)$ agrees with the rhs of
\beq
(\Del\tens {\id})\Del(w) \stackrel{!}{=} \R^{-1}_{(12),3} (w\tens w)
\eq
acting on $(V_1\tens V_2)\tens V_3$, by the first statement 
of Lemma \ref{R_l_lemma}.  Thus the lhs agree as they should.

Furthermore, since the action of $w$ on any given (finite--dimensional)
\rep can be expressed
in terms of generators of $\U$, it satisfies 
$\Del'(w) = \R \Del(w) \R^{-1}$, and together with (\ref{del_om_left})
this implies $\Del(w) = w\tens w \R_{21}$.

The action of $w_i$ resp. $w$ on the fundamental \reps can be found 
explicitely, see \cite{resh_1}. For real groups, it is essentially 
the invariant metric $g_{ij}$.

(\ref{w_2}) together with the statement that $\vareps = \pm 1$ on any irrep
now follows easily, since $\vareps \equiv v^{-1} w^2$ is grouplike as already 
pointed out, i.e. $\Del(\vareps) = \vareps\tens\vareps$, and it only remains
to check that $\vareps$ is $\pm 1$ on the fundamental representations, where
it agrees with the classical limit. This also follows from consistency with 
the tensor product.

\paragraph{A proof of $w x w^{-1} = \theta S x$ for real groups.}

Here we only consider the case of "real" groups, which are all except 
$A_n$, $D_n$ and $E_6$, i.e. those for which the 
Dynkin diagram does not have an automorphism. 

From (\ref{del_om}), (\ref{R_theta}) and (\ref{R_properties}), we know that
\beq
(w\tens w)\R (w^{-1}\tens w^{-1}) = \R_{21} = (\theta S \tens \theta S)\R
\label{autom_R}
\eq
Furthermore, it is known that the first terms of the expansion of $\R$
in powers of $X_i^{\pm}$ are 
\beq
\R = q^{\sum (\a^{-1})_{ij} h_i\tens h_j}
           \(1 + \sum_{i} c_{i}(q) q^{\frac 12 h_i} 
                     X^+_{i}\tens q^{-\frac 12 h_i}X^-_{i} + ...\) 
\eq
Now $w X_i^{\pm} w^{-1}$ certainly  has the form $\U^0 X_i^{\mp}$, as can
be seen either from the explicit formulas in \cite{kirill_resh,lev_soib}
or from the cross product algebra (\ref{smash}). On the Cartan
subalgebra, $w$ acts as classically. 
Thus by the uniqueness of $\R$ \cite{tolstoi,drinfeld_2}, 
(\ref{autom_R}) implies that 
\beq
w X_i^{\pm} w^{-1} = a_i \theta S (X_i^{\mp})
\eq
and 
\beq
w X_i^{\pm} w^{-1} = a_i^{-1} \theta S (X_i^{\mp})
\eq
with a constant $a_i$. Notice that if there exists an automorphism of
the Dynkin diagram, then this would only follow up to a corresponding
permutation of the simple root vectors. 
The constants $a_i$ can be eliminated by a redefinition
$w\rightarrow w q^{b_i H_i}$ with suitable constants $b_i$, using the
fact that the Cartan matrix is non--singular. This new $w$ now
satisfies all the equations discussed.

%\bibliographystyle{unsrt}
%\bibliography{thes}

\end{document}